\documentclass[apsrev4-1,prx,superscriptaddress,floatfix,twocolumn,longbibliography]{revtex4-1}

\usepackage{amsmath}
\usepackage{amssymb}
\usepackage[english]{babel}
\usepackage[T1]{fontenc}
\usepackage{graphics}
\usepackage{graphicx}
\usepackage[colorlinks,bookmarks=false,citecolor=blue,linkcolor=red,urlcolor=blue]{hyperref}
\usepackage{multirow}
\usepackage[caption=false]{subfig}
\usepackage{xcolor}

\begin{document}

\captionsetup[subfloat]{position=top}

\title{Competing spin liquids and hidden nematic order in spin ice with 
frustrated transverse exchange}

\author{Mathieu Taillefumier}
\affiliation{Okinawa Institute of Science and Technology Graduate University, 
Onna-son, Okinawa 904-0495, Japan}

\author{Owen Benton}
\affiliation{RIKEN Center for Emergent Matter Science (CEMS), Wako, Saitama, 351-0198,
Japan}

\author{Han Yan}
\affiliation{Okinawa Institute of Science and Technology Graduate University, 
Onna-son, Okinawa 904-0495, Japan}

\author{L. D. C. Jaubert}
\affiliation{CNRS, Universit\'e de Bordeaux, LOMA, UMR 5798, 33400 Talence, France}

\author{Nic Shannon}
\affiliation{Okinawa Institute of Science and Technology Graduate University, 
Onna-son, Okinawa 904-0495, Japan}

\begin{abstract}

Frustration in magnetic interactions can give rise 
to disordered ground states with subtle and beautiful properties.
The spin ices Ho$_2$Ti$_2$O$_7$ and Dy$_2$Ti$_2$O$_7$ 
exemplify this phenomenon, displaying a classical spin liquid state, 
with fractionalized magnetic--monopole excitations.   
Recently there has been great interest in closely--related ``quantum
spin ice'' materials, following the realization that anisotropic
exchange interactions could convert spin ice into a
massively--entangled, quantum, spin liquid, where magnetic monopoles
become the charges of an emergent quantum electrodynamics.
Here we show that even the simplest model of a quantum 
spin ice, the XXZ model on the pyrochlore lattice, can realise a still--richer 
scenario.
Using a combination of classical Monte Carlo simulation, 
semi--classical molecular--dynamics simulation, and analytic field theory, 
we explore the properties of this model for frustrated transverse exchange.
We find not one, but three, competing forms of spin liquid, 
as well as a phase with hidden, spin--nematic, order.
We explore the experimental signatures of each of these different 
states, making explicit predictions for inelastic neutron scattering. 
These results show an intriguing similarity to experiments on a 
range of pyrochlore oxides.  

\end{abstract}

\maketitle

\section{Introduction}
\label{sec:intro}


The search for spin liquids --- disordered phases of magnets which support entirely 
new forms of magnetic excitation --- has become one of 
the defining themes of modern condensed--matter physics \cite{anderson73,lee08,balents10}.
In this context, the pyrochlore lattice, a corner--sharing network of 
tetrahedra found in a wide range of naturally--occurring minerals, has 
proved an amazing gift to science.
Pyrochlore magnets play host to a variety of unconventional forms 
of magnetic order, and include systems which have not been observed to order 
at any temperature [\onlinecite{gardner10}].  
Perhaps the most celebrated of these is the  ``spin ice'' found in the 
Ising magnets  Ho$_2$Ti$_2$O$_7$ and Dy$_2$Ti$_2$O$_7$ [\onlinecite{bramwell01}]; 
a classical spin liquid, described by an emergent U(1) lattice gauge theory 
with magnetic monopole excitations 
[\onlinecite{castelnovo12}].   


As the understanding of spin ice has grown, so more attention has 
been given to the role of quantum effects.
These are of particular relevance where a spin--ice arises through
anisotropic exchange interactions in a pyrochlore magnet
\cite{curnoe07,molavian07, Onoda2010, Onoda2011a,
ross11-PRX1,mcclarty14}, and have
the potential to convert classical spin ice into a
massively--entangled, quantum, spin liquid, described by an emergent
U(1) quantum electrodynamics
[\onlinecite{hermele04,banerjee08,shannon12,benton12,savary12-PRL108,Lee12,mcclarty14,Hao2014,mcclarty15,kato15,chen16,shannon-book-chapter, savary17-PRL118,chen-arXiv.1706.04333,huang-arXiv.1707.00099}].
At the same time, great progress has been made in synthesizing
and characterizing magnetic pyrochlore oxides.
As well as revealing a number of candidates for quantum spin--ice 
behaviour \cite{Zhou2008,chang12-NatCommun3,
Fennell2012,Kimura2013,Sibille2015,Sibille16a,Anand16a,Wen2017,sibille-arXiv}, 
these experiments have turned up many unusual and 
unexpected magnetic states in systems with strongly anisotropic 
exchange~\cite{DalmasDeReotier2006,dun13,
Yaouanc2013a,taniguchi13,Hallas2014,hallas15,hallas16-PRB93.100403,Petit2016,petit16-PRB94,Takatsu16,hallas-arXiv.1708.01312}.
 
\begin{figure*}
  \centering
  \subfloat[Finite--temperature phase diagram]
  {
    \includegraphics[scale=1]{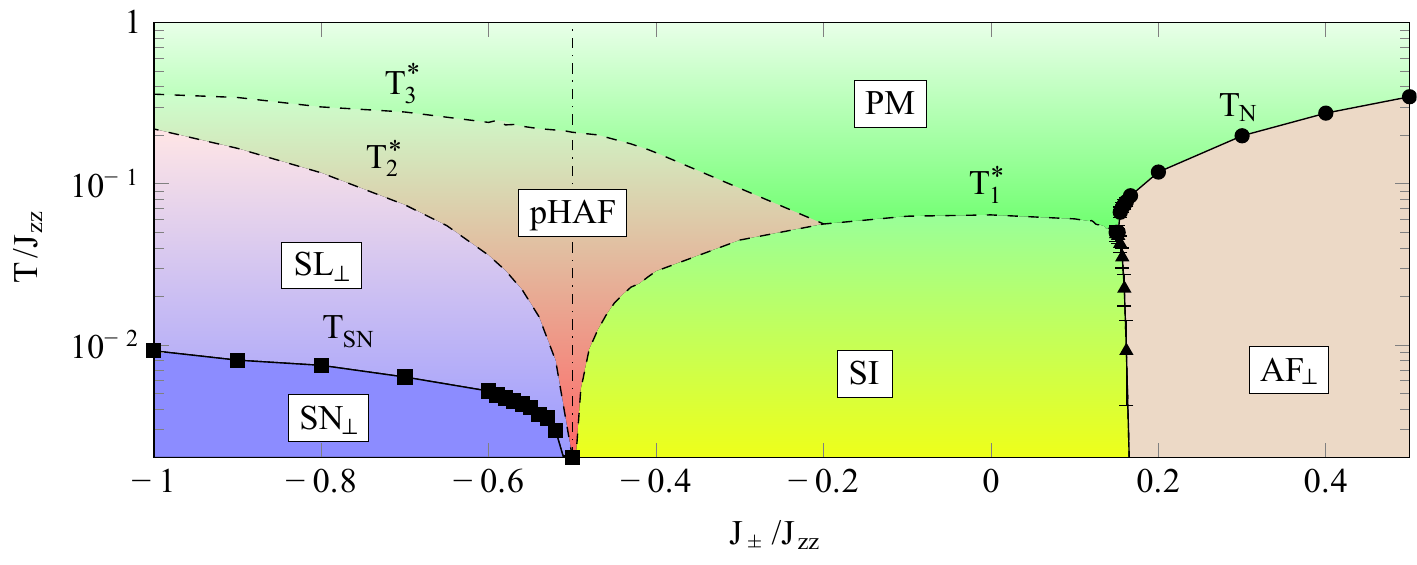}} \\
  \subfloat[Easy--plane spin--nematic ($\text{SN}_\perp$)]{\includegraphics[width=0.27\textwidth]{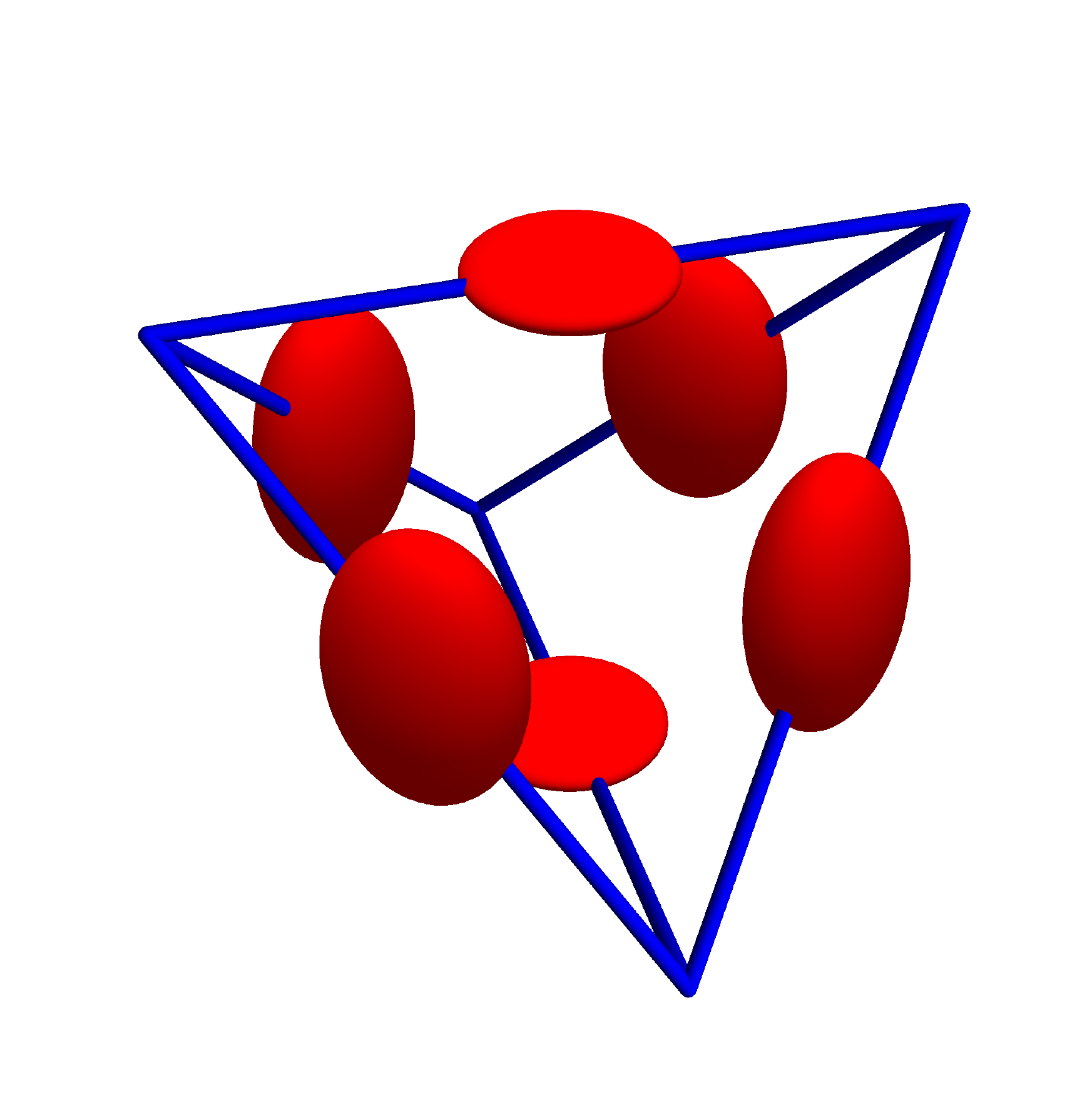}} \ \ 
  \subfloat[Spin Ice                  (SI)]
{\includegraphics[width=0.24\textwidth]{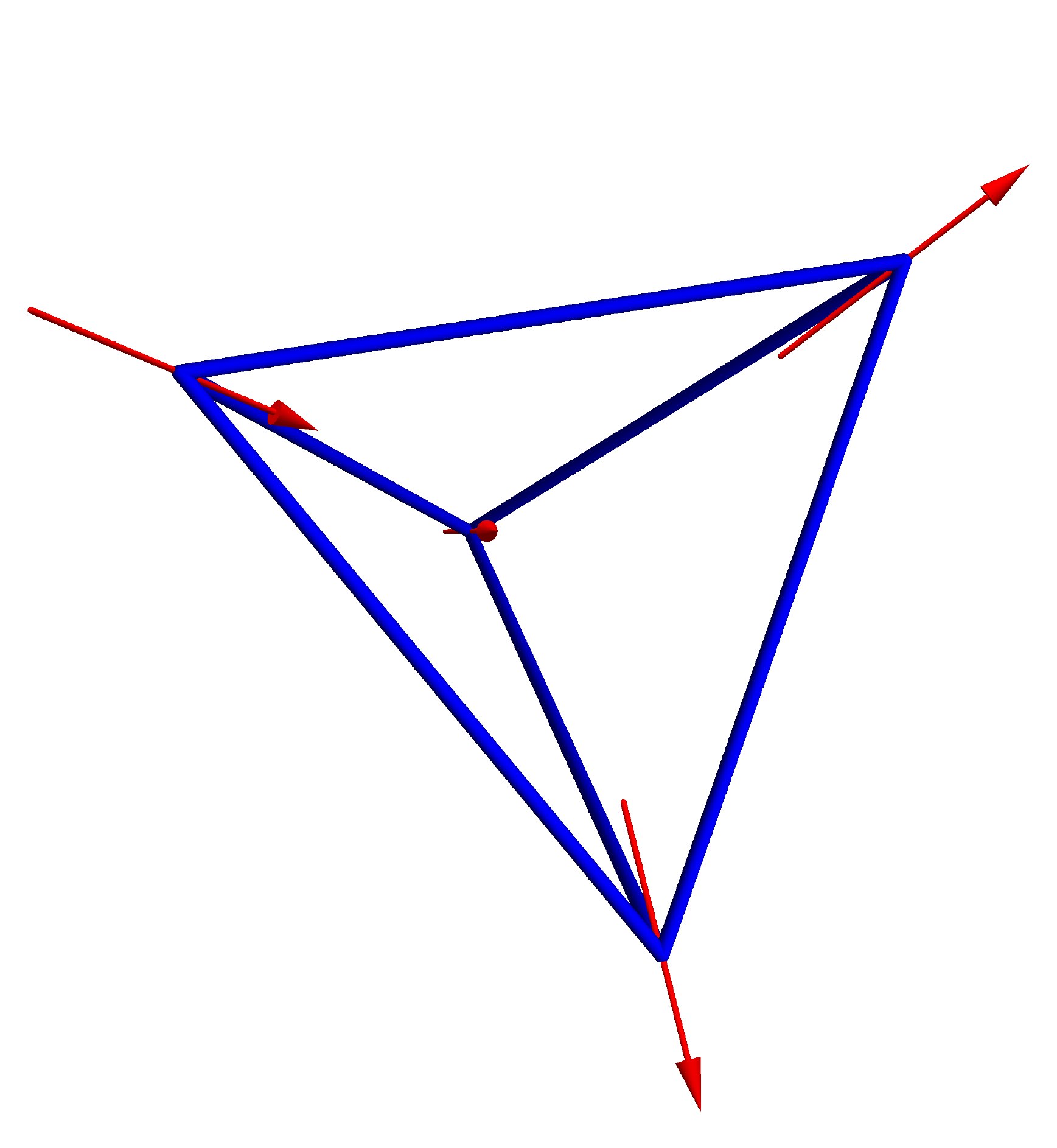}} 
 \subfloat[Easy--plane antiferromagnet (AF$_{\perp}$)]
{\includegraphics[width=0.24\textwidth]{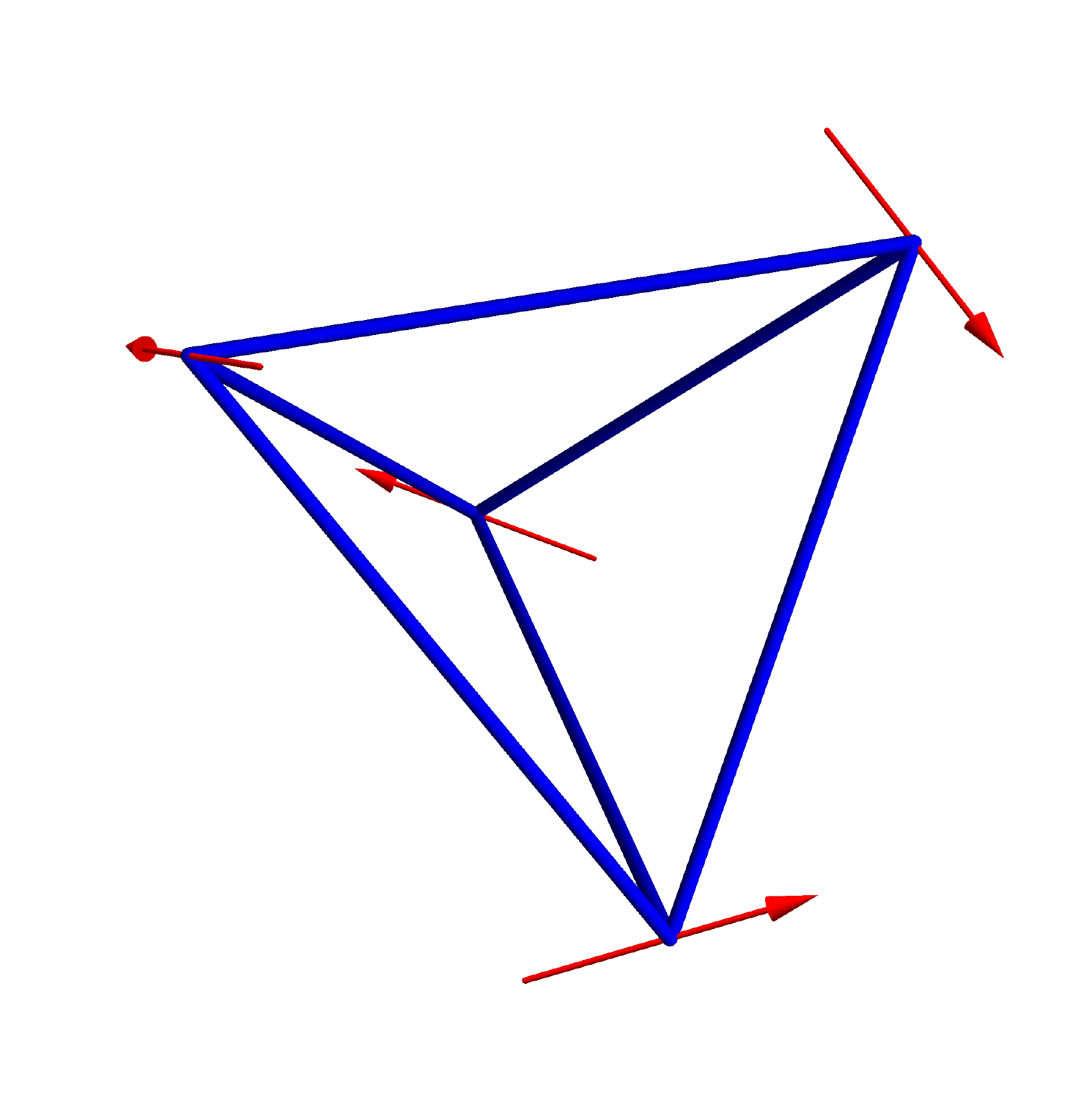}}
  \caption{
  (a) Finite--temperature phase diagram of the 
  minimal model of a quantum spin ice,  
  $\mathcal{H}_{\sf XXZ}$ [Eq.~(\ref{eq:H.xxz})], 
  as found in classical Monte Carlo simulation.
  %
  The model possesses three distinct spin--liquid regimes, 
  spin ice (SI), an easy plane spin liquid ($\text{SL}_\perp$) 
  and the pseudo--Heisenberg antiferromagnet (pHAF), 
  as well as phases with easy--plane antiferromagnetic ($\text{AF}_\perp$) 
  and easy--plane spin--nematic ($\text{SN}_\perp$) order.
  Associated crossover (phase-transition) temperatures are 
  indicated with dashed (solid) lines.
  For \mbox{$J_\pm/J_{\sf zz} = -1/2$} (dash--dotted line), the model 
  has an SU(2) symmetry and
  is thermodynamically equivalent to the Heisenberg antiferromagnet on a pyrochlore lattice.
   (b) Representation of bond quadrupolar order in the easy--plane spin--nematic 
   phase ($\text{SN}_\perp$) [cf. Appendix~\ref{app:coordinates}].
%
%
  $\text{SN}_\perp$ breaks both the $U(1)$ spin rotational symmetry of
  Eq.~(\ref{eq:H.xxz}) and the point group symmetry of the lattice.
   (c) ``Two--in, two--out'' configuration of spins in the spin ice regime (SI).
   (d) Representative configuration of spin dipoles in the easy--plane antiferromagnet 
   ($\text{AF}_\perp$)}.
\label{fig:phase.diagram}
\end{figure*}
\begin{figure*}
  \centering
  \subfloat[SI ($J_\pm = 0$, $T = 0.05 J_{\sf zz}$) \label{fig:Sq.spin.ice}]{\includegraphics[scale=0.75]{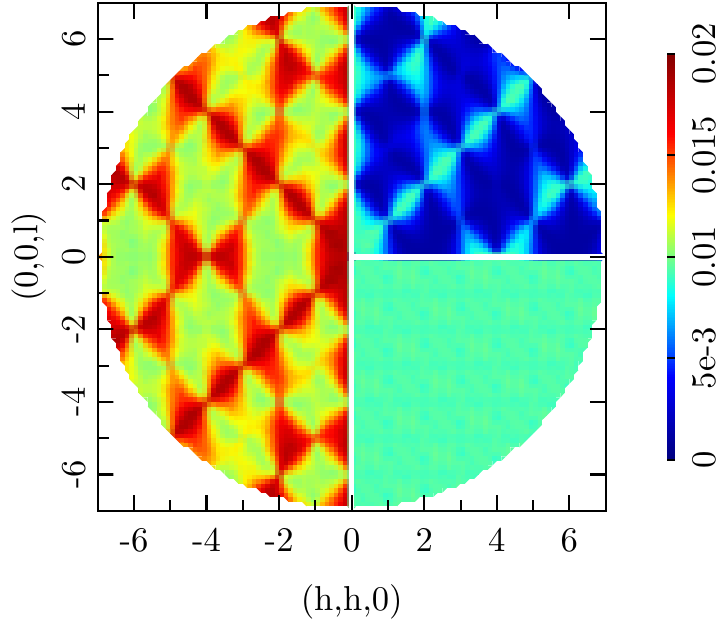}}
  \subfloat[$\text{SN}_{\perp}$, $\text{SL}_{\perp}$ ($J_\pm = - J_{\sf zz}$, $T = 0.005 J_{\sf zz}$)\label{fig:Sq.Q}.]{\includegraphics[scale=0.75]{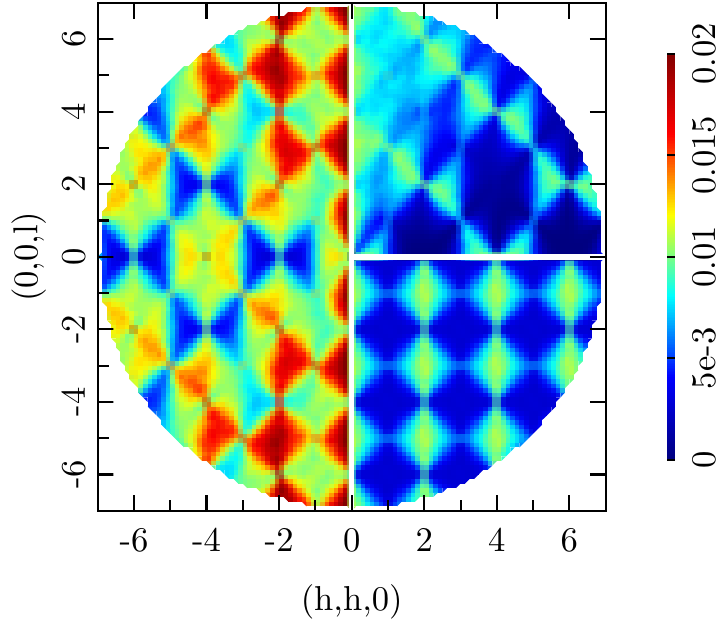}}
  \subfloat[pHAF ($J_\pm = -0.5 J_{\sf zz}$, $T = 0.05 J_{\sf zz}$\label{fig:Sq.pHAF})]{\includegraphics[scale=0.75]{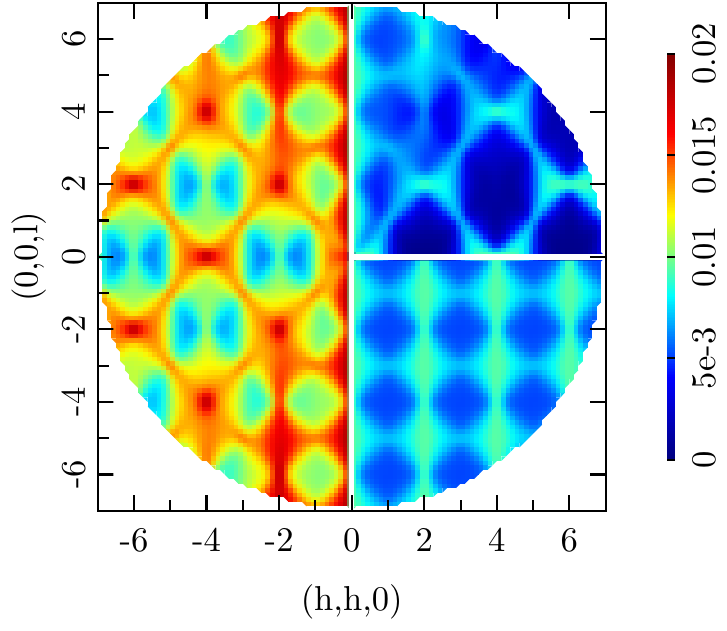}}
  \caption{
  Comparison of correlations in different spin--liquid and spin--nematic regimes.  
Left half of each panel : equal--time structure factor $S({\bf q})$, as measured in unpolarised 
  neutron--scattering experiments. 
  Right half of each panel : $S({\bf q})$ resolved into spin--flip (SF, top) and non spin--flip (NSF, bottom) 
  components, as measured in polarised neutron--scattering experiments (cf. [\onlinecite{fennell09}]).   
  a) Spin ice (SI), showing ``pinch points'' indicative of algebraic spin correlations.
  Definitions of these structure factors are given in Appendix~\ref{app:Sq}.
  b) Phase with easy--plane spin--nematic order ($\text{SN}_{\perp}$), showing the 
  absence of magnetic Bragg peaks, and persistence of algebraic spin correlations.
  The same correlations are also observed in the easy--plane spin liquid ($\text{SL}_\perp$).  
  c) Spin liquid associated with the pseudo--Heisenberg antiferomagnet (pHAF), 
  showing algebraic spin correlations distinct from those in spin ice 
  (SI) or the easy--plane spin liquid ($\text{SL}_\perp$).   
  Results are taken from Monte Carlo simulations of $\mathcal{H}_{\sf XXZ}$ [Eq.~(\ref{eq:H.xxz})], 
  for a cubic cluster of  $N=8192$ spins.   
  \label{fig:predictions.for.Sq}
  }
\end{figure*}

The main message of this Article is that even the simplest model of a 
 quantum spin ice --- the XXZ model on a pyrochlore lattice --- 
has far more to offer than spin ice alone.
Working in the classical limit, accessible to large--scale simulation, 
we find that frustrated transverse exchange gives rise to not one, 
but three, distinct, spin--liquid regimes [Fig.~\ref{fig:phase.diagram}]. 
We explore the way in which these spin liquids compete, 
identify the different gauge groups associated with each 
spin liquid, and make explicit predictions for their experimental 
signatures [Fig.~\ref{fig:predictions.for.Sq}].
We find that one of these spin liquids posses a highly--unusual
U(1)$\times$U(1) gauge structure and, as an added bonus, 
undergoes a 
phase transition into a state with hidden, spin--nematic, order.
We also use molecular dynamics simulations to characterise the 
excitations of this spin--nematic phase [Fig.~\ref{fig.predictions.for.dynamics}].
The portrait which emerges has striking similarities with the 
behavior of a number of pyrochlore materials.


The simplest model able to capture quantum effects in
a spin ice \cite{curnoe07,molavian07, Onoda2010, Onoda2011a,
ross11-PRX1,mcclarty14} is the XXZ model on 
the pyrochlore lattice
\begin{equation}
  \mathcal{H}_{\sf XXZ} = \sum_{\left<ij\right>} 
  J_{\sf zz} \mathsf{S}^z_i \mathsf{S}^z_j 
  - J_\pm \left( \mathsf{S}^+_i \mathsf{S}^-_j 
  + \mathsf{S}^-_i\mathsf{S}^+_j \right) 
    \label{eq:H.xxz}
\end{equation}
where $\mathsf{S}_i = (\mathsf{S}^x_i, \mathsf{S}^y_i, \mathsf{S}^z_i)$ 
is a (pseudo)spin--half operator describing the two states 
of the lowest energy doublet of a magnetic ion subject to a
strong crystal electric field (CEF).
The symmetry of the lattice requires that the 
quantization axis of each spin (here, $\mathsf{S}^z_i$)  
lies on a local [111] axis, as defined in Appendix~\ref{app:coordinates}.
%


Ising interactions, $J_{\sf zz}  > 0$, 
favor states obeying the ``ice rules'' in which two spins point into,
and two spins point out of, each tetrahedron on the lattice.  
The transverse term,  
$J_\pm$,  introduces dynamics about these spin--ice configurations and, 
for larger, positive, values of $J_\pm/J_{\sf zz}$, drives the system 
into a state with easy--plane order 
\cite{onoda11,Lee12,savary12-PRL108, wong13, yan17, owen-thesis}.  
The physical meaning of this easy--plane order depends on the nature 
of the magnetic ion.   
For Kramers ions like Yb$^{3+}$ and Er$^{3+}$ all components of $\sf S$
relate to a magnetic dipole moment \cite{ross11-PRX1}, and the ordered
phase is an easy--plane antiferromagnet.
However for non--Kramers ions such as Pr$^{3+}$ and Tb$^{3+}$ 
\cite{Onoda2010, petit16-PRB94}, or ``dipolar-octupolar'' Kramers ions 
like Nd$^{3+}$ or Ce$^{3+}$ \cite{huang14}, 
the easy--plane order may have quadrupolar (octupolar) character.
In what follows, we consider explicitly the case of Kramers ions.
However, suitably reinterpreted, these results also have important 
implications for non--Kramers ions.


For $J_{\pm} > 0$, $\mathcal{H}_{\sf XXZ}$ [Eq.~(\ref{eq:H.xxz})] 
is unfrustrated, in the sense that it is free of sign problems in 
Quantum Monte Carlo (QMC) simulation.   
In this case, the phase diagram is already 
well--established~\cite{banerjee08,kato15,shannon-book-chapter}.  
For \mbox{$J_{\pm}/J_{\sf zz} \lesssim 0.05$}, QMC simulations find a crossover from 
a conventional paramagnet into a classical spin--liquid (spin ice) at a temperature 
$T^*/J_{\sf zz} \sim 0.2$, and a second crossover into a quantum spin liquid (QSL) 
at a much lower temperature \mbox{$T_{QSL}^*/J_{\sf zz} \sim (J_{\pm}/J_{\sf zz})^3$}.  
In the low temperature quantum spin liquid regime, the magnetic monopoles 
of classical spin ice become dynamic, fractional, spin excitations 
(spinons), while the spectrum of the model also includes gapless photons~
\cite{hermele04,benton12}.  
For \mbox{$J_{\pm}/ J_{\sf zz} \gtrsim  0.05$}, the U(1) QSL gives way to easy--plane 
antiferromagnetic order (AF$_{\sf \perp}$), in which spins lie in the plane 
perpendicular to the local $\mathsf{S}^z$--axis~\cite{banerjee08,kato15,shannon-book-chapter}.   


Very little is known about the properties of $\mathcal{H}_{\sf XXZ}$
for $J_{\pm} < 0$ \cite{Onoda2011a,Lee12,petit16-PRB94}.
On perturbative grounds, it is expected that the ground state for 
$|J_{\pm}|/J_{\sf zz} \ll 1$ will also be a U(1) QSL \cite{hermele04}, 
albeit one with a modified spinon dispersion \cite{Lee12,chen-arXiv.1704.02734}.
Gauge mean--field calculations suggest that this QSL persists over a broad range 
of parameters, \mbox{$-4.13 \lesssim J_{\pm}/ J_{\sf zz} < 0$} [\onlinecite{Lee12}].
But the nature of competing ordered --- or disordered --- phases 
for $J_{\pm} < 0$ remains an open question.


There are many reasons to believe that the properties of the quantum spin ice 
model, $\mathcal{H}_{\sf XXZ}$ [Eq.~(\ref{eq:H.xxz})] for  frustrated coupling 
\mbox{$J_{\pm} < 0$}, could be even richer than for \mbox{$J_{\pm} > 0$}.   
In particular, for $J_{\pm}/J_{\sf zz} = -\frac{1}{2}$, $\mathcal{H}_{\sf XXZ}$ 
[Eq.~(\ref{eq:H.xxz})] is equivalent (up to a site--dependent spin--rotation), to the 
Heisenberg antiferromagnet (HAF) on a pyrochlore lattice.
Like spin ice, the HAF is known to support a classical spin liquid  
\cite{moessner98-PRL80,moessner98-PRB58,Isakov2004a,henley05},  
and it has also been argued to support a QSL ground state \cite{canals98,
canals00,burnell09,Huang2016}.
And, crucially, both the classical and quantum spin liquids in the HAF have a qualitatively 
different character from those found in spin ice.
This sets up a competition between two different kinds of spin liquid,  
namely spin ice for $J_{\pm}/J_{\sf zz} \approx 0$, 
and a state homologous to the HAF for $J_{\pm}/J_{\sf zz} \approx -\frac{1}{2}$.
It also opens the door for yet more novel magnetic phases 
for $J_{\pm}/J_{\sf zz} < -\frac{1}{2}$.


\section{Phase diagram determined by classical Monte Carlo simulations}
\label{sec:phasediagram}


Since the quantum spin ice model, $\mathcal{H}_{\sf XXZ}$ [Eq.~(\ref{eq:H.xxz})], 
is inaccessible to QMC for \mbox{$J_{\pm} < 0$}, we instead study its finite--temperature 
properties using classical Monte Carlo (MC) simulation --- the results are summarised 
in the phase diagram Fig.~\ref{fig:phase.diagram}.   
For $J_{\pm} > 0$, this phase diagram is very similar to that previously 
found in QMC simulations \cite{banerjee08,kato15,shannon-book-chapter}
--- at a qualitative level, the only significant difference is the absence 
of a QSL below \mbox{$T_{QSL}^*/J_{\sf zz} \sim (J_{\pm}/J_{\sf zz})^3 \lesssim 0.005$}.
At a quantitative level, we find changes in numerical values of the 
crossover temperature associated with the spin ice regime, $T^*_1$, 
and the
position of the
zero--temperature boundary between 
SI and $\text{AF}_\perp$.
These changes can be ascribed to the fact that
the magnetic monopoles (spinons) are not quantized in
classical simulations and do not develop phase coherence 
[\onlinecite{owen-unpublished}].   
Further details of classical MC simulations for \mbox{$J_\pm > 0$} 
will be presented  elsewhere [\onlinecite{taillefumier-in-preparation}].


We now turn to the frustrated case, \mbox{$J_{\pm} < 0$}.
At low temperatures, spin--ice correlations persist up to 
\mbox{$J_{\pm}/J_{\sf zz} = -\frac{1}{2}$} \cite{Onoda2011a,petit16-PRB94}, 
as illustrated in Fig.~\ref{fig:Sq.spin.ice}.
Upon reaching \mbox{$J_{\pm}/J_{\sf zz} = -\frac{1}{2}$} 
 the system  becomes thermodynamically equivalent to 
a HAF.
This high--symmetry point gives rise to a new form of spin liquid at finite 
temperature, 
labelled a pseudo--Heisenberg antiferromagnet (pHAF) 
in Fig.~\ref{fig:phase.diagram}.  
Once again, this spin liquid has algebraic correlations, as shown in 
Fig.~\ref{fig:Sq.pHAF}, but with qualitatively different character  
from spin ice [Fig.~\ref{fig:Sq.spin.ice}].
These correlations persist up to a crossover temperature $T^*_3$ 
associated with the Curie--law crossover (CLC) in the magnetic 
susceptibility \cite{jaubert13}.


While the correlations measured in the equal--time structure factor $S({\bf q})$
are also different from those found in the HAF 
\mbox{\cite{moessner98-PRB58,Isakov2004a,conlon09}}, 
the two models are equivalent up to a local coordinate transformation.     
And, by analogy with earlier work on the HAF \cite{Isakov2004a,henley05,henley10}, 
the spin liquid pHAF can be described by a 
U(1)$\times$U(1)$\times$U(1) gauge theory.
%


The situation for \mbox{$J_{\pm}/J_{\sf zz} < -\frac{1}{2}$} is even more interesting.
Below a second crossover scale, $T^*_2 < T^*_3$, 
identifiable by a reduction in the fluctuations of the $z$-components
of the spins [see Appendix~\ref{app:phasediagram}], 
the pHAF gives way to an easy--plane spin liquid, 
labelled $\text{SL}_\perp$ in Fig.~\ref{fig:phase.diagram}.  
Spin correlations in this regime have algebraic character, 
with pinch--points in $S({\bf q})$ [Fig.~\ref{fig:Sq.Q}].   
However these correlations are qualitatively different from 
those in either spin ice [Fig.~\ref{fig:Sq.spin.ice}], 
or the pHAF [Fig.~\ref{fig:Sq.pHAF}].  
At a still lower temperature, $T_\text{SN}  < T^*_2$, the 
system undergoes a thermodynamic phase transition, 
marked by a clear anomaly in the specific heat.
None the less, this phase transition does not give rise to any magnetic 
Bragg peaks in $S({\bf q})$ and, at least as far as dipolar spin 
correlations are concerned, the system remains disordered.


While the new phase for $T < T_\text{SN}$ --- labelled $\text{SN}_\perp$ in
Fig.~\ref{fig:phase.diagram} --- does not exhibit any conventional
magnetic order, it does posses a hidden, spin--nematic, order.
The ordered state does not break translational symmetry, but breaks 
the U(1) symmetry of $\mathcal{H}_{\sf XXZ}$ [Eq.~(\ref{eq:H.xxz})] by 
selecting an axis within the local $xy$-plane.
Such an order can be described by the bond--based order 
parameter \cite{andreev84, chubukov91, shannon06}
\begin{eqnarray}
{\bf Q}_{\perp}=\sum_{\langle ij \rangle}
 \frac{1}{3N} 
\begin{pmatrix}
{\sf S}^x_i {\sf S}^x_j-{\sf S}^y_i {\sf S}^y_j \\
{\sf S}^x_i {\sf S}^y_j+{\sf S}^y_i {\sf S}^x_j 
\label{eq:bond.nematic}
\end{pmatrix} \; ,
\end{eqnarray}
where the sum on $\langle ij \rangle$ runs over the nearest--neighbour bonds 
of the lattice, and $\mathsf{S}_i = (\mathsf{S}^x_i, \mathsf{S}^y_i, \mathsf{S}^z_i)$ 
is expressed in the local frame of site $i$ (cf. Appendix~\ref{app:coordinates}).  
%



This type of easy--plane order is formally 
identical to the spin--nematic 
phases found in a range of frustrated magnets in applied magnetic field 
[\onlinecite{shannon06,shannon10,smerald15}].
And in common with these systems, the associated Landau theory 
\begin{eqnarray}
F_{\text{$\text{SN}_\perp$}} 
   = a_2(T)\  {\bf Q}_\perp^2\ 
    + a_4\  {\bf Q}_\perp^4\  + \ldots  \; , 
\label{eq:mQlandau}
\end{eqnarray}
lacks a cubic term, and therefore permits a continuous phase transition.
Simulations suggest that the phase transition at $T=T_\text{SN}$ is indeed 
continuous for \mbox{$J_{\pm}/J_{\sf zz} \lesssim -\frac{1}{2}$}, becoming 
first-order approaching the high--symmetry point \mbox{$J_{\pm}/J_{\sf zz} \to -\frac{1}{2}$}.
Further details of the thermodynamics of this transition are given in 
Appendix~\ref{app:phasediagram}.

\section{Theory of the easy--plane spin liquid}
\label{sec:SLII}


\begin{figure}[t]
  \includegraphics[width=8cm]{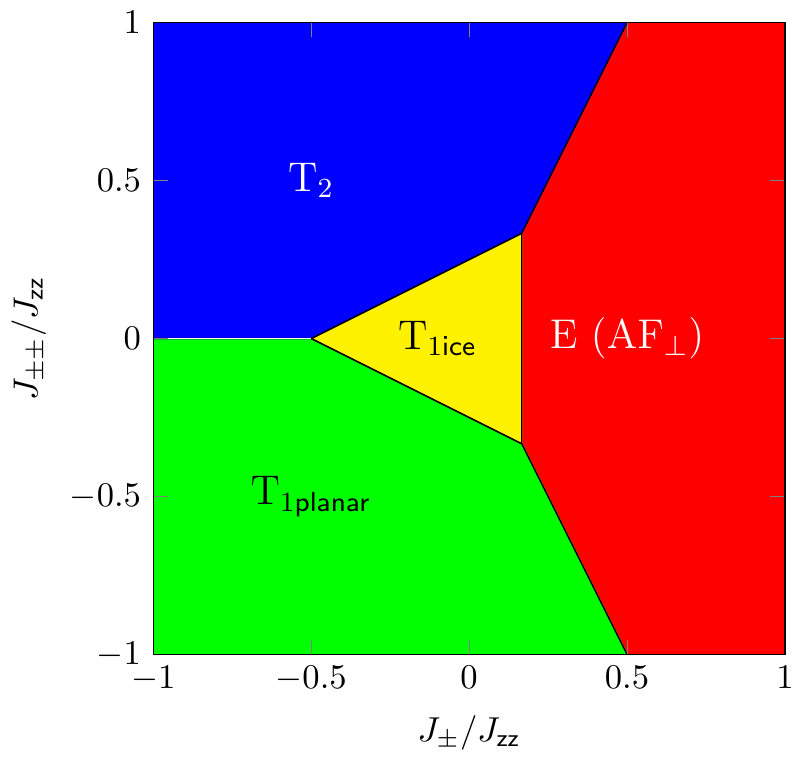}
  \caption{(Color online).
  Classical ground--state phase diagram of the anisotropic exchange 
 model ${\mathcal H}_{\sf ex}$ [Eq.~(\ref{eq:H.xxz+delta})], for 
  $J_{\sf zz} > 0$.   
  Different phases are labelled in terms of the irreps of the tetrahedral 
  symmetry group, $T_d$ [Eq.~(\ref{eq:irreps})], as described in \cite{yan17}.  
  The minimal model of a quantum spin ice $\mathcal{H}_{\sf XXZ}$ 
  [Eq.~(\ref{eq:H.xxz})] exists on the line $J_{\pm\pm} = 0$
  --- for $J_\pm < -\frac{1}{2}$ (white line), two phases with 4-sublattice easy--plane order 
  meet, 
  and the resulting enlarged ground--state manifold gives rise to the 
  easy plane spin liquid $\text{SL}_\perp$, and spin--nematic phase $\text{SN}_\perp$.
   A closely--related mean--field phase diagram for non--Kramers ions 
   is given in  \cite{Onoda2011a, petit16-PRB94}.}
  \label{fig:phase.diagram-T=0}
\end{figure}


Spin correlations in spin ice (SI) can be described using a U(1) lattice 
gauge theory  \cite{henley05, isakov05,castelnovo12}, which gives rise to 
characteristic ``pinch--points'' in the spin structure factor $S({\bf q})$ [Fig.~\ref{fig:Sq.spin.ice}].   
Meanwhile, for classical spins, spin correlations in the Heisenberg AF  
on the pyrochlore lattice --- and by extension in the pHAF --- can be 
described using a U(1)$\times$U(1)$\times$U(1) gauge theory 
\cite{moessner98-PRL80, moessner98-PRB58,Isakov2004a,henley05}.
The pHAF has qualitatively different pinch--points from spin ice,
as illustrated in Fig.~\ref{fig:Sq.pHAF}. 
It is clear  that the correlations of the easy--plane spin liquid, $\text{SL}_\perp$
[Fig.~\ref{fig:Sq.Q}] are very different from either spin ice [Fig.~\ref{fig:Sq.spin.ice}]
or the pHAF [Fig.~\ref{fig:Sq.pHAF}].
None the less, the presence of pinch points suggests that $\text{SL}_\perp$, too, 
may be described by some form of gauge theory.


We can develop a field--theory for the spin--liquid 
$\text{SL}_\perp$ by applying the 
methods developed in \cite{benton16-NatCommun7,owen-thesis}.    
The starting point of this approach is to recast the spins $\mathsf{S}_i$ 
in $\mathcal{H}_{\sf XXZ} $ [Eq.~(\ref{eq:H.xxz})] in terms of five 
order--parameter fields 
\begin{eqnarray}
\{ {\bf m}_{\lambda} \} 
=
\{
m_{\sf A_2}, {\bf m}_{\sf E} ,
{\bf m}_{\sf T_1 ice}, 
{\bf m}_{\sf T_1 planar} ,
{\bf m}_{\sf T_2}
\}
\label{eq:irreps}
\end{eqnarray}
defined on each tetrahedron ${\bf r}$.
These objects ${\bf m}_{\lambda}(\mathbf{r})$ describe the different
kinds of four-sublattice magnetic order consistent with the point
group symmetry of the pyrochlore lattice.
Definitions of each field ${\bf m}_{\lambda}$
in terms of the spins ${\sf S}_i$ are given in Appendix~\ref{app:m_lambda}.


The most general exchange Hamiltonian on the pyrochlore lattice 
can be transcribed exactly in terms of ${\bf m}_{\lambda}$ \cite{yan17}.
This greatly simplifies the determination of classical ground states and, 
where classical ground states form an extensive manifold, one can use this 
approach to determine the local constraints which control the resulting 
spin--liquid \cite{benton16-NatCommun7,owen-thesis}.
In the case of $\text{SL}_\perp$, for $T \to 0$, we have
\begin{eqnarray}
{m}_{\sf A_2}({\bf r})=0, \
{\bf m}_{\sf E}({\bf r})=0, \
{\bf m}_{\sf T_1 ice}({\bf r})=0 \ \ \forall {\bf r}
\label{eq:SLIIconstraints}
\end{eqnarray}


The spin fluctuations at low temperature are thus dominated
by the fluctuations of the remaining fields ${\bf m}_{\sf T_2}({\bf r})$
and ${\bf m}_{\sf T_1 planar}({\bf r})$.
These fields have significance as the order-parameters of the
competing four-sublattice magnetic orders which would be induced
by the symmetry--allowed perturbation 
\begin{eqnarray}
&&\delta\mathcal{H}_{\pm\pm}=\sum_{\langle ij\rangle} 
J_{\pm\pm} \left[\gamma_{ij} \mathsf{S}_i^+ \mathsf{S}_j^+ + \gamma_{ij}^*
                 \mathsf{S}_i^-\mathsf{S}_j^-\right] \; ,
\label{eq:Hpmpm}
\end{eqnarray}
where $\gamma_{ij}$ are complex phase factors 
arising from the change in coordinate 
frame between different lattice sites  
\cite{curnoe07,mcclarty09,onoda11, Onoda2010, Onoda2011a, ross11-PRX1}.   
For this reason, the spin--liquid $\text{SL}_\perp$ falls very naturally into 
the ``multiple--phase competition'' scenario for pyrochlore magnets 
\cite{yan-arXiv,jaubert15,owen-thesis,yan17}.   

 
In Fig.~\ref{fig:phase.diagram-T=0}, we show the classical ground--state 
phase diagram of anisotropic exchange model 
\begin{eqnarray}
&&\mathcal{H}_{\sf ex}=\mathcal{H}_{\sf XXZ}+\delta\mathcal{H}_{\pm\pm} \; .
\label{eq:H.xxz+delta}
\end{eqnarray}
This contains three distinct regions of 4--sublattice order~: the 
easy--plane ordered phases described by the fields ${\bf m}_{\sf E}$ 
(denoted $\text{AF}_\perp$ 
in Fig.~\ref{fig:phase.diagram}), ${\bf m}_{\sf T_1 planar}$, 
and ${\bf m}_{\sf T_2}$ (Palmer--Chalker state \cite{palmer00}).
These border a region of spin ice 
(denoted SI in Fig.~\ref{fig:phase.diagram}), dominated by 
fluctuations of ${\bf m}_{\sf T_1 ice}$.   
We note that a closely--related phase diagram has been derived  
for non--Kramers ions \cite{Onoda2011a, petit16-PRB94};  
in this case easy--plane order must be interpreted in terms 
of the quadrupole moment of the magnetic ion.   


The non-trivial correlations in the spin--liquid $\text{SL}_\perp$ arise from the
fact that neighbouring tetrahedra share a spin, so that the fields
${\bf m}_{\lambda}(\mathbf{r})$ on neighbouring tetrahedra are
not independent of one another.
This point, combined with Eq.~(\ref{eq:SLIIconstraints}), imposes
spatial constraints on the fluctuations of  ${\bf m}_{\sf T_2}({\bf r})$
and ${\bf m}_{\sf T_1 planar}({\bf r})$.
After coarse graining to extract the long wavelength physics
these constraints may be written in terms of two, independent, vector
fluxes
\begin{eqnarray}
 &&  {\bf B}_{\sf 1} =  \frac{1}{2}( 2 m_{\sf T_1 planar}^x,  -\sqrt{3}m_{\sf T_2}^y-m_{\sf T_1 planar}^y, 
\nonumber \\
&& \qquad \qquad
\sqrt{3} m_{\sf T_2}^z- m_{\sf T_1 planar}^z) \nonumber\\
 &&  {\bf B}_{\sf 2} =  \frac{1}{2}(2 m_{\sf T_1 planar}^x,  -m_{\sf T_2}^y+\sqrt{3}m_{\sf T_1 planar}^y,
\nonumber \\
&& \qquad \qquad
-m_{\sf T_2}^z - \sqrt{3}m_{\sf T_1 planar}^z) \; ,
\label{eq:fluxes}
\end{eqnarray}
which each separately obey their own Gauss' law
\begin{eqnarray}
   \nabla \cdot {\bf B}_{\sf 1}  =  0 
 \qquad , \qquad 
   \nabla \cdot {\bf B}_{\sf 2}  =  0  \; .
   \label{eq:div.m}
\end{eqnarray}


\begin{figure}
  \centering
\includegraphics[width=6cm]{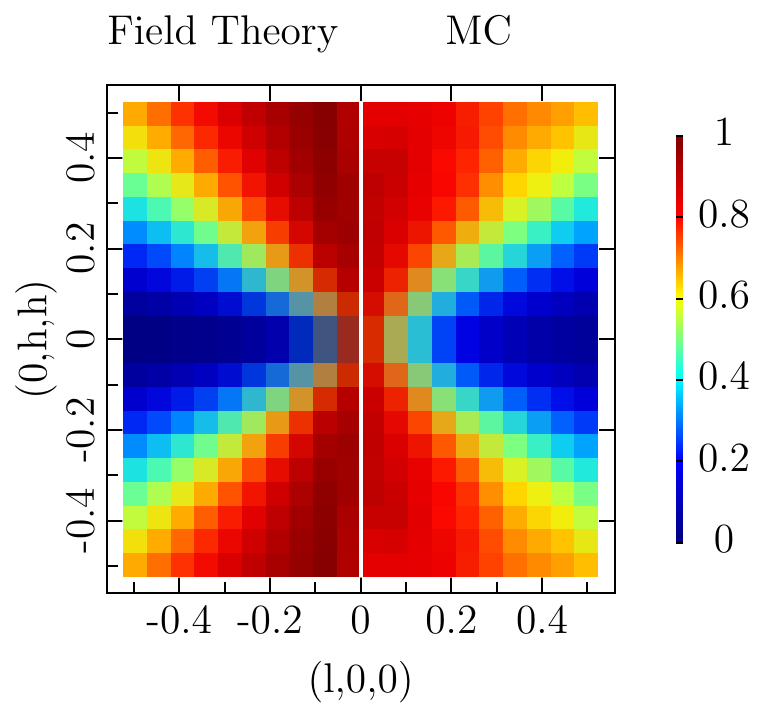}
  \caption{
   U(1)$\times$U(1) gauge structure of the easy--plane 
   spin liquid ($\text{SL}_\perp$), 
   as demonstrated by pinch--points in equal--time structure factors 
   $S^{\alpha\beta}_{{\bf B}_\mu} ({\bf q})$ [Eq.~(\ref{eq:S.mu})].    
   Left half of panel: structure factor $S^{xx}_{{\bf B}_1} ({\bf q})$
   as calculated from the theory Eq.~(\ref{eq:F}), with $\lambda=1$. 
   Right half of panel: structure factor $S^{xx}_{{\bf B}_1} ({\bf q})$ as calculated
   in classical Monte Carlo simulations of 
   $\mathcal{H}_{\sf XXZ}$ [Eq.~(\ref{eq:H.xxz})].  
   The pinch point centered on ${\bf q} = (0,0, 0)$, follows
   from the zero--divergence conditions on the fields ${\bf B}_\mu$ [Eq.~(\ref{eq:div.m})].   
   Simulations were carried out for a cubic cluster of $N=\text{8192}$ spins, 
   with $J^\pm/J_{zz} = -1$, $T=0.01 J_{\sf zz}$, as described in Appendix~\ref{app:fluxsim}.
    }
  \label{fig:pinch.point}
\end{figure}


We can therefore write 
\begin{eqnarray}
{\bf B}_{\sf 1}  =  \nabla \times{\bf A}_{\sf 1} 
\qquad , \qquad
{\bf B}_{\sf 2}  =  \nabla \times{\bf A}_{\sf 2}
\end{eqnarray}
and the theory has two, independent, U(1) gauge degrees of freedom. 


The free energy associated with the fluctuations of these fields is of 
entropic origin \cite{henley10}.
The only choice of Gaussian free-energy 
consistent with both the point group symmetry and the
$U(1)$ symmetry of $\mathcal{H}_{\sf XXZ}$ is
\begin{eqnarray}
&&{\mathcal F}_{\text{SL}_\perp} = \frac{T}{V} \int d^3r \
\lambda ({\bf B}_1^2+{\bf B}_2^2) \nonumber \\
&&\qquad
= \frac{T}{V} \int d^3r \
\lambda \left[(\nabla \times {\bf A}_1)^2+(\nabla \times {\bf A}_2)^2\right]
\label{eq:F}
\end{eqnarray}
where the coefficient $\lambda$ can be determined through fits
to simulation, or a large--N expansion \cite{Isakov2004a, benton16-NatCommun7}.  


It follows from the existence of the conserved fluxes ${\bf B}_1$
and ${\bf B}_2$ and the free-energy Eq.~(\ref{eq:F}) that $\text{SL}_\perp$ 
is a Coulomb phase with algebraic correlations \cite{henley10}.
The validity of this description is demonstrated in Fig.~\ref{fig:pinch.point}
where we compare analytic calculations of the flux structure factor
\begin{eqnarray}
S_{{\bf B}_{\mu}}^{\alpha \beta}(\mathbf{q})=
\langle
B^{\alpha}_{\mu}(-\mathbf{q})
B^{\beta}_{\mu}(\mathbf{q})
\rangle
\label{eq:S.mu}
\end{eqnarray}
based on Eq.~(\ref{eq:F}) with the results of Monte Carlo
simulation.
Pinch point singularities are clearly seen in both analytic
and numerical calculations.
It is the same fluctuations of ${\bf B}_1$
and ${\bf B}_2$ which are responsible for the characteristic pinch--point 
structures in the (spin) structure factor 
measured by neutron scattering, as shown in Fig.~\ref{fig:Sq.Q}.


At finite temperature, we anticipate that the spin liquid $\text{SL}_\perp$ 
will be perturbatively stable against terms such as $\delta\mathcal{H}_{\pm\pm}$ 
[Eq.~(\ref{eq:Hpmpm})], which retain the point--group symmetry of the lattice but lift 
the $U(1)$ symmetry of the spins.   
In this case the free energy will be modified :
\begin{eqnarray}
&&{\mathcal F}_{\text{SL}_\perp} 
   \to {\mathcal F}_{\text{SL}_\perp} + \delta{\mathcal F}_{\text{SL}_\perp} \\
&&
\delta{\mathcal F}_{\text{SL}_\perp}
= \frac{T}{V} \int d^3r \
\lambda^{\prime}
 \left\{ 
 \left( B_1^x \right)^2
 -\frac{1}{2}
   \left[ \left( B_1^y \right)^2 
 +  \left( B_1^z \right)^2 \right]
 \right.
\nonumber \\
&&
\left. 
  - \left( B_2^x \right)^2
   + \frac{1}{2} 
      \left[ \left( B_2^y \right)^2 + \left( B_2^z \right)^2 \right] 
- \sqrt{3} 
\left[  B_1^z B_2^z  -   B_1^y B_2^y \right] \right\} \nonumber \\
\label{eq:deltaF}
\end{eqnarray}
This form of free energy will still lead to pinch points in 
$S_{{\bf B}_{\mu}}^{\alpha \beta}(\mathbf{q})$ and $S({\bf q})$, 
but these will take on a more anisotropic character.

\section{Dynamics in the spin--nematic phase}
\label{sec:Qdynamics}


\begin{figure*}
  \centering
  \subfloat[Dynamical structure factor for spins, 
  $S({\bf q}, \omega)$ \label{fig:S.q.omega}]{\includegraphics[width=16cm]{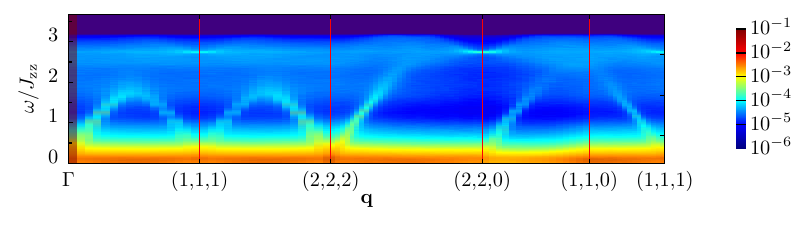}}  \\
  \subfloat[Dynamical susceptibility for quadrupoles, 
  $\chi_{Q_\perp^{\sf site}}({\bf q}, \omega)$ \label{fig:chi.Q.q.omega}]{\includegraphics[width=16cm]{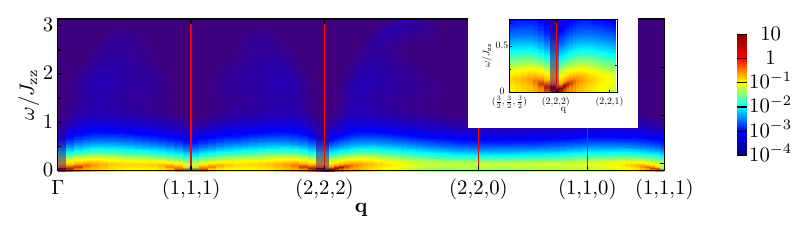}}
  \caption{
  Spin dynamics in the phase with hidden spin--nematic order ($\text{SN}_\perp$).    
  (a) Dynamical structure factor for spin degrees of freedom, 
  $S({\bf q}, \omega)$, 
  showing gapless continuum of excitations at low energies.   
  (b) Dynamical susceptibility for fluctuations of on--site quadrupole moments, 
  $\chi_{\sf Q_\perp^{\sf site}}({\bf q}, \omega)$ [cf. Appendix~\ref{app:goldstone}].     
  Inset :  details of correlations at low energies 
  near the zone center ${\bf q}_{\sf rl} = (2,2,2)$, showing a linearly--dispersing 
  Goldstone mode at low energies.
  Results are taken from molecular-dynamics simulations of $\mathcal{H}_{\sf XXZ}$ [Eq.~(\ref{eq:H.xxz})] 
  for a cluster of $N=65536$ spins, with $J_\pm/J_{\sf zz} = -1.0$, $T/J_{\sf zz} = 0.002$.}
  \label{fig.predictions.for.dynamics}.
\end{figure*}


For temperatures, $T < T_\text{SN}$ the easy--plane spin-liquid ($\text{SL}_\perp$) 
gives way to a phase with hidden spin--nematic order, 
labelled $\text{SN}_\perp$ in Fig.~\ref{fig:phase.diagram}.   
As far as the dipole moments of spins are concerned, the spin--nematic phase is disordered, 
and neutron scattering experiments would reveal algebraic correlations,  
as in the spin liquid $\text{SL}_\perp$. 
However the pinch points in $S({\bf q})$ [cf Fig.~\ref{fig:Sq.Q}] 
hide a great wealth of interesting spin excitations.   


To better understand the dynamics of the spin--nematic phase, we have calculated 
the dynamical structure factor $S({\bf q}, \omega)$, 
within a semi--classical molecular--dynamics (MD) simulation, 
using the methods described in [\onlinecite{taillefumier14}].   
Relevant definitions are given in Appendix~\ref{app:Sq}.   
For \mbox{$\omega/J_{\sf zz} \lesssim 0.2$}, $S({\bf q}, \omega)$ 
presents  a featureless, non--dispersing continuum [Fig.~\ref{fig:S.q.omega}].  
Relics of dispersing excitations are visible in $S({\bf q}, \omega)$ at higher 
energies, but these are explicitly not Goldstone modes, and have nothing to do 
with the hidden spin--nematic order.  
Examining the evolution of $S({\bf q}, \omega)$ as a function of temperature, 
we find that results for $S({\bf q}, \omega)$ in the spin--nematic phase for $T < T_\text{SN}$,   
are very similar to those found in the spin liquid $\text{SL}_\perp$ 
for $T > T_\text{SN}$.  


Incoherent, non-dispersing structure of the type shown in Fig.~\ref{fig:S.q.omega} 
is reminiscent of theoretical predictions \cite{knolle14,punk14, bieri15}
and experimental measurements \cite{han12,petit16-PRB94,paddison17}, 
for a wide range of different spin liquids.
In a quantum spin liquid the presence of a non-dispersing continuum reflects the 
fact that, unlike conventional spin waves (magnons), single elementary 
excitations of a spin liquid cannot be created by local processes.
It follows that, when a neutron scatters from a spin liquid, the energy, momentum 
and angular momentum (spin) transferred are not absorbed by a single excitation
with a well--defined energy and momentum, but rather shared between 
multiple excitations \cite{savary17-RPP80}.
In the semi--classical limit studied here, it is probably unsafe to 
attribute such a continuum to fractionalized 
excitations \cite{taillefumier-in-preparation}.
However, the fact that $S({\bf q}, \omega)$ only records dipolar spin correlations 
obscures a more important fact --- the spin--nematic order which 
is present for $T < T_\text{SN}$ which breaks a continuous, U(1), 
symmetry of the Hamiltonian.  
And, by Goldstone's theorem, 
it must, therefore, also support gapless Goldstone modes.


In order to resolve this conundrum, it is necessary to examine the dynamical correlations 
of the quadrupole moments of spin.
In Fig.~\ref{fig:chi.Q.q.omega} we present MD simulation results for the dynamical 
susceptibility $\chi_{\sf Q_\perp^{\sf site}}({\bf q}, \omega)$, 
which measures fluctuations of the on-site quadrupolar moments,
which are well defined for classical spins [cf. Appendix~\ref{app:phasediagram}].
A sharp excitation, with  dispersion
\begin{eqnarray}
\omega \approx v_Q |{\bf q} - {\bf q}_{\sf rl}|
\label{eq:goldstone.mode}
\end{eqnarray}
can now be resolved near to 
the zone centers with \mbox{${\bf q}_{\sf rl} = (0,0,0), (1,1,1), (2,2,2)$}.   
These are the same zone centers for which the Bragg peaks associated with 
the hidden spin--nematic order $\text{SN}_\perp$  would occur in a quadrupolar 
structure factor, which might, in principle, be measured in resonant X--ray 
experiments \cite{Savary15a}.


At present, relatively little is known about the dynamical properties of spin--nematic 
states.
Field--theoretic analysis [\onlinecite{smerald13,starykh14,smerald15,smerald16}], 
based on the symmetry of the order parameter, predicts that spin--nematics support 
gapless Goldstone modes, visible in $\chi_{Q_\perp}({\bf q}, \omega)$.  
This Goldstone mode has dispersion $\omega \propto |{\bf q}|$ [cf. Eq.~(\ref{eq:goldstone.mode})], 
and at zero temperature the associated intensity diverges as 
$\sim 1/\omega$ for $\omega \to 0$ [\onlinecite{smerald15}].  
The same behaviour is seen in ``flavour--wave'' calculations 
and QMC simulations of spin--1 models constructed to support quadrupolar 
order \cite{tsunetsugu06,laeuchli06,voell15}.
The dynamics of the spin--1/2 frustrated ferromagnetic spin chain have also been studied
using DMRG, and reveal a broad continuum of excitations at high energies \cite{onishi15}.
However, because of the absence of long--range order, no Goldstone modes can be 
resolved.
A continuum of excitations at high energies is also found in calculations for two--dimensional 
frustrated ferromagnets, within a slave--particle mean--field picture \cite{shindou13}.    


Our MD simulations of $\text{SN}_\perp$ clearly reveal a linearly--dispersing Goldstone 
mode, 
with intensity which diverges for $\omega \to 0$ [cf. inset to Fig.~\ref{fig:chi.Q.q.omega}].    
The form of this divergence is $\sim 1/\omega^2$, rather than $\sim 1/\omega$.
This follows from the fact that simulations are carried out at finite temperature, and 
probe thermal rather than quantum fluctuations.
Most striking, however, is the broad continuum of excitations visible in both 
spin-- and quadrupole structure factors.
And it is also interesting to note that the way in which the Goldstone mode ``dissolves'' 
into this continuum bears some resemblence to what is seen in QMC simulations 
of a spin--1 model, at higher temperature \cite{voell15}.
Overall, the picture which emerges from MD simulation is consistent with all known 
facts about spin--nematics, and should provide a reliable guide for comparison 
with experiment.


Further details of the spin dynamics in the spin--nematic phase,
and specifically the characterization of the Goldstone mode are given in
Appendix~\ref{app:goldstone}.
We note that the U(1) symmetry of 
$\mathcal{H}_{\sf XXZ}$ [Eq.~(\ref{eq:H.xxz})] is not a necessary 
condition for spin--nematic order to exist.
However, if this symmetry were broken, the low--energy 
\mbox{(pseudo--)Goldstone} mode associated with $\text{SN}_\perp$ 
would acquire a small gap.
\\

\section{Discussion}
\label{sec:discussion}

Spin liquids \cite{anderson73,lee08,balents10} and spin nematics 
\cite{andreev84, chubukov91, shannon06} are prime examples of 
unconventional states of matter, and have many unusual and interesting properties.
The experimental search for these exotic states has a long history, 
with many twists and turns, and not a few dead ends.
Given this, finding both in one simple, canonical, 
and experimentally--motivated model is remarkable.
It is therefore worth considering the possibilities for observing the unconventional 
spin liquid $\text{SL}_\perp$, and the spin--nematic $\text{SN}_\perp$, 
in the type of rare--earth pyrochlore magnet 
described by Eq.~(\ref{eq:H.xxz}).  


In the case of $\text{SN}_\perp$, it is important to make a distinction 
between the type of spin--nematic order considered in this manuscript, 
which is driven by fluctuations, 
and the quadrupolar or octupolar order which can arise directly 
from the ground states of rare--earth ions.
Here we particularly have in mind the non--Kramers ions Pr$^{3+}$ 
\cite{Onoda2010,Lee12,petit16-PRB94} and Tb$^{3+}$ 
\cite{Onoda2010,Guitteny2013,taniguchi13,Takatsu16}, 
and Kramers doublets of dipolar--octupolar character, 
such as Nd$^{3+}$ \cite{huang14}, and Ce$^{3+}$ \cite{Li17}.
For these ions,
quadrupolar or octupolar order may occur as a``classical'' order 
of the transverse part of the pseudospins ${\sf S}_i$.
The multipolar character of the order follows from the symmetry of the crystal--field 
ground--state (doublet) of the magnetic ion, which is described by ${\sf S}_i$.
Where multipolar order of this kind occurs, experiments which probe the dynamics
of dipoles will see a gapped response and a sharp excitation spectrum.
In contrast, in the easy--plane spin--nematic $\text{SN}_\perp$, 
dipole moments remain in a spin--liquid like state, 
with strong fluctuations at low temperature and a broad, gapless response 
coexisting with the hidden nematic order [Fig.~\ref{fig:S.q.omega}].


Where, then, might we observe these unusual magnetic states? 
%
Further experimental work will be necessary to definitively 
answer this question, but there are already a few trails to follow. 
In particular, the Pr-based pyrochlores have the 
recommended single-ion and interaction 
anisotropies \cite{Onoda2010,Lee12,petit16-PRB94}. 
Coupling parameters of Pr$_{2}$Zr$_{2}$O$_{7}$ for example have been suggested 
to sit in the AF$_{\perp}$ phase of Fig.~\ref{fig:phase.diagram} \cite{petit16-PRB94},
although it seems that the coupling of structural disorder to the non-Kramers doublets
plays a significant role \cite{savary17-PRL118,Wen2017,benton-arXiv17,martin-arXiv17}.
Since chemical pressure has already proven to be a useful tool to move a 
family of compounds across a phase diagram \cite{dun14a,wiebe15a,jaubert15,hallas16-PRB93.100403,yan17}, 
Pr$_{2}$X$_{2}$O$_{7}$ (X=Sn,Hf,Pb) are promising candidates to investigate, 
with ferromagnetic correlations consistent with positive $J_{\sf zz}$ and no dipole 
order yet observed  \cite{Matsuhira2004,Zhou2008,hallas15,Sibille16a,Anand16a}.


The notion of hidden order also resonates with the elusive physics of 
Yb--based pyrochlores. 
As far as we know, Yb pyrochlores lie in a different regime of magnetic interactions
 than the $\mathcal{H}_{\sf XXZ}$ model of Eq.~(\ref{eq:H.xxz}).
Specifically, experiments on Yb$_2$Ti$_2$O$_7$ point to an unfrustrated 
value of $J_{\pm}>0$ \cite{ross11-PRX1,Robert2015,Thompson17}, and to an  
important role for other competing exchange interactions.
In the light of this, the properties of that particular material seem to be connected
with a different phase boundary from that associated with 
$\text{SL}_\perp$ \cite{jaubert15, yan17,hallas-arXiv.1708.01312}.
That being said, some of the similarities between our results 
and the Yb--pyrochlores are striking: a gapless continuum of spin excitations, 
oblivious to thermodynamic phase transitions 
\cite{Ross2009,Maisuradze2015,hallas16-PRB93.100403,gaudet16} 
[Fig.~\ref{fig.predictions.for.dynamics}(a)], and robust in temperature 
up to a broad feature in specific heat \cite{hallas16-PRB93.100403} 
(in the present Article, between pHAF and $\text{SL}_\perp$). 
And while the magnetic order in 
Yb-pyrochlores is, at least partially, an order of dipolar moments 
\cite{Yasui2003a,chang12-NatCommun3,Yaouanc2013a,dun13,lhotel14,hallas16-PRB93.104405}, 
recent experiments have indicated that the primary order parameter may be ``hidden'', 
and distinct from a standard dipole order \cite{hallas16-PRB93.100403}. 
Thus, while the specific case developed in this Article 
probably does not apply to the Yb--pyrochlores, related physics may be at play.


Furthermore, since the spin--nematic phase $\text{SN}_\perp$ is found 
within the spin liquid $\text{SL}_\perp$ [Fig.~\ref{fig:phase.diagram}], this work 
provides a prototype for the peaceful co-existence of emergent 
gauge fields and long--range order. 
In this sense, $\text{SN}_\perp$ is an interesting 
new addition to the other phases where gauge fluctuations and broken symmetries 
co-exist, such as the Coulombic ferromagnet~\cite{savary12-PRL108,Powell2015}, 
and states with magnetic moment fragmentation \cite{Brooks-Bartlett2014}, 
as recently observed in Nd$_2$Zr$_2$O$_7$ \cite{Petit2016,Benton16} 
and Ho$_2$Ir$_2$O$_7$ \cite{Lefrancois17}.


We also note that many other magnetic systems outside the rare earth oxides R$_2$X$_2$O$_7$
feature moments located on a pyrochlore lattice.
Of particular interest are materials such as NaCaCo$_2$F$_7$ and NaSrCo$_2$F$_7$
\cite{ross16, ross17} which boast XY like interactions with much higher energy 
scales than observed in the rare-earth oxides.
If such a case could be found with frustrated transverse coupling $J_{\pm}<0$ 
then it would render the physics discussed here accessible at a
much more amenable temperature range.


In almost all  spin-liquid candidates, the role
of quenched structural and chemical disorder is an important issue
\cite{ross12,Wen2017,benton-arXiv17,mostaed17,taniguchi13,kermarrec14,han16,norman16,li17-PRL118,martin-arXiv17}.
Depending on the type and strength of disorder, its consequences can vary.
It is worth noting however, that disorder is not necessarily deleterious to
spin liquid physics.
It is known, for example, that weak disorder in non-Kramers pyrochlores,
which leads to splittings in the low energy non-Kramers doublet can
actually play a role in promoting a $U(1)$ spin liquid ground state \cite{savary17-PRL118}.
The spin--liquid states discussed in this manuscript do not
depend on the translational symmetry of the Hamiltonian, but rather 
on the emergent gauge symmetries which arise from the local constraints
in the ground state [Eq.~(\ref{eq:SLIIconstraints})].
Thus, as long as the disorder is not so strong that these constraints
are strongly violated, the essence of the spin liquids should be maintained 
in the presence of disorder, at least at finite temperature.
For sufficiently strong disorder, or sufficiently low temperature,
disorder may lead to order--by--disorder or glassiness \cite{bellier01, saunders07}.
A quantitative study of the effects of disorder on the phase diagram
in Fig.~\ref{fig:phase.diagram} is a large undertaking and is beyond the
scope of the present work but may be an interesting direction for future
consideration.

\section{Summary and conclusions}
\label{sec:conclusion}


``Quantum spin ice'', in which magnetic ions on a pyrochlore lattice interact through 
highly--anisotropic exchange interactions, have become an important paradigm in 
the search for quantum spin liquids.
In this Article we have used large--scale classical
Monte--Carlo simulation to explore the physics of the canonical 
model of a quantum spin ice,  
the XXZ model on a pyrochlore lattice $\mathcal{H}_{\sf XXZ}$ 
[Eq.~(\ref{eq:H.xxz})].
We find that this model has far more to offer than spin ice alone, supporting 
three distinct types of spin liquid, each with a different emergent gauge symmetry.
Each of these spin liquids has a different signature in neutron 
scattering [Fig.~\ref{fig:predictions.for.Sq}].
And the states found include a completely new form 
of spin liquid, described by a U(1)$\times$U(1) gauge theory.
At low temperatures this novel spin--liquid undergoes a phase transition to a state with 
hidden spin--nematic order [Fig.~\ref{fig:phase.diagram}], but retaining algebraic
correlations of the spin dipoles.
We have studied the excitations of this phase using state--of--the--art dynamical
simulations, revealing a sharply defined Goldstone mode which would be 
hidden from conventional neutron scattering techniques.


So far as experiment is concerned, the main lesson 
of these results is that ``quantum spin--ice'' materials, 
can play host to a great many different spin--liquid and (hidden--)order 
phases, even where they are described by a Hamiltonian as simple as 
$\mathcal{H}_{\sf XXZ}$ [Eq.~(\ref{eq:H.xxz})].  
This reinforces the point that pinch--points in pyrochlore magnets
need not imply spin ice \cite{henley10,benton16-NatCommun7}.  
The existence of a sharp Goldstone mode in the nematic phase $\text{SN}_\perp$
also serves as a salutary reminder that broad, non--dispersing continua of 
excitations can hide a multitude of secrets [Fig.~\ref{fig.predictions.for.dynamics}].


From the theoretical point of view, this work identifies a new spin liquid, 
a novel spin nematic phase, and opens an interesting new perspective on the way in 
which different spin liquids can compete.  
The effect of quantum fluctuations on the 
phase diagram shown in Fig.~\ref{fig:phase.diagram} 
for  $J_\pm < 0$ remains a subject for future study.   
However, experience with QMC simulation of 
$\mathcal{H}_{\sf XXZ}$ [Eq.~(\ref{eq:H.xxz})] for $J_\pm > 0$ suggests that 
quantitative values of the crossover temperature $T^*_2$ and $T^*_3$ may 
be substantially renormalized, but that the qualitative structure of the phase 
diagram should remain the same down to very low temperatures 
\cite{banerjee08,kato15,shannon-book-chapter,huang-arXiv.1707.00099}.
The high--symmetry point, $J_\pm/J_{\sf zz} = -1/2$ is also a high--symmetry point
for quantum spins, and so remains the anchor for the spin liquid pHAF.
%
None the less, the fate of this U(1)$\times$U(1)$\times$U(1) spin liquid for quantum 
spins at $T=0$ remains an open question \cite{canals98,canals00,tsunetsugu01,Huang2016}.
And, to the best of our knowledge, quantum analogues of the new spin liquid, 
$\text{SL}_\perp$, which has a U(1)$\times$U(1) gauge structure, 
remain unexplored \cite{ueda16-PRA93}.  
However it seems reasonable to speculate that quantum effects will
enhance, rather than suppress, the fluctuations which drive 
$\text{SL}_\perp$ and pHAF, 
and that the phase $\text{SN}_\perp$ will survive as hidden quantum 
spin--nematic order, within a quantum spin liquid.
And preliminary numerical results for the spin--1/2 model 
at high temperatures are entirely consistent with the topology of the phase 
diagram shown in Fig.~\ref{fig:phase.diagram} \cite{rajiv-unpublished}.      
All of these questions open exciting avenues for future research.

\section*{Acknowledgements}

We are pleased to acknowledge many helpful discussions with Karim Essafi, Jaan Oitmaa 
and Rajiv Singh.  
This work was supported by the Theory of Quantum Matter Unit of the Okinawa Institute 
of Science and Technology Graduate University (OIST).   
Numerical calculations were carried out using HPC facilities provided by OIST.  

\begin{appendix}

\section{Definition of local coordinate frame}
\label{app:coordinates}

We describe the local-coordinate frame which is defined for
four spins on a pyrochlore tetrahedron
${\sf S_0}$,
${\sf S_1}$,
${\sf S_2}$,
${\sf S_3}$ occupying positions 
\begin{eqnarray}
&{\bf r}_0 =  \frac{a}{8} \left( 1, 1, 1 \right) 
\qquad 
&{\bf r}_1 =  \frac{a}{8} \left( 1,-1,-1 \right) 
\nonumber \\
&{\bf r}_2 =  \frac{a}{8} \left( -1,1,-1 \right) 
\qquad 
&{\bf r}_3 = \frac{a}{8} \left( -1,-1,1 \right) 
\; ,
\label{eq:r}
\end{eqnarray}
where $a$ is the length of a cubic, 16-site unit 
cell of the pyrochlore lattice.    


The pseudospins in the global, crystal, coordinate frame ${\bf S}_i$
relate to the pseudospins  in the local frame ${\sf S}_i$ [Eq.~(\ref{eq:H.xxz})]
as 
\begin{eqnarray}
{\bf S}_i=
\mathbf{x}^{\sf local}_i {\sf S}^x_i
+
\mathbf{y}^{\sf local}_i {\sf S}^y_i
+
\mathbf{z}^{\sf local}_i {\sf S}^z_i
\end{eqnarray}


Where
\begin{eqnarray}
&\mathbf{z}^{\sf local}_0 = \dfrac{1}{\sqrt{3}}(1,1,1)
\qquad
&\mathbf{z}^{\sf local}_1 = \dfrac{1}{\sqrt{3}}(1,-1,-1) 
\nonumber \\
&\mathbf{z}^{\sf local}_2 = \dfrac{1}{\sqrt{3}}(-1,1,-1)
\qquad
&\mathbf{z}^{\sf local}_3 = \dfrac{1}{\sqrt{3}}(-1,-1,1) 
\; , \nonumber
\label{eq:local-111-axis}\\
\end{eqnarray}


\begin{eqnarray}
&\mathbf{x}^{\sf local}_0 = \dfrac{1}{\sqrt{6}}(-2,1,1)
\qquad
&\mathbf{x}^{\sf local}_1 = \dfrac{1}{\sqrt{6}}(-2,-1,-1)
\nonumber \\
&\mathbf{x}^{\sf local}_2 = \dfrac{1}{\sqrt{6}}(2,1,-1)
\qquad
&\mathbf{x}^{\sf local}_3 = \dfrac{1}{\sqrt{6}}(2,-1,1)
\; , \nonumber
\label{eq:local-easy-plane-x}\\
\end{eqnarray}
and
\begin{eqnarray}
&\mathbf{y}^{\sf local}_0 = \dfrac{1}{\sqrt{2}}(0,-1,1)
\qquad
&\mathbf{y}^{\sf local}_1 = \dfrac{1}{\sqrt{2}}(0,1,-1)
\nonumber \\
&\mathbf{y}^{\sf local}_2 = \dfrac{1}{\sqrt{2}}(0,-1,-1)
\qquad
&\mathbf{y}^{\sf local}_3 = \dfrac{1}{\sqrt{2}}(0,1,1)
\; . \nonumber
\label{eq:local-easy-plane-y}\\
\end{eqnarray}


We have used this relationship between the local
coordinate frame of Eq.~(\ref{eq:H.xxz}) and the crystal coordinate
frame to plot a representation of the bond quadrupolar
order in real space in Fig.~\ref{fig:phase.diagram}(b).
The ellipsoid on each bond $ij$ in Fig.~\ref{fig:phase.diagram}(b)
has principal axes aligned along the cubic axes of the pyrochlore
lattice, with the length of each principal axis given by
\begin{eqnarray}
l_{\alpha}=4(c+\langle S^{\alpha}_i S^{\alpha}_j \rangle)
\end{eqnarray}
where $c=0.08$ is chosen to make the figure readable 
and $S^{\alpha}_i$ is the component of spin $i$ in the
global frame, along crystal axis $\alpha=x,y,z$.

\section{Details of the numerical determination of the phase diagram}
\label{app:phasediagram}


\begin{figure}[b]
  \includegraphics[width=8cm]{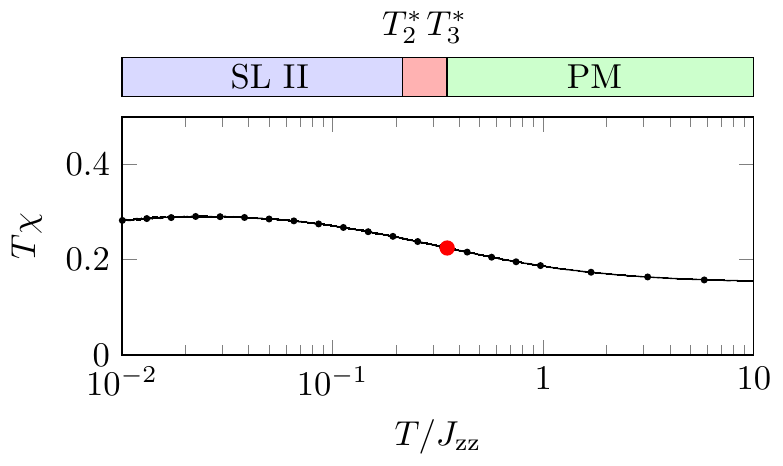}
  \caption{(Color online).
  Crossover in the magnetic susceptibility $\chi(T)$, 
  from a traditional Curie law in the high--temperature paramagnet, 
  to a low--temperature Curie law in the spin liquids, as seen by different plateaux 
  in the function $T\chi$, plotted as a function of $\log (T)$.
 The crossover temperature $T_{3}^{\ast}/J_{\sf zz} \approx 0.3$ (red dot) is estimated 
 from the point of inflection of $T\chi$. 
 The extraction of the crossover temperature $T_{2}^{\ast}$ is explained in Fig.~\ref{thermo}.
 Results are taken from classical Monte Carlo simulations of $\mathcal{H}_{\sf XXZ}$ 
 [Eq (\ref{eq:H.xxz})], for a cubic cluster of $N=8192$ spins, with
 $J_{\pm}/J_{\sf zz} = -1$.}
  \label{fig:T.CLC}
\end{figure}


\begin{figure}
  \subfloat[Specific heat \label{fig:cV}]{\includegraphics[width=6cm]{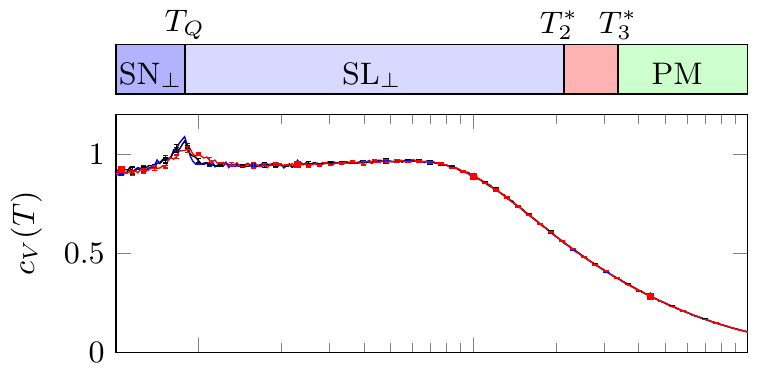}} \\
  \subfloat[ $T \chi_{\sf T_1 ice}(T)$ \label{fig:TChiIce}]{\includegraphics[width=6cm]{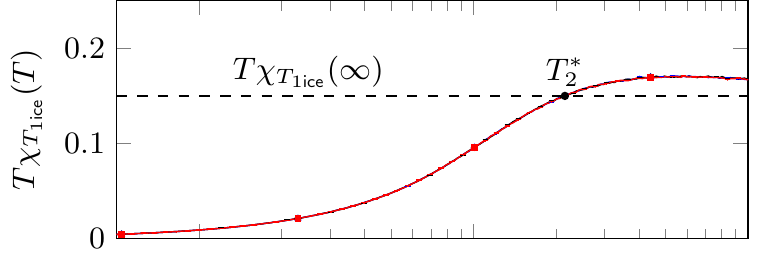}} \\
  \subfloat[ Spin--nematic bond order parameter \label{fig:Qbond}]
{\includegraphics[width=6cm]{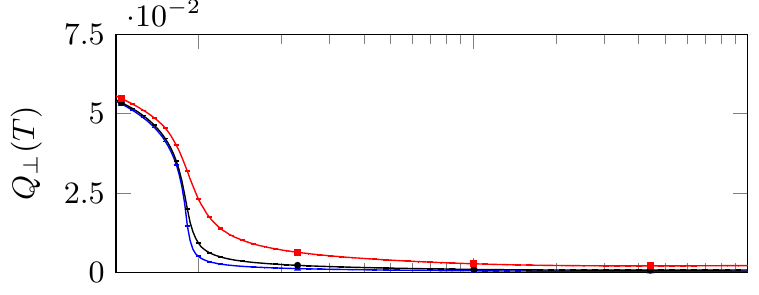}} \\
  \subfloat[Spin--nematic site order parameter  \label{fig:Q}]
{\includegraphics[width=6cm]{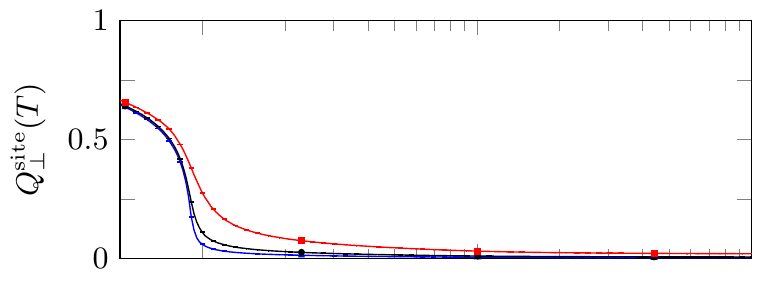}}\\
  \subfloat[Order--parameter susceptibility \label{fig:chi.Q}]{\includegraphics[width=6cm]{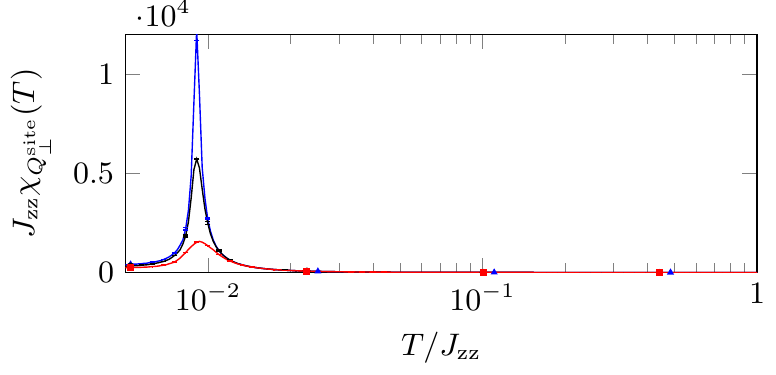}}\\
  \caption{
    Thermodynamics of the QSI in the region of spin--nematic order.     
    (a) Specific heat $c_V(T)$, showing an upturn followed by a 
shallow maximum in the region of the 
    crossover into the spin liquid $\text{SL}_\perp$ at $T_2^*/J_{\sf zz} \sim 10^{-1}$, 
    and small peak associated with the onset of spin--nematic order at 
    $T_\text{SN}/J_{\sf zz} \approx 10^{-2}$.
(b) Correlation function $T \chi_{\sf T_1 ice}(T)$ used to determine the crossover
temperature into the spin-liquid $\text{SL}_\perp$.
$\chi_{\sf T_1 ice}$ is the susceptibility of the field ${\bf m}_{\sf T_1 ice}$
defined in Appendix~\ref{app:m_lambda}.
The crossover temperature $T_2^{\ast}$ is defined as the point at which
$T \chi_{\sf T_1 ice}$ drops below its infinite temperature limit
[Eq. \ref{eq:T2criterion}]
(c) Norm of the bond order parameter ${\bf Q}_\perp (T)$ [cf. Eq.~(\ref{eq:Q})], 
    showing a continuous phase transition into the phase 
    with hidden spin--nematic order at $T_\text{SN}/J_{\sf zz} \approx 10^{-2}$.
    (d) Norm of the site order parameter ${\bf Q}^{\sf site}_\perp (T)$ [cf. Eq.~(\ref{eq:Q})], 
    also showing the phase transition into the spin--nematic phase.
    (e) Order--parameter susceptibility $\chi_{\sf Q_\perp^{\sf site}}(T)$ [Eq.~(\ref{eq:chiQ})], 
    used to estimate the transition temperature $T_\text{SN}$.
    %
    %
    Results are taken from classical Monte Carlo simulation of
    $\mathcal{H}_{\sf XXZ}$ [Eq~(\ref{eq:H.xxz})], for cubic clusters
    of $N=1024, 8192$ and $27648$ spins, with $J_{\pm}/J_{\sf zz}=-1$.}
  \label{thermo}
\end{figure}


The phase diagram shown in Fig.~\ref{fig:phase.diagram} was extracted from 
classical Monte Carlo (MC) simulations of the quantum spin ice model, 
$\mathcal{H}_{\sf XXZ}$ [Eq.~(\ref{eq:H.xxz})].  
Spins were treated as classical vectors with fixed length $|{\sf S}_i|=\frac{1}{2}$.
These simulations were carried out for a cubic cluster of 8192 spins,
using a single spin flip algorithm combined with simulated annealing,
parallel tempering and over--relaxation. The phase diagram is obtained
using 75000 simulated annealing steps using a stepwise decrease of
temperature starting from $T=10 J_{\sf zz}$ down to the target
temperature. 
Each annealing step consists of 10 Monte Carlo
steps (a Monte Carlo step consists of a full sweep of the lattice
combined with over--relaxation). The simulated annealing is followed by
1000 parallel tempering steps with 500 Monte Carlo steps in between,
and then by 200000 Monte Carlo steps for thermalization at fixed
temperature. Measurements consist of 200000 samples separated by 10
Monte Carlo steps and combined to parallel tempering every 50
measures. We use 256 different replicas with temperature set in linear
scale for $J_{\pm}/J_{\sf zz}>-\frac12$ and 256 temperatures in
logarithmic scale for $J_{\pm}/J_{\sf zz}\le -\frac12$.


The phase boundary of the antiferromagnetically ordered (AF$_{\perp}$) phase,
$T_N$, was extracted from the susceptibility of the relevant order
parameter, ${\bf m}_{\sf E}$, as defined in Appendix~\ref{app:m_lambda}.


The crossover scale for the spin--ice regime (SI), $T_1^*$, was
estimated from the Schottky-like peak in the heat capacity.


The crossover scale $T_3^{\ast}$ for the spin--liquid pHAF was 
estimated from the Curie-Law crossover shown in Fig.~\ref{fig:T.CLC}.


For \mbox{$J_\pm < -\frac{1}{2}$}, the crossover scale $T_2^*$ is associated
with a weakening of the correlations of the local $z$-components of the spins. 
This can be observed by measuring the susceptibility, $\chi_{\sf T_1 Ice}(T)$, of the field
${\bf m}_{\sf T_1 Ice}$, defined in Appendix~\ref{app:m_lambda}].
Decreasing the temperature for $-1<\frac{J_{\pm}}{J_{zz}}<-0.5$
the quantity $T\chi_{\sf T_1 Ice}(T)$ first increases during the crossover
from the paramagnet to pHAF and then drops as the system 
enters $\text{SL}_\perp$.
We define the crossover temperature $T_2^{\ast}$ as the point
at which the quantity $T \chi_{\sf T_1 ice}(T)$ drops below its infinite 
temperature value
\begin{eqnarray}
T_2^{\ast}\chi_{\sf T_1 ice}(T_2^{\ast})=\lim_{T\to \infty}T \chi_{\sf T_1 ice}(T)
\label{eq:T2criterion}
\end{eqnarray}
This is illustrated in Fig.  \ref{fig:TChiIce}.


For quantum, spin--1/2 moments, the onset of spin--nematic order is 
heralded by the bond--based order parameter Eq.~(\ref{eq:bond.nematic}).
However for the purpose of classical simulation, it is sufficient to consider 
the simpler, site--based order parameter
 \begin{eqnarray}
   {\bf Q}_\perp^{\sf site} = 
\frac{4}{N} \sum_i \begin{pmatrix}
   {\mathsf{S}^x_i}^2 -{\mathsf{S}^y_i}^2 \\
    2 \mathsf{S}^x_i \mathsf{S}^y_i
    \end{pmatrix}  \; .
 \label{eq:Q}
 \end{eqnarray}
Please note the prefactor 
of 4 in the definition of the spin nematic site order parameter 
[Eq.~(\ref{eq:Q})] to account for the spin length $|{\sf S}_i|=\frac{1}{2}$.  
The onset of spin--nematic order in simulations can be observed in either 
the site--based [cf. Fig.~\ref{fig:Q}] or bond--based order parameters 
[cf. Fig.~\ref{fig:Qbond}].
However, for simplicity, the values of the spin--nematic ordering 
temperature $T_\text{SN}$ shown in Fig.~\ref{fig:phase.diagram}, were 
extracted from the peak in the order--parameter susceptibility 
\begin{equation}
  \label{eq:chiQ}
  \chi_{Q_{\perp}^{\sf site}} 
  = \frac{N}{T} \left( \langle {{\bf Q}^{\sf site} _\perp}^2 \rangle 
  - \langle| {\bf Q}^{\sf site}_\perp |\rangle^2 \right).
\end{equation}
associated with the site--based order parameter, Eq.~(\ref{eq:Q}) [cf.~Fig.~\ref{fig:chi.Q}]%


Fig.~\ref{thermo} is obtained using 300 temperatures in logarithmic
scale coverring 3 orders of magnitude, parallel tempering every 100
Monte Carlo steps, simulated annealing and thermalization at
temperature $T$ for 100000 Monte Carlo steps each. Measurements
consist of 100000 different samples with 10 Monte Carlo steps between
each sample. 
Error bars were estimated by comparing the results of three independent
runs of the simulation.

\section{Definitions of dynamical structure factors}
\label{app:Sq}

In Fig.~\ref{fig:predictions.for.Sq} we show predictions for  neutron scattering experiments,
based on the equal--time (i.e. energy--integrated) structure factor
\begin{equation}
  S({\bf q}) = \int d\omega\ S({\bf q},\omega)  \; ,
  \label{eq:S.q}
\end{equation}
where the dynamical structure factor $S({\bf q},\omega)$ is defined
through
\begin{eqnarray}
  &&S({\bf q},\omega) 
  =  \sum_{\alpha\beta} \left( 
\delta_{\alpha\beta}-\frac{q_\alpha q_\beta}{q^2}  \right) 
  \langle m^\alpha(-{\bf q},\omega) m^\beta({\bf q} \;,\omega) \rangle 
  \nonumber\\ 
&&m^\alpha({\bf q},\omega) 
  = \sum_{i \beta \gamma} R^{\alpha \beta}_i g^{\beta\gamma}_i  
  \left(\int  \mathsf{S}^\gamma_i(t) e^{i\omega t} dt\right)  e^{i {\bf q}\cdot {\bf r}_i} \;,
\end{eqnarray}
and the  $g^{\beta\gamma}_i$ 
is the $g$-tensor written in the local coordinate frame \cite{yan17}.
For simplicity, we have here taken 
$g^{\beta\gamma}_i = 2 \delta_{\beta\gamma}$ for all of 
the calculations in this paper.
$R^{\alpha\beta}_i$ is a rotation matrix which rotates
from the local coordinate frame on site $i$, to the global,
crystal coordinate frame.
The definition of the local coordinate frame is given in
Appendix~\ref{app:coordinates}.
Results for $S({\bf q})$ are shown in the left half--panels of Fig.~\ref{fig:predictions.for.Sq}.
These results were taken from classical MC simulations of
$\mathcal{H}_{\sf XXZ}$ at a given temperature, with further averaging
provided by numerically integrating the semi--classical equations of
motion for the spins.
This secondary molecular--dynamics (MD) simulation was carried out
using methods described in Ref.~[\onlinecite{taillefumier14}].


It is also useful to decompose the structure factor into the
spin--flip (SF) and non spin--flip (NSF) channels measured in
polarised neutron--scattering experiments.
\begin{eqnarray}
  \label{eq:S.5}
&&  S_{\text{SF}}({\bf q}) 
  = \frac{1}{q^2}\int d\omega\ \langle|{\bf m}({\bf q},\omega) 
      \cdot (\hat{\bf n}\times {\bf q})|^2\rangle 
\nonumber \\
&&  S_{\text{NSF}}({\bf q}) 
  = \int d\omega\ \langle|{\bf m}({\bf q},\omega) 
  \cdot \hat{\bf n}|^2\rangle \; ,
\end{eqnarray}
where $\hat{\bf n}$ is the direction of polarization of the neutron
magnetic moment.
Following Fennell {\it et al.} [\onlinecite{fennell09}], we take 
$\hat{\bf n} = (1,-1,0)/\sqrt{2}$.
Simulation results for $S_{\text{SF}}({\bf q})$ 
and $S_{\text{NSF}}({\bf q})$ 
are shown in the right half--panels of Fig.~\ref{fig:predictions.for.Sq}.


We have also used MD simulation to calculate the dynamical structure
factor $S({\bf q},\omega)$.
Results for $S({\bf q},\omega)$ within the spin--nematic phase of 
the quantum spin ice model are shown in Fig.~\ref{fig:S.q.omega}.
Further details of the calculation of dynamical properties can
be found in Appendix~\ref{app:goldstone}.

\section{Definitions of local order parameter fields}
\label{app:m_lambda}

The definitions of the local order parameter fields
${\bf m}_{\lambda}$ which appear in the theory
of the spin liquid $\text{SL}_\perp$ [Section~\ref{sec:SLII}] are
given in Table~\ref{table:m.lambda}.


Here we give the definitions in terms of the spins
written in the local coordinate frame ${\sf S}_i$ (defined in 
Appendix~\ref{app:coordinates}), cf. Ref.~\onlinecite{yan17} 
where definitions are given in the global, crystal basis.\\


\begin{table}
\resizebox{\columnwidth}{!}{%
\begin{tabular}{ | c | c |  }
\hline
\multirow{2}{*}{}
     & 
   Definition in terms  of spins within tetrahedron
\\
\hline
\multirow{1}{*}{} 
$ m_{\sf A_2}$ & $\frac{1}{2} (\mathsf{S}_0^z+\mathsf{S}_1^z+\mathsf{S}_2^z+\mathsf{S}_3^z)$\\
\hline
\multirow{1}{*}{} 
   ${\bf m}_{\sf E}$ & 
   $\dfrac{1}{2}
   \begin{pmatrix}
        \mathsf{S}_0^x+\mathsf{S}_1^x+\mathsf{S}_2^x+\mathsf{S}_3^x\\
        \mathsf{S}_0^y+\mathsf{S}_1^y+\mathsf{S}_2^y+\mathsf{S}_3^y 
     \end{pmatrix}$  \\ 
\hline
\multirow{1}{*}{}
   ${\bf m}_{\sf T_{1, ice}}$  & 
   $\dfrac{1}{2}\begin{pmatrix}
       \mathsf{S}^z_0 + \mathsf{S}^z_1 - \mathsf{S}^z_2  - \mathsf{S}^z_3  \\  
       \mathsf{S}^z_0 - \mathsf{S}^z_1 + \mathsf{S}^z_2  - \mathsf{S}^z_3\\
       \mathsf{S}^z_0 - \mathsf{S}^z_1 - \mathsf{S}^z_2  + \mathsf{S}^z_3
    \end{pmatrix} $ 
    \\
\hline
\multirow{1}{*}{}
   ${\bf m}_{\sf T_{1, planar}}$  & 
   $\begin{pmatrix}
        \frac{1}{2} \left(  \mathsf{S}^x_0 + \mathsf{S}^x_1 - \mathsf{S}^x_2  - \mathsf{S}^x_3 \right)  \\
        \frac{1}{4} \left( - \mathsf{S}^x_0 + \sqrt{3} \mathsf{S}^y_0 +  \mathsf{S}^x_1 - \sqrt{3} \mathsf{S}^y_1
                              -  \mathsf{S}^x_2 + \sqrt{3} \mathsf{S}^y_2 +  \mathsf{S}^x_3 - \sqrt{3} \mathsf{S}^y_3 \right)   \\
        \frac{1}{4} \left(  - \mathsf{S}^x_0 - \sqrt{3} \mathsf{S}^y_0 +  \mathsf{S}^x_1 + \sqrt{3} \mathsf{S}^y_1
                              +  \mathsf{S}^x_2 + \sqrt{3} \mathsf{S}^y_2 -  \mathsf{S}^x_3 - \sqrt{3} \mathsf{S}^y_3 \right)  
   \end{pmatrix}$ 
\\
\hline
\multirow{1}{*}{}
   ${\bf m}_{\sf T_2} $ & 
   $\begin{pmatrix}
        \frac{1}{2} 
        \left(
                -\mathsf{S}_0^y-\mathsf{S}_1^y+\mathsf{S}_2^y+\mathsf{S}_3^y 
        \right) 
        \\
        \frac{1}{4} 
        \left(
        \sqrt 3  \mathsf{S}_0^x + \mathsf{S}_0^y - \sqrt 3 \mathsf{S}_1^x - \mathsf{S}_1^y + \sqrt 3 \mathsf{S}_2^x + \mathsf{S}_2^y - 
 \sqrt 3 \mathsf{S}_3^x - \mathsf{S}_3^y
        \right) \\
        \frac{1}{4} 
        \left(
       - \sqrt 3  \mathsf{S}_0^x + \mathsf{S}_0^y + \sqrt 3 \mathsf{S}_1^x - \mathsf{S}_1^y + \sqrt 3 \mathsf{S}_2^x - \mathsf{S}_2^y - 
 \sqrt 3 \mathsf{S}_3^x + \mathsf{S}_3^y
        \right)
      \end{pmatrix} $    
\\ 
\hline
\end{tabular}}
\caption{
Order--parameter fields ${\bf m}_\lambda$, derived from irreducible representations (irreps)
of the tetrahedral point-group T$_d$.   
Spin components $\mathsf{S}_i = (\mathsf{S}^x_i, \mathsf{S}^y_i, \mathsf{S}^z_i)$ 
are written in the local frame of the magnetic ions, see Appendix~\ref{app:coordinates}
for a definition of this coordinate frame.
The convention for the labelling of the spins with an tetrahedron is 
given in  Appendix~\ref{app:coordinates}.
}
\label{table:m.lambda}
\end{table}

\section{Numerical simulation of the correlations of the flux}

\label{app:fluxsim}

Values of the flux field ${\bf B}_{\mu}(\mathbf{r})$
are calculated for each tetrahedron ${\bf r}$
according to Eq.~(\ref{eq:fluxes})
and the definitions of ${\bf m}_{\lambda}$ given in Table
\ref{table:m.lambda}.


The tetrahedra of the pyrochlore lattice may be divided into
two sets A and B.
The centres of each set of tetrahedra each form an FCC lattice.


\begin{figure}[h!]
  \centering
  \subfloat[Dispersion relation of the Goldstone mode]{\includegraphics[width=8cm]{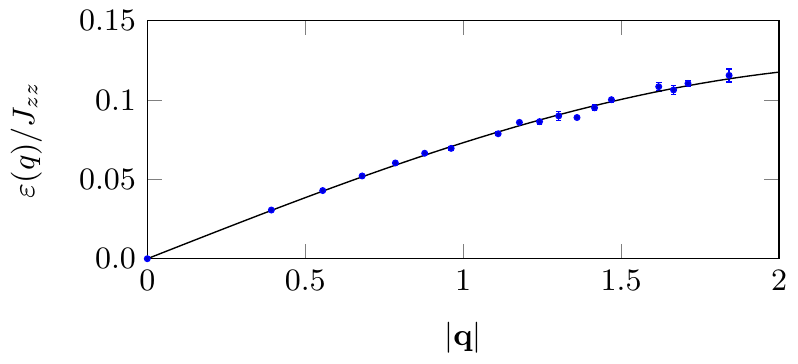}}\\
  \subfloat[Intensity of the quadrupolar waves]{\includegraphics[width=8cm]{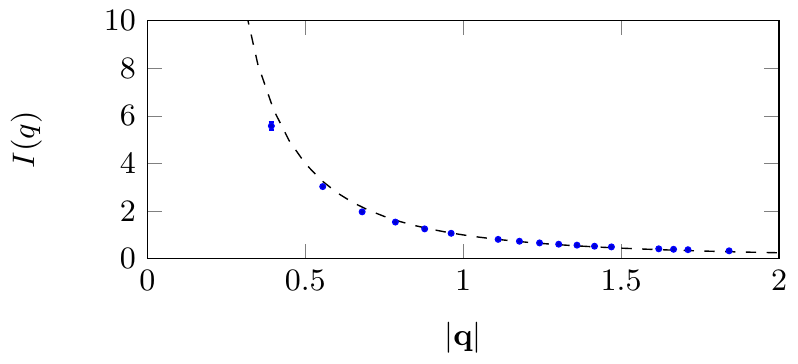}}
  \caption{
    Dispersion and intensity of the Goldstone mode \
    in the phase with hidden spin--nematic order, as shown 
    in Fig.~\ref{fig:chi.Q.q.omega}.   
  (a) Dispersion $\epsilon (q)$ of low--energy peak in 
  $\chi_{Q_{\perp}}({\bf q}, \omega)$ showing the expected behaviour $\epsilon (q) =  q$ 
  at small $q$.
  (b) Intensity $I(q)$ of the peak as a function of 
  momentum $q$. 
  The dashed line shows the expected behaviour at finite temperature, 
  $I(q) \propto 1/q^2$.  
  Results are taken from molecular--dynamics simulations 
  of a cluster of $N=65536$ spins, 
  for $J_\pm/J_{\sf zz} = -1.0$, $T = 0.002 J_{\sf zz}$.
  Momentum $q$ is measured relative to ${\bf q}=(0,0,0)$.}
  \label{fig:Goldstone.mode}
\end{figure}


To calculate ${\bf S}_{{\bf B}_{\mu}}^{\alpha \beta}$ we use Eq.~(\ref{eq:S.mu})
where ${\bf B}_{\mu}(\mathbf{q})$ is defined as
the lattice Fourier transform of  ${\bf B}_{\mu}(\mathbf{r})$ over
only the A sublattice of tetrahedra.
\begin{eqnarray}
{\bf B}_{\mu}(\mathbf{q})=\sqrt{\frac{1}{N_{\sf uc}}}
\sum_{{\bf r} \in {\bf r}_A} \exp(-i \mathbf{q} \cdot {\mathbf{r}}_A)
{\bf B}_{\mu}(\mathbf{r})
\end{eqnarray}
where $N_{\sf u.c.}$ is the number of unit cells in the system.


Simulations were carried out using local spin updates, augmented by over--relaxation, 
within a parallel tempering scheme with 300 temperatures distributed on a log scale 
between $T= 0.003\  J_{\sf zz}$ and $T=  0.1 \  J_{\sf zz}$.   
Thermalisation was accomplished through a process of simulated annealing,
with 10$^4$ Monte Carlo steps (MCs) of annealing from high temperature to temperature $T$, 
followed by 10$^4$ MCs of thermalization at temperature $T$, and 10$^5$ MCs of measurements 
at temperature $T$.
Spin configurations were sampled every 100 MCs during the measurements, 
giving an ensemble of 1000 samples.

\section{Dynamics of excitations in the spin--nematic phase}
\label{app:goldstone}

To study the Goldstone mode associated to the
development of spin--nematic order, we calculate the dynamical
correlation function.
\begin{equation}
  \chi_{Q_\perp^{\sf site}} ({\bf q},\omega) = \langle| \delta{\bf Q}^{\sf site}_\perp({\bf q},\omega) |^2\rangle \; ,
\end{equation}
where fluctuations of spin--nematic order are given by 
\begin{eqnarray}
  &&\delta{\bf Q}^{\sf site}_\perp({\bf q},\omega) 
  = \sum_{i} \int dt\ \left[ {\bf Q}^{\sf site}_\perp({\bf r}_i, t) 
   - \bar{\bf Q}^{\sf site}_\perp(t)  \right] 
  e^{i  \omega t} e^{i {\bf q}\cdot {\bf r}_i} 
   \nonumber \\
&&
   \bar{\bf Q}^{\sf site}_\perp(t)
   = \frac{1}{N} \sum_i {\bf Q}^{\sf site}_\perp({\bf r}_i, t) \; ,
\end{eqnarray}
and the order parameter ${\bf Q}^{\sf site}_\perp({\bf r}_i, t)$  is defined 
through Eq.~(\ref{eq:Q}).


$\chi_{Q_{\perp}^{\sf site}}(q,\omega)$ is calculated numerically from 200 sample
configurations extracted
from Monte Carlo simulations on a system of linear size $L=16$. 
We used
20000 steps for the simulated annealing spaced by 10 Monte Carlo steps
between each simulated annealing step. 
The other parameters for the
thermalization and parallel tempering are identical to the parameters
used to calculate the phase diagram [Appendix~\ref{app:phasediagram}]. 


The ensemble of configurations obtained from Monte Carlo is then evolved in time
according to the equation of motion, 
\begin{eqnarray}
\frac{d {\sf S}_i}{dt}={\sf S}_i(t)\times {\sf H}_i(t) 
\end{eqnarray}
where
\begin{eqnarray}
{\sf H}_i(t)=\sum_{j \in {\sf nn} \ i} {\sf J}_{ij} \cdot {\sf S}_j(t)
\label{eq:exchangefield}
\end{eqnarray}
is the effective exchange field acting on site $i$, ${\sf J}_{ij}$ is the anisotropic 
exchange interaction tensor and the sum in Eq.~(\ref{eq:exchangefield})
runs over the neighbors of $i$.
The numerical integration of this nonlinear equation of motion proceeds
as described in Ref. 
[\onlinecite{taillefumier14}].


In Fig~\ref{fig:Goldstone.mode} we plot the dispersion of the Goldstone mode 
found in molecular dynamics (MD) simulations of $\mathcal{H}_{\sf XXZ}$ 
within the spin--nematic phase for $J_\pm/J_{\sf zz} = -1$.
The dispersion was extracted from the position of low--energy dispersing peak 
in $\chi_{Q_\perp}({\bf q},\omega)$, as shown in the inset to Fig.~\ref{fig:chi.Q.q.omega}.

\end{appendix}

\bibliography{article}

\begin{thebibliography}{125}%
\makeatletter
\providecommand \@ifxundefined [1]{%
 \@ifx{#1\undefined}
}%
\providecommand \@ifnum [1]{%
 \ifnum #1\expandafter \@firstoftwo
 \else \expandafter \@secondoftwo
 \fi
}%
\providecommand \@ifx [1]{%
 \ifx #1\expandafter \@firstoftwo
 \else \expandafter \@secondoftwo
 \fi
}%
\providecommand \natexlab [1]{#1}%
\providecommand \enquote  [1]{{\it #1}}%
\providecommand \bibnamefont  [1]{#1}%
\providecommand \bibfnamefont [1]{#1}%
\providecommand \citenamefont [1]{#1}%
\providecommand \href@noop [0]{\@secondoftwo}%
\providecommand \href [0]{\begingroup \@sanitize@url \@href}%
\providecommand \@href[1]{\@@startlink{#1}\@@href}%
\providecommand \@@href[1]{\endgroup#1\@@endlink}%
\providecommand \@sanitize@url [0]{\catcode `\\12\catcode `\$12\catcode
  `\&12\catcode `\#12\catcode `\^12\catcode `\_12\catcode `\%12\relax}%
\providecommand \@@startlink[1]{}%
\providecommand \@@endlink[0]{}%
\providecommand \url  [0]{\begingroup\@sanitize@url \@url }%
\providecommand \@url [1]{\endgroup\@href {#1}{\urlprefix }}%
\providecommand \urlprefix  [0]{URL }%
\providecommand \Eprint [0]{\href }%
\providecommand \doibase [0]{http://dx.doi.org/}%
\providecommand \selectlanguage [0]{\@gobble}%
\providecommand \bibinfo  [0]{\@secondoftwo}%
\providecommand \bibfield  [0]{\@secondoftwo}%
\providecommand \translation [1]{[#1]}%
\providecommand \BibitemOpen [0]{}%
\providecommand \bibitemStop [0]{}%
\providecommand \bibitemNoStop [0]{.\EOS\space}%
\providecommand \EOS [0]{\spacefactor3000\relax}%
\providecommand \BibitemShut  [1]{\csname bibitem#1\endcsname}%
\let\auto@bib@innerbib\@empty
\bibitem [{\citenamefont {Anderson}(1973)}]{anderson73}%
  \BibitemOpen
  \bibfield  {author} {\bibinfo {author} {\bibfnamefont {P.~W.}\ \bibnamefont
  {Anderson}},\ }\bibfield  {title} {\enquote {\bibinfo {title} {{Resonating
  valence bonds: A new kind of insulator?}}}\ }\href {\doibase
  10.1016/0025-5408(73)90167-0} {\bibfield  {journal} {\bibinfo  {journal}
  {Mater. Res. Bull.}\ }\textbf {\bibinfo {volume} {8}},\ \bibinfo {pages}
  {153--160} (\bibinfo {year} {1973})}\BibitemShut {NoStop}%
\bibitem [{\citenamefont {Lee}(2008)}]{lee08}%
  \BibitemOpen
  \bibfield  {author} {\bibinfo {author} {\bibfnamefont {P.~A.}\ \bibnamefont
  {Lee}},\ }\bibfield  {title} {\enquote {\bibinfo {title} {{An End to the
  Drought of Quantum Spin Liquids}},}\ }\href
  {http://science.sciencemag.org/content/321/5894/1306.abstract} {\bibfield
  {journal} {\bibinfo  {journal} {Science}\ }\textbf {\bibinfo {volume}
  {321}},\ \bibinfo {pages} {1306--1307} (\bibinfo {year} {2008})}\BibitemShut
  {NoStop}%
\bibitem [{\citenamefont {Balents}(2010)}]{balents10}%
  \BibitemOpen
  \bibfield  {author} {\bibinfo {author} {\bibfnamefont {L.}~\bibnamefont
  {Balents}},\ }\bibfield  {title} {\enquote {\bibinfo {title} {{Spin liquids
  in frustrated magnets}},}\ }\href
  {http://www.nature.com/nature/journal/v464/n7286/full/nature08917.html}
  {\bibfield  {journal} {\bibinfo  {journal} {Nature (London)}\ }\textbf
  {\bibinfo {volume} {464}},\ \bibinfo {pages} {199} (\bibinfo {year}
  {2010})}\BibitemShut {NoStop}%
\bibitem [{\citenamefont {Gardner}\ \emph {et~al.}(2010)\citenamefont
  {Gardner}, \citenamefont {Gingras},\ and\ \citenamefont
  {Greedan}}]{gardner10}%
  \BibitemOpen
  \bibfield  {author} {\bibinfo {author} {\bibfnamefont {Jason~S.}\
  \bibnamefont {Gardner}}, \bibinfo {author} {\bibfnamefont {Michel J.~P.}\
  \bibnamefont {Gingras}}, \ and\ \bibinfo {author} {\bibfnamefont {John~E.}\
  \bibnamefont {Greedan}},\ }\bibfield  {title} {\enquote {\bibinfo {title}
  {{Magnetic pyrochlore oxides}},}\ }\href
  {http://link.aps.org/doi/10.1103/RevModPhys.82.53} {\bibfield  {journal}
  {\bibinfo  {journal} {Rev. Mod. Phys.}\ }\textbf {\bibinfo {volume} {82}},\
  \bibinfo {pages} {53--107} (\bibinfo {year} {2010})}\BibitemShut {NoStop}%
\bibitem [{\citenamefont {Bramwell}\ and\ \citenamefont
  {Gingras}(2001)}]{bramwell01}%
  \BibitemOpen
  \bibfield  {author} {\bibinfo {author} {\bibfnamefont {S.~T.}\ \bibnamefont
  {Bramwell}}\ and\ \bibinfo {author} {\bibfnamefont {M.~J.~P.}\ \bibnamefont
  {Gingras}},\ }\bibfield  {title} {\enquote {\bibinfo {title} {{Spin Ice State
  in Frustrated Magnetic Pyrochlore Materials}},}\ }\href
  {http://science.sciencemag.org/content/294/5546/1495} {\bibfield  {journal}
  {\bibinfo  {journal} {Science}\ }\textbf {\bibinfo {volume} {294}} (\bibinfo
  {year} {2001})}\BibitemShut {NoStop}%
\bibitem [{\citenamefont {Castelnovo}\ \emph {et~al.}(2012)\citenamefont
  {Castelnovo}, \citenamefont {Moessner},\ and\ \citenamefont
  {Sondhi}}]{castelnovo12}%
  \BibitemOpen
  \bibfield  {author} {\bibinfo {author} {\bibfnamefont {C.}~\bibnamefont
  {Castelnovo}}, \bibinfo {author} {\bibfnamefont {R.}~\bibnamefont
  {Moessner}}, \ and\ \bibinfo {author} {\bibfnamefont {S.L.}\ \bibnamefont
  {Sondhi}},\ }\bibfield  {title} {\enquote {\bibinfo {title} {{Spin Ice,
  Fractionalization, and Topological Order}},}\ }\href
  {http://www.annualreviews.org/doi/10.1146/annurev-conmatphys-020911-125058}
  {\bibfield  {journal} {\bibinfo  {journal} {Annu. Rev. Condens. Matter
  Phys.}\ }\textbf {\bibinfo {volume} {3}},\ \bibinfo {pages} {35} (\bibinfo
  {year} {2012})}\BibitemShut {NoStop}%
\bibitem [{\citenamefont {Curnoe}(2007)}]{curnoe07}%
  \BibitemOpen
  \bibfield  {author} {\bibinfo {author} {\bibfnamefont {S.~H.}\ \bibnamefont
  {Curnoe}},\ }\bibfield  {title} {\enquote {\bibinfo {title} {{Quantum spin
  configurations in $\mathrm{{Tb}}_2\mathrm{{Ti}}_2\mathrm{{O}}_7$}},}\ }\href
  {\doibase 10.1103/PhysRevB.75.212404} {\bibfield  {journal} {\bibinfo
  {journal} {Phys. Rev. B}\ }\textbf {\bibinfo {volume} {75}},\ \bibinfo
  {pages} {212404} (\bibinfo {year} {2007})}\BibitemShut {NoStop}%
\bibitem [{\citenamefont {Molavian}\ \emph {et~al.}(2007)\citenamefont
  {Molavian}, \citenamefont {Gingras},\ and\ \citenamefont
  {Canals}}]{molavian07}%
  \BibitemOpen
  \bibfield  {author} {\bibinfo {author} {\bibfnamefont {H.~R.}\ \bibnamefont
  {Molavian}}, \bibinfo {author} {\bibfnamefont {M.~J.~P.}\ \bibnamefont
  {Gingras}}, \ and\ \bibinfo {author} {\bibfnamefont {B.}~\bibnamefont
  {Canals}},\ }\bibfield  {title} {\enquote {\bibinfo {title} {{Dynamically
  Induced Frustration as a Route to a Quantum Spin Ice State in
  ${\mathrm{Tb}}_{2}{\mathrm{Ti}}_{2}{\mathrm{O}}_{7}$ via Virtual Crystal
  Field Excitations and Quantum Many-Body Effects}},}\ }\href {\doibase
  10.1103/PhysRevLett.98.157204} {\bibfield  {journal} {\bibinfo  {journal}
  {Phys. Rev. Lett.}\ }\textbf {\bibinfo {volume} {98}},\ \bibinfo {pages}
  {157204} (\bibinfo {year} {2007})}\BibitemShut {NoStop}%
\bibitem [{\citenamefont {Onoda}\ and\ \citenamefont
  {Tanaka}(2010)}]{Onoda2010}%
  \BibitemOpen
  \bibfield  {author} {\bibinfo {author} {\bibfnamefont {S.}~\bibnamefont
  {Onoda}}\ and\ \bibinfo {author} {\bibfnamefont {Y.}~\bibnamefont {Tanaka}},\
  }\bibfield  {title} {\enquote {\bibinfo {title} {{Quantum Melting of Spin
  Ice: Emergent Cooperative Quadrupole and Chirality}},}\ }\href {\doibase
  10.1103/PhysRevLett.105.047201} {\bibfield  {journal} {\bibinfo  {journal}
  {Phys. Rev. Lett.}\ }\textbf {\bibinfo {volume} {105}},\ \bibinfo {pages}
  {047201} (\bibinfo {year} {2010})}\BibitemShut {NoStop}%
\bibitem [{\citenamefont {Onoda}\ and\ \citenamefont
  {Tanaka}(2011)}]{Onoda2011a}%
  \BibitemOpen
  \bibfield  {author} {\bibinfo {author} {\bibfnamefont {S.}~\bibnamefont
  {Onoda}}\ and\ \bibinfo {author} {\bibfnamefont {Y.}~\bibnamefont {Tanaka}},\
  }\bibfield  {title} {\enquote {\bibinfo {title} {{Quantum fluctuations in the
  effective pseudospin-$\frac{1}{2}$ model for magnetic pyrochlore oxides}},}\
  }\href {\doibase 10.1103/PhysRevB.83.094411} {\bibfield  {journal} {\bibinfo
  {journal} {Phys. Rev. B}\ }\textbf {\bibinfo {volume} {83}},\ \bibinfo
  {pages} {094411} (\bibinfo {year} {2011})}\BibitemShut {NoStop}%
\bibitem [{\citenamefont {Ross}\ \emph {et~al.}(2011)\citenamefont {Ross},
  \citenamefont {Savary}, \citenamefont {Gaulin},\ and\ \citenamefont
  {Balents}}]{ross11-PRX1}%
  \BibitemOpen
  \bibfield  {author} {\bibinfo {author} {\bibfnamefont {K.~A.}\ \bibnamefont
  {Ross}}, \bibinfo {author} {\bibfnamefont {L.}~\bibnamefont {Savary}},
  \bibinfo {author} {\bibfnamefont {B.~D.}\ \bibnamefont {Gaulin}}, \ and\
  \bibinfo {author} {\bibfnamefont {L.}~\bibnamefont {Balents}},\ }\bibfield
  {title} {\enquote {\bibinfo {title} {{Quantum Excitations in Quantum Spin
  Ice}},}\ }\href {\doibase 10.1103/PhysRevX.1.021002} {\bibfield  {journal}
  {\bibinfo  {journal} {Phys. Rev. X}\ }\textbf {\bibinfo {volume} {1}},\
  \bibinfo {pages} {021002} (\bibinfo {year} {2011})}\BibitemShut {NoStop}%
\bibitem [{\citenamefont {Gingras}\ and\ \citenamefont
  {McClarty}(2014)}]{mcclarty14}%
  \BibitemOpen
  \bibfield  {author} {\bibinfo {author} {\bibfnamefont {M.~J.~P.}\
  \bibnamefont {Gingras}}\ and\ \bibinfo {author} {\bibfnamefont {P.~A.}\
  \bibnamefont {McClarty}},\ }\bibfield  {title} {\enquote {\bibinfo {title}
  {{Quantum spin ice: a search for gapless quantum spin liquids in pyrochlore
  magnets}},}\ }\href {http://stacks.iop.org/0034-4885/77/i=5/a=056501}
  {\bibfield  {journal} {\bibinfo  {journal} {Rep. Prog. Phys.}\ }\textbf
  {\bibinfo {volume} {77}},\ \bibinfo {pages} {056501} (\bibinfo {year}
  {2014})}\BibitemShut {NoStop}%
\bibitem [{\citenamefont {Hermele}\ \emph {et~al.}(2004)\citenamefont
  {Hermele}, \citenamefont {Fisher},\ and\ \citenamefont
  {Balents}}]{hermele04}%
  \BibitemOpen
  \bibfield  {author} {\bibinfo {author} {\bibfnamefont {M.}~\bibnamefont
  {Hermele}}, \bibinfo {author} {\bibfnamefont {M.~P.~A.}\ \bibnamefont
  {Fisher}}, \ and\ \bibinfo {author} {\bibfnamefont {L.}~\bibnamefont
  {Balents}},\ }\bibfield  {title} {\enquote {\bibinfo {title} {{Pyrochlore
  photons: The \protect{U(1)} spin liquid in a \protect{S=$\frac{1}{2}$}
  three-dimensional frustrated magnet}},}\ }\href
  {http://link.aps.org/doi/10.1103/PhysRevB.69.064404} {\bibfield  {journal}
  {\bibinfo  {journal} {Phys. Rev. B}\ }\textbf {\bibinfo {volume} {69}},\
  \bibinfo {pages} {064404} (\bibinfo {year} {2004})}\BibitemShut {NoStop}%
\bibitem [{\citenamefont {Banerjee}\ \emph {et~al.}(2008)\citenamefont
  {Banerjee}, \citenamefont {Isakov}, \citenamefont {Damle},\ and\
  \citenamefont {Kim}}]{banerjee08}%
  \BibitemOpen
  \bibfield  {author} {\bibinfo {author} {\bibfnamefont {A.}~\bibnamefont
  {Banerjee}}, \bibinfo {author} {\bibfnamefont {S.~V.}\ \bibnamefont
  {Isakov}}, \bibinfo {author} {\bibfnamefont {K.}~\bibnamefont {Damle}}, \
  and\ \bibinfo {author} {\bibfnamefont {Y.~B.}\ \bibnamefont {Kim}},\
  }\bibfield  {title} {\enquote {\bibinfo {title} {{Unusual Liquid State of
  Hard-Core Bosons on the Pyrochlore Lattice}},}\ }\href
  {http://link.aps.org/doi/10.1103/PhysRevLett.100.047208} {\bibfield
  {journal} {\bibinfo  {journal} {Phys. Rev. Lett.}\ }\textbf {\bibinfo
  {volume} {100}},\ \bibinfo {pages} {047208} (\bibinfo {year}
  {2008})}\BibitemShut {NoStop}%
\bibitem [{\citenamefont {Shannon}\ \emph {et~al.}(2012)\citenamefont
  {Shannon}, \citenamefont {Sikora}, \citenamefont {Pollmann}, \citenamefont
  {Penc},\ and\ \citenamefont {Fulde}}]{shannon12}%
  \BibitemOpen
  \bibfield  {author} {\bibinfo {author} {\bibfnamefont {N.}~\bibnamefont
  {Shannon}}, \bibinfo {author} {\bibfnamefont {O.}~\bibnamefont {Sikora}},
  \bibinfo {author} {\bibfnamefont {F.}~\bibnamefont {Pollmann}}, \bibinfo
  {author} {\bibfnamefont {K.}~\bibnamefont {Penc}}, \ and\ \bibinfo {author}
  {\bibfnamefont {P.}~\bibnamefont {Fulde}},\ }\bibfield  {title} {\enquote
  {\bibinfo {title} {{Quantum Ice: A Quantum Monte Carlo Study}},}\ }\href
  {\doibase 10.1103/PhysRevLett.108.067204} {\bibfield  {journal} {\bibinfo
  {journal} {Phys. Rev. Lett.}\ }\textbf {\bibinfo {volume} {108}},\ \bibinfo
  {pages} {067204} (\bibinfo {year} {2012})}\BibitemShut {NoStop}%
\bibitem [{\citenamefont {Benton}\ \emph {et~al.}(2012)\citenamefont {Benton},
  \citenamefont {Sikora},\ and\ \citenamefont {Shannon}}]{benton12}%
  \BibitemOpen
  \bibfield  {author} {\bibinfo {author} {\bibfnamefont {O.}~\bibnamefont
  {Benton}}, \bibinfo {author} {\bibfnamefont {O.}~\bibnamefont {Sikora}}, \
  and\ \bibinfo {author} {\bibfnamefont {N.}~\bibnamefont {Shannon}},\
  }\bibfield  {title} {\enquote {\bibinfo {title} {{Seeing the light:
  Experimental signatures of emergent electromagnetism in a quantum spin
  ice}},}\ }\href
  {https://journals.aps.org/prb/abstract/10.1103/PhysRevB.86.075154} {\bibfield
   {journal} {\bibinfo  {journal} {Phys. Rev. B}\ }\textbf {\bibinfo {volume}
  {86}},\ \bibinfo {pages} {075154} (\bibinfo {year} {2012})}\BibitemShut
  {NoStop}%
\bibitem [{\citenamefont {Savary}\ and\ \citenamefont
  {Balents}(2012)}]{savary12-PRL108}%
  \BibitemOpen
  \bibfield  {author} {\bibinfo {author} {\bibfnamefont {L.}~\bibnamefont
  {Savary}}\ and\ \bibinfo {author} {\bibfnamefont {L.}~\bibnamefont
  {Balents}},\ }\bibfield  {title} {\enquote {\bibinfo {title} {{Coulombic
  Quantum Liquids in Spin-$1/2$ Pyrochlores}},}\ }\href {\doibase
  10.1103/PhysRevLett.108.037202} {\bibfield  {journal} {\bibinfo  {journal}
  {Phys. Rev. Lett.}\ }\textbf {\bibinfo {volume} {108}},\ \bibinfo {pages}
  {037202} (\bibinfo {year} {2012})}\BibitemShut {NoStop}%
\bibitem [{\citenamefont {Lee}\ \emph {et~al.}(2012)\citenamefont {Lee},
  \citenamefont {Onoda},\ and\ \citenamefont {Balents}}]{Lee12}%
  \BibitemOpen
  \bibfield  {author} {\bibinfo {author} {\bibfnamefont {S.~B.}\ \bibnamefont
  {Lee}}, \bibinfo {author} {\bibfnamefont {S.}~\bibnamefont {Onoda}}, \ and\
  \bibinfo {author} {\bibfnamefont {L.}~\bibnamefont {Balents}},\ }\bibfield
  {title} {\enquote {\bibinfo {title} {{Generic quantum spin ice}},}\ }\href
  {http://link.aps.org/doi/10.1103/PhysRevB.86.104412} {\bibfield  {journal}
  {\bibinfo  {journal} {Phys. Rev. B}\ }\textbf {\bibinfo {volume} {86}},\
  \bibinfo {pages} {104412} (\bibinfo {year} {2012})}\BibitemShut {NoStop}%
\bibitem [{\citenamefont {Hao}\ \emph {et~al.}(2014)\citenamefont {Hao},
  \citenamefont {Day},\ and\ \citenamefont {Gingras}}]{Hao2014}%
  \BibitemOpen
  \bibfield  {author} {\bibinfo {author} {\bibfnamefont {Z.}~\bibnamefont
  {Hao}}, \bibinfo {author} {\bibfnamefont {A.~G.~R.}\ \bibnamefont {Day}}, \
  and\ \bibinfo {author} {\bibfnamefont {M.~J.~P.}\ \bibnamefont {Gingras}},\
  }\bibfield  {title} {\enquote {\bibinfo {title} {{Bosonic many-body theory of
  quantum spin ice}},}\ }\href {\doibase 10.1103/PhysRevB.90.214430} {\bibfield
   {journal} {\bibinfo  {journal} {Phys. Rev. B}\ }\textbf {\bibinfo {volume}
  {90}},\ \bibinfo {pages} {214430} (\bibinfo {year} {2014})}\BibitemShut
  {NoStop}%
\bibitem [{\citenamefont {McClarty}\ \emph {et~al.}(2015)\citenamefont
  {McClarty}, \citenamefont {Sikora}, \citenamefont {Moessner}, \citenamefont
  {Penc}, \citenamefont {Pollmann},\ and\ \citenamefont
  {Shannon}}]{mcclarty15}%
  \BibitemOpen
  \bibfield  {author} {\bibinfo {author} {\bibfnamefont {P.~A.}\ \bibnamefont
  {McClarty}}, \bibinfo {author} {\bibfnamefont {O.}~\bibnamefont {Sikora}},
  \bibinfo {author} {\bibfnamefont {R.}~\bibnamefont {Moessner}}, \bibinfo
  {author} {\bibfnamefont {K.}~\bibnamefont {Penc}}, \bibinfo {author}
  {\bibfnamefont {F.}~\bibnamefont {Pollmann}}, \ and\ \bibinfo {author}
  {\bibfnamefont {N.}~\bibnamefont {Shannon}},\ }\bibfield  {title} {\enquote
  {\bibinfo {title} {{Chain-based order and quantum spin liquids in dipolar
  spin ice}},}\ }\href {\doibase 10.1103/PhysRevB.92.094418} {\bibfield
  {journal} {\bibinfo  {journal} {Phys. Rev. B}\ }\textbf {\bibinfo {volume}
  {92}},\ \bibinfo {pages} {094418} (\bibinfo {year} {2015})}\BibitemShut
  {NoStop}%
\bibitem [{\citenamefont {Kato}\ and\ \citenamefont {Onoda}(2015)}]{kato15}%
  \BibitemOpen
  \bibfield  {author} {\bibinfo {author} {\bibfnamefont {Y.}~\bibnamefont
  {Kato}}\ and\ \bibinfo {author} {\bibfnamefont {S.}~\bibnamefont {Onoda}},\
  }\bibfield  {title} {\enquote {\bibinfo {title} {{Numerical Evidence of
  Quantum Melting of Spin Ice: Quantum-to-Classical Crossover}},}\ }\href
  {\doibase 10.1103/PhysRevLett.115.077202} {\bibfield  {journal} {\bibinfo
  {journal} {Phys. Rev. Lett.}\ }\textbf {\bibinfo {volume} {115}},\ \bibinfo
  {pages} {077202} (\bibinfo {year} {2015})}\BibitemShut {NoStop}%
\bibitem [{\citenamefont {Chen}(2016)}]{chen16}%
  \BibitemOpen
  \bibfield  {author} {\bibinfo {author} {\bibfnamefont {G.}~\bibnamefont
  {Chen}},\ }\bibfield  {title} {\enquote {\bibinfo {title} {{ ``Magnetic
  monopole'' condensation of the pyrochlore ice U(1) quantum spin liquid:
  Application to ${\mathrm{Pr}}_{2}{\mathrm{Ir}}_{2}{\mathrm{O}}_{7}$ and
  ${\mathrm{Yb}}_{2}{\mathrm{Ti}}_{2}{\mathrm{O}}_{7}$}},}\ }\href {\doibase
  10.1103/PhysRevB.94.205107} {\bibfield  {journal} {\bibinfo  {journal} {Phys.
  Rev. B}\ }\textbf {\bibinfo {volume} {94}},\ \bibinfo {pages} {205107}
  (\bibinfo {year} {2016})}\BibitemShut {NoStop}%
\bibitem [{\citenamefont {Shannon}(2017)}]{shannon-book-chapter}%
  \BibitemOpen
  \bibfield  {author} {\bibinfo {author} {\bibfnamefont {N.}~\bibnamefont
  {Shannon}},\ }\enquote {\bibinfo {title} {{Spin Ice}},}\ \ (\bibinfo
  {publisher} {Springer},\ \bibinfo {year} {2017})\ Chap.\ \bibinfo {chapter}
  {``Quantum Monte Carlo simulations of quantum spin ice''}\BibitemShut
  {NoStop}%
\bibitem [{\citenamefont {Savary}\ and\ \citenamefont
  {Balents}(2017{\natexlab{a}})}]{savary17-PRL118}%
  \BibitemOpen
  \bibfield  {author} {\bibinfo {author} {\bibfnamefont {L.}~\bibnamefont
  {Savary}}\ and\ \bibinfo {author} {\bibfnamefont {L.}~\bibnamefont
  {Balents}},\ }\bibfield  {title} {\enquote {\bibinfo {title}
  {{Disorder-Induced Quantum Spin Liquid in Spin Ice Pyrochlores}},}\ }\href
  {\doibase 10.1103/PhysRevLett.118.087203} {\bibfield  {journal} {\bibinfo
  {journal} {Phys. Rev. Lett.}\ }\textbf {\bibinfo {volume} {118}},\ \bibinfo
  {pages} {087203} (\bibinfo {year} {2017}{\natexlab{a}})}\BibitemShut
  {NoStop}%
\bibitem [{\citenamefont {Chen}(2017{\natexlab{a}})}]{chen-arXiv.1706.04333}%
  \BibitemOpen
  \bibfield  {author} {\bibinfo {author} {\bibfnamefont {Gang}\ \bibnamefont
  {Chen}},\ }\bibfield  {title} {\enquote {\bibinfo {title} {{Dirac's
  ``magnetic monopoles'' in pyrochlore ice $U(1)$ spin liquids: Spectrum and
  classification}},}\ }\href {\doibase 10.1103/PhysRevB.96.195127} {\bibfield
  {journal} {\bibinfo  {journal} {Phys. Rev. B}\ }\textbf {\bibinfo {volume}
  {96}},\ \bibinfo {pages} {195127} (\bibinfo {year}
  {2017}{\natexlab{a}})}\BibitemShut {NoStop}%
\bibitem [{\citenamefont {Huang}\ \emph {et~al.}()\citenamefont {Huang},
  \citenamefont {Deng}, \citenamefont {Wan},\ and\ \citenamefont
  {Meng}}]{huang-arXiv.1707.00099}%
  \BibitemOpen
  \bibfield  {author} {\bibinfo {author} {\bibfnamefont {C.-J.}\ \bibnamefont
  {Huang}}, \bibinfo {author} {\bibfnamefont {Y.}~\bibnamefont {Deng}},
  \bibinfo {author} {\bibfnamefont {Y.}~\bibnamefont {Wan}}, \ and\ \bibinfo
  {author} {\bibfnamefont {Z.~Y.}\ \bibnamefont {Meng}},\ }\bibfield  {title}
  {\enquote {\bibinfo {title} {{Dynamics of topological excitations in a model
  quantum spin ice}},}\ }\href {https://arxiv.org/abs/1707.00099} {\bibinfo
  {journal} {arXiv:1707.00099}\ }\BibitemShut {NoStop}%
\bibitem [{\citenamefont {Zhou}\ \emph {et~al.}(2008)\citenamefont {Zhou},
  \citenamefont {Wiebe}, \citenamefont {Janik}, \citenamefont {Balicas},
  \citenamefont {Yo}, \citenamefont {Qiu}, \citenamefont {Copley},\ and\
  \citenamefont {Gardner}}]{Zhou2008}%
  \BibitemOpen
\bibfield  {journal} {  }\bibfield  {author} {\bibinfo {author} {\bibfnamefont
  {H.~D.}\ \bibnamefont {Zhou}}, \bibinfo {author} {\bibfnamefont {C.~R.}\
  \bibnamefont {Wiebe}}, \bibinfo {author} {\bibfnamefont {J.~A.}\ \bibnamefont
  {Janik}}, \bibinfo {author} {\bibfnamefont {L.}~\bibnamefont {Balicas}},
  \bibinfo {author} {\bibfnamefont {Y.~J.}\ \bibnamefont {Yo}}, \bibinfo
  {author} {\bibfnamefont {Y.}~\bibnamefont {Qiu}}, \bibinfo {author}
  {\bibfnamefont {J.~R.~D.}\ \bibnamefont {Copley}}, \ and\ \bibinfo {author}
  {\bibfnamefont {J.~S.}\ \bibnamefont {Gardner}},\ }\bibfield  {title}
  {\enquote {\bibinfo {title} {{Dynamic Spin Ice:
  ${\mathrm{Pr}}_{2}{\mathrm{Sn}}_{2}{\mathrm{O}}_{7}$}},}\ }\href {\doibase
  10.1103/PhysRevLett.101.227204} {\bibfield  {journal} {\bibinfo  {journal}
  {Phys. Rev. Lett.}\ }\textbf {\bibinfo {volume} {101}},\ \bibinfo {pages}
  {227204} (\bibinfo {year} {2008})}\BibitemShut {NoStop}%
\bibitem [{\citenamefont {Chang}\ \emph {et~al.}(2012)\citenamefont {Chang},
  \citenamefont {Onoda}, \citenamefont {Su}, \citenamefont {Kao}, \citenamefont
  {Tsuei}, \citenamefont {Yasui}, \citenamefont {Kakurai},\ and\ \citenamefont
  {Lees}}]{chang12-NatCommun3}%
  \BibitemOpen
  \bibfield  {author} {\bibinfo {author} {\bibfnamefont {L.~J.}\ \bibnamefont
  {Chang}}, \bibinfo {author} {\bibfnamefont {S.}~\bibnamefont {Onoda}},
  \bibinfo {author} {\bibfnamefont {Y.}~\bibnamefont {Su}}, \bibinfo {author}
  {\bibfnamefont {Y.~J.}\ \bibnamefont {Kao}}, \bibinfo {author} {\bibfnamefont
  {K.~D.}\ \bibnamefont {Tsuei}}, \bibinfo {author} {\bibfnamefont
  {Y.}~\bibnamefont {Yasui}}, \bibinfo {author} {\bibfnamefont
  {K.}~\bibnamefont {Kakurai}}, \ and\ \bibinfo {author} {\bibfnamefont
  {M.~R.}\ \bibnamefont {Lees}},\ }\bibfield  {title} {\enquote {\bibinfo
  {title} {{Higgs transition from a magnetic Coulomb liquid to a ferromagnet in
  $\mathrm{{Yb}}_2\mathrm{{Ti}}_2\mathrm{{O}}_7$}},}\ }\href
  {http://www.nature.com/articles/ncomms1989#a9} {\bibfield  {journal}
  {\bibinfo  {journal} {Nat. Commun.}\ }\textbf {\bibinfo {volume} {3}},\
  \bibinfo {pages} {992} (\bibinfo {year} {2012})}\BibitemShut {NoStop}%
\bibitem [{\citenamefont {Fennell}\ \emph {et~al.}(2012)\citenamefont
  {Fennell}, \citenamefont {Kenzelmann}, \citenamefont {Roessli}, \citenamefont
  {Haas},\ and\ \citenamefont {Cava}}]{Fennell2012}%
  \BibitemOpen
  \bibfield  {author} {\bibinfo {author} {\bibfnamefont {T.}~\bibnamefont
  {Fennell}}, \bibinfo {author} {\bibfnamefont {M.}~\bibnamefont {Kenzelmann}},
  \bibinfo {author} {\bibfnamefont {B.}~\bibnamefont {Roessli}}, \bibinfo
  {author} {\bibfnamefont {M.~K.}\ \bibnamefont {Haas}}, \ and\ \bibinfo
  {author} {\bibfnamefont {R.~J.}\ \bibnamefont {Cava}},\ }\bibfield  {title}
  {\enquote {\bibinfo {title} {{Power-Law Spin Correlations in the Pyrochlore
  Antiferromagnet ${\mathrm{Tb}}_{2}{\mathrm{Ti}}_{2}{\mathrm{O}}_{7}$}},}\
  }\href {\doibase 10.1103/PhysRevLett.109.017201} {\bibfield  {journal}
  {\bibinfo  {journal} {Phys. Rev. Lett.}\ }\textbf {\bibinfo {volume} {109}},\
  \bibinfo {pages} {017201} (\bibinfo {year} {2012})}\BibitemShut {NoStop}%
\bibitem [{\citenamefont {Kimura}\ \emph {et~al.}(2013)\citenamefont {Kimura},
  \citenamefont {Nakatsuji}, \citenamefont {Wen}, \citenamefont {Broholm},
  \citenamefont {Stone}, \citenamefont {Nishibori},\ and\ \citenamefont
  {Sawa}}]{Kimura2013}%
  \BibitemOpen
  \bibfield  {author} {\bibinfo {author} {\bibfnamefont {K.}~\bibnamefont
  {Kimura}}, \bibinfo {author} {\bibfnamefont {S.}~\bibnamefont {Nakatsuji}},
  \bibinfo {author} {\bibfnamefont {J-J.}\ \bibnamefont {Wen}}, \bibinfo
  {author} {\bibfnamefont {C.}~\bibnamefont {Broholm}}, \bibinfo {author}
  {\bibfnamefont {M.~B.}\ \bibnamefont {Stone}}, \bibinfo {author}
  {\bibfnamefont {E.}~\bibnamefont {Nishibori}}, \ and\ \bibinfo {author}
  {\bibfnamefont {H.}~\bibnamefont {Sawa}},\ }\bibfield  {title} {\enquote
  {\bibinfo {title} {{Quantum fluctuations in spin-ice-like
  $\mathrm{{Pr}}_2\mathrm{{Zr}}_2\mathrm{{O}}_7$}},}\ }\href
  {http://dx.doi.org/10.1038/ncomms2914} {\bibfield  {journal} {\bibinfo
  {journal} {Nat. Commun.}\ }\textbf {\bibinfo {volume} {4}},\ \bibinfo {pages}
  {1934} (\bibinfo {year} {2013})}\BibitemShut {NoStop}%
\bibitem [{\citenamefont {Sibille}\ \emph {et~al.}(2015)\citenamefont
  {Sibille}, \citenamefont {Lhotel}, \citenamefont {Pomjakushin}, \citenamefont
  {Baines}, \citenamefont {Fennell},\ and\ \citenamefont
  {Kenzelmann}}]{Sibille2015}%
  \BibitemOpen
  \bibfield  {author} {\bibinfo {author} {\bibfnamefont {R.}~\bibnamefont
  {Sibille}}, \bibinfo {author} {\bibfnamefont {E.}~\bibnamefont {Lhotel}},
  \bibinfo {author} {\bibfnamefont {V.}~\bibnamefont {Pomjakushin}}, \bibinfo
  {author} {\bibfnamefont {C.}~\bibnamefont {Baines}}, \bibinfo {author}
  {\bibfnamefont {T.}~\bibnamefont {Fennell}}, \ and\ \bibinfo {author}
  {\bibfnamefont {M.}~\bibnamefont {Kenzelmann}},\ }\bibfield  {title}
  {\enquote {\bibinfo {title} {{Candidate Quantum Spin Liquid in the
  ${\mathrm{Ce}}^{3+}$ Pyrochlore Stannate
  ${\mathrm{Ce}}_{2}{\mathrm{Sn}}_{2}{\mathrm{O}}_{7}$}},}\ }\href {\doibase
  10.1103/PhysRevLett.115.097202} {\bibfield  {journal} {\bibinfo  {journal}
  {Phys. Rev. Lett.}\ }\textbf {\bibinfo {volume} {115}},\ \bibinfo {pages}
  {097202} (\bibinfo {year} {2015})}\BibitemShut {NoStop}%
\bibitem [{\citenamefont {Sibille}\ \emph {et~al.}(2016)\citenamefont
  {Sibille}, \citenamefont {Lhotel}, \citenamefont {Hatnean}, \citenamefont
  {Balakrishnan}, \citenamefont {F\aa{}k}, \citenamefont {Gauthier},
  \citenamefont {Fennell},\ and\ \citenamefont {Kenzelmann}}]{Sibille16a}%
  \BibitemOpen
  \bibfield  {author} {\bibinfo {author} {\bibfnamefont {R.}~\bibnamefont
  {Sibille}}, \bibinfo {author} {\bibfnamefont {E.}~\bibnamefont {Lhotel}},
  \bibinfo {author} {\bibfnamefont {M.~C.}\ \bibnamefont {Hatnean}}, \bibinfo
  {author} {\bibfnamefont {G.}~\bibnamefont {Balakrishnan}}, \bibinfo {author}
  {\bibfnamefont {B.}~\bibnamefont {F\aa{}k}}, \bibinfo {author} {\bibfnamefont
  {N.}~\bibnamefont {Gauthier}}, \bibinfo {author} {\bibfnamefont
  {T.}~\bibnamefont {Fennell}}, \ and\ \bibinfo {author} {\bibfnamefont
  {M.}~\bibnamefont {Kenzelmann}},\ }\bibfield  {title} {\enquote {\bibinfo
  {title} {{Candidate quantum spin ice in the pyrochlore
  ${\mathrm{Pr}}_{2}{\mathrm{Hf}}_{2}{\mathrm{O}}_{7}$}},}\ }\href {\doibase
  10.1103/PhysRevB.94.024436} {\bibfield  {journal} {\bibinfo  {journal} {Phys.
  Rev. B}\ }\textbf {\bibinfo {volume} {94}},\ \bibinfo {pages} {024436}
  (\bibinfo {year} {2016})}\BibitemShut {NoStop}%
\bibitem [{\citenamefont {Anand}\ \emph {et~al.}(2016)\citenamefont {Anand},
  \citenamefont {Opherden}, \citenamefont {Xu}, \citenamefont {Adroja},
  \citenamefont {Islam}, \citenamefont {Herrmannsd\"orfer}, \citenamefont
  {Hornung}, \citenamefont {Sch\"onemann}, \citenamefont {Uhlarz},
  \citenamefont {Walker}, \citenamefont {Casati},\ and\ \citenamefont
  {Lake}}]{Anand16a}%
  \BibitemOpen
  \bibfield  {author} {\bibinfo {author} {\bibfnamefont {V.~K.}\ \bibnamefont
  {Anand}}, \bibinfo {author} {\bibfnamefont {L.}~\bibnamefont {Opherden}},
  \bibinfo {author} {\bibfnamefont {J.}~\bibnamefont {Xu}}, \bibinfo {author}
  {\bibfnamefont {D.~T.}\ \bibnamefont {Adroja}}, \bibinfo {author}
  {\bibfnamefont {A.~T. M.~N.}\ \bibnamefont {Islam}}, \bibinfo {author}
  {\bibfnamefont {T.}~\bibnamefont {Herrmannsd\"orfer}}, \bibinfo {author}
  {\bibfnamefont {J.}~\bibnamefont {Hornung}}, \bibinfo {author} {\bibfnamefont
  {R.}~\bibnamefont {Sch\"onemann}}, \bibinfo {author} {\bibfnamefont
  {M.}~\bibnamefont {Uhlarz}}, \bibinfo {author} {\bibfnamefont {H.~C.}\
  \bibnamefont {Walker}}, \bibinfo {author} {\bibfnamefont {N.}~\bibnamefont
  {Casati}}, \ and\ \bibinfo {author} {\bibfnamefont {B.}~\bibnamefont
  {Lake}},\ }\bibfield  {title} {\enquote {\bibinfo {title} {{Physical
  properties of the candidate quantum spin-ice system
  ${\mathrm{Pr}}_{2}{\mathrm{Hf}}_{2}{\mathrm{O}}_{7}$}},}\ }\href {\doibase
  10.1103/PhysRevB.94.144415} {\bibfield  {journal} {\bibinfo  {journal} {Phys.
  Rev. B}\ }\textbf {\bibinfo {volume} {94}},\ \bibinfo {pages} {144415}
  (\bibinfo {year} {2016})}\BibitemShut {NoStop}%
\bibitem [{\citenamefont {Wen}\ \emph {et~al.}(2017)\citenamefont {Wen},
  \citenamefont {Koohpayeh}, \citenamefont {Ross}, \citenamefont {Trump},
  \citenamefont {McQueen}, \citenamefont {Kimura}, \citenamefont {Nakatsuji},
  \citenamefont {Qiu}, \citenamefont {Pajerowski}, \citenamefont {Copley},\
  and\ \citenamefont {Broholm}}]{Wen2017}%
  \BibitemOpen
  \bibfield  {author} {\bibinfo {author} {\bibfnamefont {J.-J.}\ \bibnamefont
  {Wen}}, \bibinfo {author} {\bibfnamefont {S.~M.}\ \bibnamefont {Koohpayeh}},
  \bibinfo {author} {\bibfnamefont {K.~A.}\ \bibnamefont {Ross}}, \bibinfo
  {author} {\bibfnamefont {B.~A.}\ \bibnamefont {Trump}}, \bibinfo {author}
  {\bibfnamefont {T.~M.}\ \bibnamefont {McQueen}}, \bibinfo {author}
  {\bibfnamefont {K.}~\bibnamefont {Kimura}}, \bibinfo {author} {\bibfnamefont
  {S.}~\bibnamefont {Nakatsuji}}, \bibinfo {author} {\bibfnamefont
  {Y.}~\bibnamefont {Qiu}}, \bibinfo {author} {\bibfnamefont {D.~M.}\
  \bibnamefont {Pajerowski}}, \bibinfo {author} {\bibfnamefont {J.~R.~D.}\
  \bibnamefont {Copley}}, \ and\ \bibinfo {author} {\bibfnamefont {C.~L.}\
  \bibnamefont {Broholm}},\ }\bibfield  {title} {\enquote {\bibinfo {title}
  {{Disordered Route to the Coulomb Quantum Spin Liquid: Random Transverse
  Fields on Spin Ice in
  ${\mathrm{Pr}}_{2}{\mathrm{Zr}}_{2}{\mathrm{O}}_{7}$}},}\ }\href {\doibase
  10.1103/PhysRevLett.118.107206} {\bibfield  {journal} {\bibinfo  {journal}
  {Phys. Rev. Lett.}\ }\textbf {\bibinfo {volume} {118}},\ \bibinfo {pages}
  {107206} (\bibinfo {year} {2017})}\BibitemShut {NoStop}%
\bibitem [{\citenamefont {Sibille}\ \emph {et~al.}()\citenamefont {Sibille},
  \citenamefont {Gauthier}, \citenamefont {Yan}, \citenamefont {Hatnean},
  \citenamefont {Ollivier}, \citenamefont {Winn}, \citenamefont {Balakrishnan},
  \citenamefont {Kenzelmann}, \citenamefont {Shannon},\ and\ \citenamefont
  {Fennell}}]{sibille-arXiv}%
  \BibitemOpen
  \bibfield  {author} {\bibinfo {author} {\bibfnamefont {R.}~\bibnamefont
  {Sibille}}, \bibinfo {author} {\bibfnamefont {N.}~\bibnamefont {Gauthier}},
  \bibinfo {author} {\bibfnamefont {H.}~\bibnamefont {Yan}}, \bibinfo {author}
  {\bibfnamefont {M.~C.}\ \bibnamefont {Hatnean}}, \bibinfo {author}
  {\bibfnamefont {J.}~\bibnamefont {Ollivier}}, \bibinfo {author}
  {\bibfnamefont {B.}~\bibnamefont {Winn}}, \bibinfo {author} {\bibfnamefont
  {G.}~\bibnamefont {Balakrishnan}}, \bibinfo {author} {\bibfnamefont
  {M.}~\bibnamefont {Kenzelmann}}, \bibinfo {author} {\bibfnamefont
  {N.}~\bibnamefont {Shannon}}, \ and\ \bibinfo {author} {\bibfnamefont
  {T.}~\bibnamefont {Fennell}},\ }\bibfield  {title} {\enquote {\bibinfo
  {title} {{Experimental signatures of emergent quantum electrodynamics in a
  quantum spin ice}},}\ }\href {http://arxiv.org/abs/1706.03604} {\bibinfo
  {journal} {arXiv:1706.03604}\ }\BibitemShut {NoStop}%
\bibitem [{\citenamefont {Dalmas~de R\'eotier}\ \emph
  {et~al.}(2006)\citenamefont {Dalmas~de R\'eotier}, \citenamefont {Yaouanc},
  \citenamefont {Keller}, \citenamefont {Cervellino}, \citenamefont {Roessli},
  \citenamefont {Baines}, \citenamefont {Forget}, \citenamefont {Vaju},
  \citenamefont {Gubbens}, \citenamefont {Amato},\ and\ \citenamefont
  {King}}]{DalmasDeReotier2006}%
  \BibitemOpen
\bibfield  {journal} {  }\bibfield  {author} {\bibinfo {author} {\bibfnamefont
  {P.}~\bibnamefont {Dalmas~de R\'eotier}}, \bibinfo {author} {\bibfnamefont
  {A.}~\bibnamefont {Yaouanc}}, \bibinfo {author} {\bibfnamefont
  {L.}~\bibnamefont {Keller}}, \bibinfo {author} {\bibfnamefont
  {A.}~\bibnamefont {Cervellino}}, \bibinfo {author} {\bibfnamefont
  {B.}~\bibnamefont {Roessli}}, \bibinfo {author} {\bibfnamefont
  {C.}~\bibnamefont {Baines}}, \bibinfo {author} {\bibfnamefont
  {A.}~\bibnamefont {Forget}}, \bibinfo {author} {\bibfnamefont
  {C.}~\bibnamefont {Vaju}}, \bibinfo {author} {\bibfnamefont {P.~C.~M.}\
  \bibnamefont {Gubbens}}, \bibinfo {author} {\bibfnamefont {A.}~\bibnamefont
  {Amato}}, \ and\ \bibinfo {author} {\bibfnamefont {P.~J.~C.}\ \bibnamefont
  {King}},\ }\bibfield  {title} {\enquote {\bibinfo {title} {{Spin Dynamics and
  Magnetic Order in Magnetically Frustrated
  ${\mathrm{Tb}}_{2}{\mathrm{Sn}}_{2}{\mathrm{O}}_{7}$}},}\ }\href {\doibase
  10.1103/PhysRevLett.96.127202} {\bibfield  {journal} {\bibinfo  {journal}
  {Phys. Rev. Lett.}\ }\textbf {\bibinfo {volume} {96}},\ \bibinfo {pages}
  {127202} (\bibinfo {year} {2006})}\BibitemShut {NoStop}%
\bibitem [{\citenamefont {Dun}\ \emph {et~al.}(2013)\citenamefont {Dun},
  \citenamefont {Choi}, \citenamefont {Zhou}, \citenamefont {Hallas},
  \citenamefont {Silverstein}, \citenamefont {Qiu}, \citenamefont {Copley},
  \citenamefont {Gardner},\ and\ \citenamefont {Wiebe}}]{dun13}%
  \BibitemOpen
  \bibfield  {author} {\bibinfo {author} {\bibfnamefont {Z.~L.}\ \bibnamefont
  {Dun}}, \bibinfo {author} {\bibfnamefont {E.~S.}\ \bibnamefont {Choi}},
  \bibinfo {author} {\bibfnamefont {H.~D.}\ \bibnamefont {Zhou}}, \bibinfo
  {author} {\bibfnamefont {A.~M.}\ \bibnamefont {Hallas}}, \bibinfo {author}
  {\bibfnamefont {H.~J.}\ \bibnamefont {Silverstein}}, \bibinfo {author}
  {\bibfnamefont {Y.}~\bibnamefont {Qiu}}, \bibinfo {author} {\bibfnamefont
  {J.~R.~D.}\ \bibnamefont {Copley}}, \bibinfo {author} {\bibfnamefont {J.~S.}\
  \bibnamefont {Gardner}}, \ and\ \bibinfo {author} {\bibfnamefont {C.~R.}\
  \bibnamefont {Wiebe}},\ }\bibfield  {title} {\enquote {\bibinfo {title}
  {{$\mathrm{{Yb}}_2\mathrm{{Sn}}_2\mathrm{{O}}_7$: A magnetic Coulomb liquid
  at a quantum critical point}},}\ }\href {\doibase 10.1103/PhysRevB.87.134408}
  {\bibfield  {journal} {\bibinfo  {journal} {Phys. Rev. B}\ }\textbf {\bibinfo
  {volume} {87}},\ \bibinfo {pages} {134408} (\bibinfo {year}
  {2013})}\BibitemShut {NoStop}%
\bibitem [{\citenamefont {Yaouanc}\ \emph {et~al.}(2013)\citenamefont
  {Yaouanc}, \citenamefont {Dalmas~de R\'eotier}, \citenamefont {Bonville},
  \citenamefont {Hodges}, \citenamefont {Glazkov}, \citenamefont {Keller},
  \citenamefont {Sikolenko}, \citenamefont {Bartkowiak}, \citenamefont {Amato},
  \citenamefont {Baines}, \citenamefont {King}, \citenamefont {Gubbens},\ and\
  \citenamefont {Forget}}]{Yaouanc2013a}%
  \BibitemOpen
  \bibfield  {author} {\bibinfo {author} {\bibfnamefont {A.}~\bibnamefont
  {Yaouanc}}, \bibinfo {author} {\bibfnamefont {P.}~\bibnamefont {Dalmas~de
  R\'eotier}}, \bibinfo {author} {\bibfnamefont {P.}~\bibnamefont {Bonville}},
  \bibinfo {author} {\bibfnamefont {J.~A.}\ \bibnamefont {Hodges}}, \bibinfo
  {author} {\bibfnamefont {V.}~\bibnamefont {Glazkov}}, \bibinfo {author}
  {\bibfnamefont {L.}~\bibnamefont {Keller}}, \bibinfo {author} {\bibfnamefont
  {V.}~\bibnamefont {Sikolenko}}, \bibinfo {author} {\bibfnamefont
  {M.}~\bibnamefont {Bartkowiak}}, \bibinfo {author} {\bibfnamefont
  {A.}~\bibnamefont {Amato}}, \bibinfo {author} {\bibfnamefont
  {C.}~\bibnamefont {Baines}}, \bibinfo {author} {\bibfnamefont {P.~J.~C.}\
  \bibnamefont {King}}, \bibinfo {author} {\bibfnamefont {P.~C.~M.}\
  \bibnamefont {Gubbens}}, \ and\ \bibinfo {author} {\bibfnamefont
  {A.}~\bibnamefont {Forget}},\ }\bibfield  {title} {\enquote {\bibinfo {title}
  {{Dynamical Splayed Ferromagnetic Ground State in the Quantum Spin Ice
  ${\mathrm{Yb}}_{2}{\mathrm{Sn}}_{2}{\mathrm{O}}_{7}$}},}\ }\href {\doibase
  10.1103/PhysRevLett.110.127207} {\bibfield  {journal} {\bibinfo  {journal}
  {Phys. Rev. Lett.}\ }\textbf {\bibinfo {volume} {110}},\ \bibinfo {pages}
  {127207} (\bibinfo {year} {2013})}\BibitemShut {NoStop}%
\bibitem [{\citenamefont {Taniguchi}\ \emph {et~al.}(2013)\citenamefont
  {Taniguchi}, \citenamefont {Kadowaki}, \citenamefont {Takatsu}, \citenamefont
  {F\aa{}k}, \citenamefont {Ollivier}, \citenamefont {Yamazaki}, \citenamefont
  {Sato}, \citenamefont {Yoshizawa}, \citenamefont {Shimura}, \citenamefont
  {Sakakibara}, \citenamefont {Hong}, \citenamefont {Goto}, \citenamefont
  {Yaraskavitch},\ and\ \citenamefont {Kycia}}]{taniguchi13}%
  \BibitemOpen
  \bibfield  {author} {\bibinfo {author} {\bibfnamefont {T.}~\bibnamefont
  {Taniguchi}}, \bibinfo {author} {\bibfnamefont {H.}~\bibnamefont {Kadowaki}},
  \bibinfo {author} {\bibfnamefont {H.}~\bibnamefont {Takatsu}}, \bibinfo
  {author} {\bibfnamefont {B.}~\bibnamefont {F\aa{}k}}, \bibinfo {author}
  {\bibfnamefont {J.}~\bibnamefont {Ollivier}}, \bibinfo {author}
  {\bibfnamefont {T.}~\bibnamefont {Yamazaki}}, \bibinfo {author}
  {\bibfnamefont {T.~J.}\ \bibnamefont {Sato}}, \bibinfo {author}
  {\bibfnamefont {H.}~\bibnamefont {Yoshizawa}}, \bibinfo {author}
  {\bibfnamefont {Y.}~\bibnamefont {Shimura}}, \bibinfo {author} {\bibfnamefont
  {T.}~\bibnamefont {Sakakibara}}, \bibinfo {author} {\bibfnamefont
  {T.}~\bibnamefont {Hong}}, \bibinfo {author} {\bibfnamefont {K.}~\bibnamefont
  {Goto}}, \bibinfo {author} {\bibfnamefont {L.~R.}\ \bibnamefont
  {Yaraskavitch}}, \ and\ \bibinfo {author} {\bibfnamefont {J.~B.}\
  \bibnamefont {Kycia}},\ }\bibfield  {title} {\enquote {\bibinfo {title}
  {{Long-range order and spin-liquid states of polycrystalline
  $\mathrm{{Tb}}_{2+x}\mathrm{{Ti}}_{2-x}\mathrm{{O}}_{7+y}$}},}\ }\href
  {\doibase 10.1103/PhysRevB.87.060408} {\bibfield  {journal} {\bibinfo
  {journal} {Phys. Rev. B}\ }\textbf {\bibinfo {volume} {87}},\ \bibinfo
  {pages} {060408} (\bibinfo {year} {2013})}\BibitemShut {NoStop}%
\bibitem [{\citenamefont {Hallas}\ \emph {et~al.}(2014)\citenamefont {Hallas},
  \citenamefont {Cheng}, \citenamefont {Arevalo-Lopez}, \citenamefont
  {Silverstein}, \citenamefont {Su}, \citenamefont {Sarte}, \citenamefont
  {Zhou}, \citenamefont {Choi}, \citenamefont {Attfield}, \citenamefont
  {Luke},\ and\ \citenamefont {Wiebe}}]{Hallas2014}%
  \BibitemOpen
  \bibfield  {author} {\bibinfo {author} {\bibfnamefont {A.~M.}\ \bibnamefont
  {Hallas}}, \bibinfo {author} {\bibfnamefont {J.~G.}\ \bibnamefont {Cheng}},
  \bibinfo {author} {\bibfnamefont {A.~M.}\ \bibnamefont {Arevalo-Lopez}},
  \bibinfo {author} {\bibfnamefont {H.~J.}\ \bibnamefont {Silverstein}},
  \bibinfo {author} {\bibfnamefont {Y.}~\bibnamefont {Su}}, \bibinfo {author}
  {\bibfnamefont {P.~M.}\ \bibnamefont {Sarte}}, \bibinfo {author}
  {\bibfnamefont {H.~D.}\ \bibnamefont {Zhou}}, \bibinfo {author}
  {\bibfnamefont {E.~S.}\ \bibnamefont {Choi}}, \bibinfo {author}
  {\bibfnamefont {J.~P.}\ \bibnamefont {Attfield}}, \bibinfo {author}
  {\bibfnamefont {G.~M.}\ \bibnamefont {Luke}}, \ and\ \bibinfo {author}
  {\bibfnamefont {C.~R.}\ \bibnamefont {Wiebe}},\ }\bibfield  {title} {\enquote
  {\bibinfo {title} {{Incipient Ferromagnetism in
  ${\mathrm{Tb}}_{2}{\mathrm{Ge}}_{2}{\mathrm{O}}_{7}$: Application of Chemical
  Pressure to the Enigmatic Spin-Liquid Compound
  ${\mathrm{Tb}}_{2}{\mathrm{Ti}}_{2}{\mathrm{O}}_{7}$}},}\ }\href {\doibase
  10.1103/PhysRevLett.113.267205} {\bibfield  {journal} {\bibinfo  {journal}
  {Phys. Rev. Lett.}\ }\textbf {\bibinfo {volume} {113}},\ \bibinfo {pages}
  {267205} (\bibinfo {year} {2014})}\BibitemShut {NoStop}%
\bibitem [{\citenamefont {Hallas}\ \emph {et~al.}(2015)\citenamefont {Hallas},
  \citenamefont {Arevalo-Lopez}, \citenamefont {Sharma}, \citenamefont
  {Munsie}, \citenamefont {Attfield}, \citenamefont {Wiebe},\ and\
  \citenamefont {Luke}}]{hallas15}%
  \BibitemOpen
  \bibfield  {author} {\bibinfo {author} {\bibfnamefont {A.~M.}\ \bibnamefont
  {Hallas}}, \bibinfo {author} {\bibfnamefont {A.~M.}\ \bibnamefont
  {Arevalo-Lopez}}, \bibinfo {author} {\bibfnamefont {A.~Z.}\ \bibnamefont
  {Sharma}}, \bibinfo {author} {\bibfnamefont {T.}~\bibnamefont {Munsie}},
  \bibinfo {author} {\bibfnamefont {J.~P.}\ \bibnamefont {Attfield}}, \bibinfo
  {author} {\bibfnamefont {C.~R.}\ \bibnamefont {Wiebe}}, \ and\ \bibinfo
  {author} {\bibfnamefont {G.~M.}\ \bibnamefont {Luke}},\ }\bibfield  {title}
  {\enquote {\bibinfo {title} {{Magnetic frustration in lead pyrochlores}},}\
  }\href {\doibase 10.1103/PhysRevB.91.104417} {\bibfield  {journal} {\bibinfo
  {journal} {Phys. Rev. B}\ }\textbf {\bibinfo {volume} {91}},\ \bibinfo
  {pages} {104417} (\bibinfo {year} {2015})}\BibitemShut {NoStop}%
\bibitem [{\citenamefont {Hallas}\ \emph
  {et~al.}(2016{\natexlab{a}})\citenamefont {Hallas}, \citenamefont {Gaudet},
  \citenamefont {Butch}, \citenamefont {Tachibana}, \citenamefont {Freitas},
  \citenamefont {Luke}, \citenamefont {Wiebe},\ and\ \citenamefont
  {Gaulin}}]{hallas16-PRB93.100403}%
  \BibitemOpen
  \bibfield  {author} {\bibinfo {author} {\bibfnamefont {A.~M.}\ \bibnamefont
  {Hallas}}, \bibinfo {author} {\bibfnamefont {J.}~\bibnamefont {Gaudet}},
  \bibinfo {author} {\bibfnamefont {N.~P.}\ \bibnamefont {Butch}}, \bibinfo
  {author} {\bibfnamefont {M.}~\bibnamefont {Tachibana}}, \bibinfo {author}
  {\bibfnamefont {R.~S.}\ \bibnamefont {Freitas}}, \bibinfo {author}
  {\bibfnamefont {G.~M.}\ \bibnamefont {Luke}}, \bibinfo {author}
  {\bibfnamefont {C.~R.}\ \bibnamefont {Wiebe}}, \ and\ \bibinfo {author}
  {\bibfnamefont {B.~D.}\ \bibnamefont {Gaulin}},\ }\bibfield  {title}
  {\enquote {\bibinfo {title} {{Universal dynamic magnetism in Yb pyrochlores
  with disparate ground states}},}\ }\href {\doibase
  10.1103/PhysRevB.93.100403} {\bibfield  {journal} {\bibinfo  {journal} {Phys.
  Rev. B}\ }\textbf {\bibinfo {volume} {93}},\ \bibinfo {pages} {100403}
  (\bibinfo {year} {2016}{\natexlab{a}})}\BibitemShut {NoStop}%
\bibitem [{\citenamefont {Petit}\ \emph
  {et~al.}(2016{\natexlab{a}})\citenamefont {Petit}, \citenamefont {Lhotel},
  \citenamefont {Canals}, \citenamefont {Hatnean}, \citenamefont {Ollivier},
  \citenamefont {Mutka}, \citenamefont {Ressouche}, \citenamefont {Wildes},
  \citenamefont {Lees},\ and\ \citenamefont {Balakrishnan}}]{Petit2016}%
  \BibitemOpen
  \bibfield  {author} {\bibinfo {author} {\bibfnamefont {S.}~\bibnamefont
  {Petit}}, \bibinfo {author} {\bibfnamefont {E.}~\bibnamefont {Lhotel}},
  \bibinfo {author} {\bibfnamefont {B.}~\bibnamefont {Canals}}, \bibinfo
  {author} {\bibfnamefont {M.~C.}\ \bibnamefont {Hatnean}}, \bibinfo {author}
  {\bibfnamefont {J.}~\bibnamefont {Ollivier}}, \bibinfo {author}
  {\bibfnamefont {H.}~\bibnamefont {Mutka}}, \bibinfo {author} {\bibfnamefont
  {E.}~\bibnamefont {Ressouche}}, \bibinfo {author} {\bibfnamefont {A.~R.}\
  \bibnamefont {Wildes}}, \bibinfo {author} {\bibfnamefont {M.~R.}\
  \bibnamefont {Lees}}, \ and\ \bibinfo {author} {\bibfnamefont
  {G.}~\bibnamefont {Balakrishnan}},\ }\bibfield  {title} {\enquote {\bibinfo
  {title} {{Observation of magnetic fragmentation in spin ice}},}\ }\href
  {http://dx.doi.org/10.1038/nphys3710} {\bibfield  {journal} {\bibinfo
  {journal} {Nat. Phys.}\ }\textbf {\bibinfo {volume} {12}},\ \bibinfo {pages}
  {746--750} (\bibinfo {year} {2016}{\natexlab{a}})}\BibitemShut {NoStop}%
\bibitem [{\citenamefont {Petit}\ \emph
  {et~al.}(2016{\natexlab{b}})\citenamefont {Petit}, \citenamefont {Lhotel},
  \citenamefont {Guitteny}, \citenamefont {Florea}, \citenamefont {Robert},
  \citenamefont {Bonville}, \citenamefont {Mirebeau}, \citenamefont {Ollivier},
  \citenamefont {Mutka}, \citenamefont {Ressouche}, \citenamefont {Decorse},
  \citenamefont {Ciomaga~Hatnean},\ and\ \citenamefont
  {Balakrishnan}}]{petit16-PRB94}%
  \BibitemOpen
  \bibfield  {author} {\bibinfo {author} {\bibfnamefont {S.}~\bibnamefont
  {Petit}}, \bibinfo {author} {\bibfnamefont {E.}~\bibnamefont {Lhotel}},
  \bibinfo {author} {\bibfnamefont {S.}~\bibnamefont {Guitteny}}, \bibinfo
  {author} {\bibfnamefont {O.}~\bibnamefont {Florea}}, \bibinfo {author}
  {\bibfnamefont {J.}~\bibnamefont {Robert}}, \bibinfo {author} {\bibfnamefont
  {P.}~\bibnamefont {Bonville}}, \bibinfo {author} {\bibfnamefont
  {I.}~\bibnamefont {Mirebeau}}, \bibinfo {author} {\bibfnamefont
  {J.}~\bibnamefont {Ollivier}}, \bibinfo {author} {\bibfnamefont
  {H.}~\bibnamefont {Mutka}}, \bibinfo {author} {\bibfnamefont
  {E.}~\bibnamefont {Ressouche}}, \bibinfo {author} {\bibfnamefont
  {C.}~\bibnamefont {Decorse}}, \bibinfo {author} {\bibfnamefont
  {M.}~\bibnamefont {Ciomaga~Hatnean}}, \ and\ \bibinfo {author} {\bibfnamefont
  {G.}~\bibnamefont {Balakrishnan}},\ }\bibfield  {title} {\enquote {\bibinfo
  {title} {{Antiferroquadrupolar correlations in the quantum spin ice candidate
  ${\mathrm{Pr}}_{2}{\mathrm{Zr}}_{2}{\mathrm{O}}_{7}$}},}\ }\href {\doibase
  10.1103/PhysRevB.94.165153} {\bibfield  {journal} {\bibinfo  {journal} {Phys.
  Rev. B}\ }\textbf {\bibinfo {volume} {94}},\ \bibinfo {pages} {165153}
  (\bibinfo {year} {2016}{\natexlab{b}})}\BibitemShut {NoStop}%
\bibitem [{\citenamefont {Takatsu}\ \emph {et~al.}(2016)\citenamefont
  {Takatsu}, \citenamefont {Onoda}, \citenamefont {Kittaka}, \citenamefont
  {Kasahara}, \citenamefont {Kono}, \citenamefont {Sakakibara}, \citenamefont
  {Kato}, \citenamefont {F\aa{}k}, \citenamefont {Ollivier}, \citenamefont
  {Lynn}, \citenamefont {Taniguchi}, \citenamefont {Wakita},\ and\
  \citenamefont {Kadowaki}}]{Takatsu16}%
  \BibitemOpen
  \bibfield  {author} {\bibinfo {author} {\bibfnamefont {H.}~\bibnamefont
  {Takatsu}}, \bibinfo {author} {\bibfnamefont {S.}~\bibnamefont {Onoda}},
  \bibinfo {author} {\bibfnamefont {S.}~\bibnamefont {Kittaka}}, \bibinfo
  {author} {\bibfnamefont {A.}~\bibnamefont {Kasahara}}, \bibinfo {author}
  {\bibfnamefont {Y.}~\bibnamefont {Kono}}, \bibinfo {author} {\bibfnamefont
  {T.}~\bibnamefont {Sakakibara}}, \bibinfo {author} {\bibfnamefont
  {Y.}~\bibnamefont {Kato}}, \bibinfo {author} {\bibfnamefont {B.}~\bibnamefont
  {F\aa{}k}}, \bibinfo {author} {\bibfnamefont {J.}~\bibnamefont {Ollivier}},
  \bibinfo {author} {\bibfnamefont {J.~W.}\ \bibnamefont {Lynn}}, \bibinfo
  {author} {\bibfnamefont {T.}~\bibnamefont {Taniguchi}}, \bibinfo {author}
  {\bibfnamefont {M.}~\bibnamefont {Wakita}}, \ and\ \bibinfo {author}
  {\bibfnamefont {H.}~\bibnamefont {Kadowaki}},\ }\bibfield  {title} {\enquote
  {\bibinfo {title} {{Quadrupole Order in the Frustrated Pyrochlore
  $\mathrm{{Tb}}_{2+x}\mathrm{{Ti}}_{2-x}\mathrm{{O}}_{7+y}$}},}\ }\href
  {\doibase 10.1103/PhysRevLett.116.217201} {\bibfield  {journal} {\bibinfo
  {journal} {Phys. Rev. Lett.}\ }\textbf {\bibinfo {volume} {116}},\ \bibinfo
  {pages} {217201} (\bibinfo {year} {2016})}\BibitemShut {NoStop}%
\bibitem [{\citenamefont {Hallas}\ \emph {et~al.}()\citenamefont {Hallas},
  \citenamefont {Gaudet},\ and\ \citenamefont
  {Gaulin}}]{hallas-arXiv.1708.01312}%
  \BibitemOpen
  \bibfield  {author} {\bibinfo {author} {\bibfnamefont {A.~M.}\ \bibnamefont
  {Hallas}}, \bibinfo {author} {\bibfnamefont {J.}~\bibnamefont {Gaudet}}, \
  and\ \bibinfo {author} {\bibfnamefont {B.~D.}\ \bibnamefont {Gaulin}},\
  }\bibfield  {title} {\enquote {\bibinfo {title} {{Experimental Insights into
  Ground State Selection of Quantum XY Pyrochlores}},}\ }\href
  {https://arxiv.org/abs/1708.01312} {\bibinfo  {journal} {accepted for
  publication in Annu. Rev. Condens. Matter Phys.; arXiv:1708.01312}\
  }\BibitemShut {NoStop}%
\bibitem [{\citenamefont {Onoda}(2011)}]{onoda11}%
  \BibitemOpen
\bibfield  {journal} {  }\bibfield  {author} {\bibinfo {author} {\bibfnamefont
  {S.}~\bibnamefont {Onoda}},\ }\bibfield  {title} {\enquote {\bibinfo {title}
  {{Effective quantum pseudospin-1/2 model for Yb pyrochlore oxides}},}\ }\href
  {http://stacks.iop.org/1742-6596/320/i=1/a=012065} {\bibfield  {journal}
  {\bibinfo  {journal} {J. Phys.: Conf. Ser.}\ }\textbf {\bibinfo {volume}
  {320}},\ \bibinfo {pages} {012065} (\bibinfo {year} {2011})}\BibitemShut
  {NoStop}%
\bibitem [{\citenamefont {Wong}\ \emph {et~al.}(2013)\citenamefont {Wong},
  \citenamefont {Hao},\ and\ \citenamefont {Gingras}}]{wong13}%
  \BibitemOpen
  \bibfield  {author} {\bibinfo {author} {\bibfnamefont {A.~W.~C.}\
  \bibnamefont {Wong}}, \bibinfo {author} {\bibfnamefont {Z.}~\bibnamefont
  {Hao}}, \ and\ \bibinfo {author} {\bibfnamefont {M.~J.~P.}\ \bibnamefont
  {Gingras}},\ }\bibfield  {title} {\enquote {\bibinfo {title} {{Ground state
  phase diagram of generic $XY$ pyrochlore magnets with quantum
  fluctuations}},}\ }\href {\doibase 10.1103/PhysRevB.88.144402} {\bibfield
  {journal} {\bibinfo  {journal} {Phys. Rev. B}\ }\textbf {\bibinfo {volume}
  {88}},\ \bibinfo {pages} {144402} (\bibinfo {year} {2013})}\BibitemShut
  {NoStop}%
\bibitem [{\citenamefont {Yan}\ \emph {et~al.}(2017)\citenamefont {Yan},
  \citenamefont {Benton}, \citenamefont {Jaubert},\ and\ \citenamefont
  {Shannon}}]{yan17}%
  \BibitemOpen
  \bibfield  {author} {\bibinfo {author} {\bibfnamefont {H.}~\bibnamefont
  {Yan}}, \bibinfo {author} {\bibfnamefont {O.}~\bibnamefont {Benton}},
  \bibinfo {author} {\bibfnamefont {L.}~\bibnamefont {Jaubert}}, \ and\
  \bibinfo {author} {\bibfnamefont {N.}~\bibnamefont {Shannon}},\ }\bibfield
  {title} {\enquote {\bibinfo {title} {{Theory of multiple-phase competition in
  pyrochlore magnets with anisotropic exchange with application to
  ${\mathrm{Yb}}_{2}{\mathrm{Ti}}_{2}{\mathrm{O}}_{7},
  {\mathrm{Er}}_{2}{\mathrm{Ti}}_{2}{\mathrm{O}}_{7}$, and
  ${\mathrm{Er}}_{2}{\mathrm{Sn}}_{2}{\mathrm{O}}_{7}$}},}\ }\href {\doibase
  10.1103/PhysRevB.95.094422} {\bibfield  {journal} {\bibinfo  {journal} {Phys.
  Rev. B}\ }\textbf {\bibinfo {volume} {95}},\ \bibinfo {pages} {094422}
  (\bibinfo {year} {2017})}\BibitemShut {NoStop}%
\bibitem [{\citenamefont {Benton}(2015)}]{owen-thesis}%
  \BibitemOpen
  \bibfield  {author} {\bibinfo {author} {\bibfnamefont {O.}~\bibnamefont
  {Benton}},\ }\emph {\bibinfo {title} {{Classical and quantum spin liquids on
  the pyrochlore lattice}}},\ \href@noop {} {Ph.D. thesis},\ \bibinfo  {school}
  {University of Bristol} (\bibinfo {year} {2015})\BibitemShut {NoStop}%
\bibitem [{\citenamefont {Huang}\ \emph {et~al.}(2014)\citenamefont {Huang},
  \citenamefont {Chen},\ and\ \citenamefont {Hermele}}]{huang14}%
  \BibitemOpen
  \bibfield  {author} {\bibinfo {author} {\bibfnamefont {Y.-P.}\ \bibnamefont
  {Huang}}, \bibinfo {author} {\bibfnamefont {G.}~\bibnamefont {Chen}}, \ and\
  \bibinfo {author} {\bibfnamefont {M.}~\bibnamefont {Hermele}},\ }\bibfield
  {title} {\enquote {\bibinfo {title} {{Quantum Spin Ices and Topological
  Phases from Dipolar-Octupolar Doublets on the Pyrochlore Lattice}},}\ }\href
  {\doibase 10.1103/PhysRevLett.112.167203} {\bibfield  {journal} {\bibinfo
  {journal} {Phys. Rev. Lett.}\ }\textbf {\bibinfo {volume} {112}},\ \bibinfo
  {pages} {167203} (\bibinfo {year} {2014})}\BibitemShut {NoStop}%
\bibitem [{\citenamefont {Chen}(2017{\natexlab{b}})}]{chen-arXiv.1704.02734}%
  \BibitemOpen
  \bibfield  {author} {\bibinfo {author} {\bibfnamefont {G.}~\bibnamefont
  {Chen}},\ }\bibfield  {title} {\enquote {\bibinfo {title} {Spectral
  periodicity of the spinon continuum in quantum spin ice},}\ }\href {\doibase
  10.1103/PhysRevB.96.085136} {\bibfield  {journal} {\bibinfo  {journal} {Phys.
  Rev. B}\ }\textbf {\bibinfo {volume} {96}},\ \bibinfo {pages} {085136}
  (\bibinfo {year} {2017}{\natexlab{b}})}\BibitemShut {NoStop}%
\bibitem [{\citenamefont {Moessner}\ and\ \citenamefont
  {Chalker}(1998{\natexlab{a}})}]{moessner98-PRL80}%
  \BibitemOpen
  \bibfield  {author} {\bibinfo {author} {\bibfnamefont {R.}~\bibnamefont
  {Moessner}}\ and\ \bibinfo {author} {\bibfnamefont {J.~T.}\ \bibnamefont
  {Chalker}},\ }\bibfield  {title} {\enquote {\bibinfo {title} {{Properties of
  a Classical Spin Liquid: The Heisenberg Pyrochlore Antiferromagnet}},}\
  }\href {http://link.aps.org/doi/10.1103/PhysRevLett.80.2929} {\bibfield
  {journal} {\bibinfo  {journal} {Phys. Rev. Lett.}\ }\textbf {\bibinfo
  {volume} {80}},\ \bibinfo {pages} {2929--2932} (\bibinfo {year}
  {1998}{\natexlab{a}})}\BibitemShut {NoStop}%
\bibitem [{\citenamefont {Moessner}\ and\ \citenamefont
  {Chalker}(1998{\natexlab{b}})}]{moessner98-PRB58}%
  \BibitemOpen
  \bibfield  {author} {\bibinfo {author} {\bibfnamefont {R.}~\bibnamefont
  {Moessner}}\ and\ \bibinfo {author} {\bibfnamefont {J.~T.}\ \bibnamefont
  {Chalker}},\ }\bibfield  {title} {\enquote {\bibinfo {title}
  {{Low-temperature properties of classical geometrically frustrated
  antiferromagnets}},}\ }\href
  {http://link.aps.org/doi/10.1103/PhysRevB.58.12049} {\bibfield  {journal}
  {\bibinfo  {journal} {Phys. Rev. B}\ }\textbf {\bibinfo {volume} {58}},\
  \bibinfo {pages} {12049--12062} (\bibinfo {year}
  {1998}{\natexlab{b}})}\BibitemShut {NoStop}%
\bibitem [{\citenamefont {Isakov}\ \emph {et~al.}(2004)\citenamefont {Isakov},
  \citenamefont {Gregor}, \citenamefont {Moessner},\ and\ \citenamefont
  {Sondhi}}]{Isakov2004a}%
  \BibitemOpen
  \bibfield  {author} {\bibinfo {author} {\bibfnamefont {S.~V.}\ \bibnamefont
  {Isakov}}, \bibinfo {author} {\bibfnamefont {K.}~\bibnamefont {Gregor}},
  \bibinfo {author} {\bibfnamefont {R.}~\bibnamefont {Moessner}}, \ and\
  \bibinfo {author} {\bibfnamefont {S.~L.}\ \bibnamefont {Sondhi}},\ }\bibfield
   {title} {\enquote {\bibinfo {title} {{Dipolar Spin Correlations in Classical
  Pyrochlore Magnets}},}\ }\href {\doibase 10.1103/PhysRevLett.93.167204}
  {\bibfield  {journal} {\bibinfo  {journal} {Phys. Rev. Lett.}\ }\textbf
  {\bibinfo {volume} {93}},\ \bibinfo {pages} {167204} (\bibinfo {year}
  {2004})}\BibitemShut {NoStop}%
\bibitem [{\citenamefont {Henley}(2005)}]{henley05}%
  \BibitemOpen
  \bibfield  {author} {\bibinfo {author} {\bibfnamefont {C.~L.}\ \bibnamefont
  {Henley}},\ }\bibfield  {title} {\enquote {\bibinfo {title} {{Power-law spin
  correlations in pyrochlore antiferromagnets}},}\ }\href {\doibase
  10.1103/PhysRevB.71.014424} {\bibfield  {journal} {\bibinfo  {journal} {Phys.
  Rev. B}\ }\textbf {\bibinfo {volume} {71}},\ \bibinfo {pages} {014424}
  (\bibinfo {year} {2005})}\BibitemShut {NoStop}%
\bibitem [{\citenamefont {Canals}\ and\ \citenamefont
  {Lacroix}(1998)}]{canals98}%
  \BibitemOpen
  \bibfield  {author} {\bibinfo {author} {\bibfnamefont {B.}~\bibnamefont
  {Canals}}\ and\ \bibinfo {author} {\bibfnamefont {C.}~\bibnamefont
  {Lacroix}},\ }\bibfield  {title} {\enquote {\bibinfo {title} {{Pyrochlore
  Antiferromagnet: A Three-Dimensional Quantum Spin Liquid}},}\ }\href
  {https://journals.aps.org/prl/abstract/10.1103/PhysRevLett.80.2933}
  {\bibfield  {journal} {\bibinfo  {journal} {Phys. Rev. Lett.}\ }\textbf
  {\bibinfo {volume} {80}},\ \bibinfo {pages} {2933} (\bibinfo {year}
  {1998})}\BibitemShut {NoStop}%
\bibitem [{\citenamefont {Canals}\ and\ \citenamefont
  {Lacroix}(2000)}]{canals00}%
  \BibitemOpen
  \bibfield  {author} {\bibinfo {author} {\bibfnamefont {B.}~\bibnamefont
  {Canals}}\ and\ \bibinfo {author} {\bibfnamefont {C.}~\bibnamefont
  {Lacroix}},\ }\bibfield  {title} {\enquote {\bibinfo {title} {{Quantum spin
  liquid: The Heisenberg antiferromagnet on the three-dimensional pyrochlore
  lattice}},}\ }\href
  {https://journals.aps.org/prb/abstract/10.1103/PhysRevB.61.1149} {\bibfield
  {journal} {\bibinfo  {journal} {Phys. Rev. B}\ }\textbf {\bibinfo {volume}
  {61}},\ \bibinfo {pages} {1149} (\bibinfo {year} {2000})}\BibitemShut
  {NoStop}%
\bibitem [{\citenamefont {Burnell}\ \emph {et~al.}(2009)\citenamefont
  {Burnell}, \citenamefont {Chakravarty},\ and\ \citenamefont
  {Sondhi}}]{burnell09}%
  \BibitemOpen
  \bibfield  {author} {\bibinfo {author} {\bibfnamefont {F.~J.}\ \bibnamefont
  {Burnell}}, \bibinfo {author} {\bibfnamefont {S.}~\bibnamefont
  {Chakravarty}}, \ and\ \bibinfo {author} {\bibfnamefont {S.~L.}\ \bibnamefont
  {Sondhi}},\ }\bibfield  {title} {\enquote {\bibinfo {title} {{Monopole flux
  state on the pyrochlore lattice}},}\ }\href {\doibase
  10.1103/PhysRevB.79.144432} {\bibfield  {journal} {\bibinfo  {journal} {Phys.
  Rev. B}\ }\textbf {\bibinfo {volume} {79}},\ \bibinfo {pages} {144432}
  (\bibinfo {year} {2009})}\BibitemShut {NoStop}%
\bibitem [{\citenamefont {Huang}\ \emph {et~al.}(2016)\citenamefont {Huang},
  \citenamefont {Chen}, \citenamefont {Deng}, \citenamefont {Prokof'ev},\ and\
  \citenamefont {Svistunov}}]{Huang2016}%
  \BibitemOpen
  \bibfield  {author} {\bibinfo {author} {\bibfnamefont {Y.}~\bibnamefont
  {Huang}}, \bibinfo {author} {\bibfnamefont {K.}~\bibnamefont {Chen}},
  \bibinfo {author} {\bibfnamefont {Y.}~\bibnamefont {Deng}}, \bibinfo {author}
  {\bibfnamefont {N.}~\bibnamefont {Prokof'ev}}, \ and\ \bibinfo {author}
  {\bibfnamefont {B.}~\bibnamefont {Svistunov}},\ }\bibfield  {title} {\enquote
  {\bibinfo {title} {{Spin-Ice State of the Quantum Heisenberg Antiferromagnet
  on the Pyrochlore Lattice}},}\ }\href {\doibase
  10.1103/PhysRevLett.116.177203} {\bibfield  {journal} {\bibinfo  {journal}
  {Phys. Rev. Lett.}\ }\textbf {\bibinfo {volume} {116}},\ \bibinfo {pages}
  {177203} (\bibinfo {year} {2016})}\BibitemShut {NoStop}%
\bibitem [{\citenamefont {Fennell}\ \emph {et~al.}(2009)\citenamefont
  {Fennell}, \citenamefont {Deen}, \citenamefont {Wildes}, \citenamefont
  {Schmalzl}, \citenamefont {Prabhakaran}, \citenamefont {Boothroyd},
  \citenamefont {Aldus}, \citenamefont {McMorrow},\ and\ \citenamefont
  {Bramwell}}]{fennell09}%
  \BibitemOpen
  \bibfield  {author} {\bibinfo {author} {\bibfnamefont {T.}~\bibnamefont
  {Fennell}}, \bibinfo {author} {\bibfnamefont {P.~P.}\ \bibnamefont {Deen}},
  \bibinfo {author} {\bibfnamefont {A.~R.}\ \bibnamefont {Wildes}}, \bibinfo
  {author} {\bibfnamefont {K.}~\bibnamefont {Schmalzl}}, \bibinfo {author}
  {\bibfnamefont {D.}~\bibnamefont {Prabhakaran}}, \bibinfo {author}
  {\bibfnamefont {A.~T.}\ \bibnamefont {Boothroyd}}, \bibinfo {author}
  {\bibfnamefont {R.~J.}\ \bibnamefont {Aldus}}, \bibinfo {author}
  {\bibfnamefont {D.~F.}\ \bibnamefont {McMorrow}}, \ and\ \bibinfo {author}
  {\bibfnamefont {S.~T.}\ \bibnamefont {Bramwell}},\ }\bibfield  {title}
  {\enquote {\bibinfo {title} {{Magnetic Coulomb Phase in the Spin Ice
  $\mathrm{{Ho}}_2\mathrm{{Ti}}_2\mathrm{{O}}_7$}},}\ }\href
  {http://science.sciencemag.org/content/326/5951/415.abstract} {\bibfield
  {journal} {\bibinfo  {journal} {Science}\ }\textbf {\bibinfo {volume}
  {326}},\ \bibinfo {pages} {415--417} (\bibinfo {year} {2009})}\BibitemShut
  {NoStop}%
\bibitem [{\citenamefont {Benton}()}]{owen-unpublished}%
  \BibitemOpen
  \bibfield  {author} {\bibinfo {author} {\bibfnamefont {O.}~\bibnamefont
  {Benton}},\ }\href@noop {} {\bibinfo  {journal} {unpublished}\ }\BibitemShut
  {NoStop}%
\bibitem [{\citenamefont {Taillefumier}\ \emph {et~al.}()\citenamefont
  {Taillefumier}, \citenamefont {Benton},\ and\ \citenamefont
  {Shannon}}]{taillefumier-in-preparation}%
  \BibitemOpen
\bibfield  {journal} {  }\bibfield  {author} {\bibinfo {author} {\bibfnamefont
  {M.}~\bibnamefont {Taillefumier}}, \bibinfo {author} {\bibfnamefont
  {O.}~\bibnamefont {Benton}}, \ and\ \bibinfo {author} {\bibfnamefont
  {N.}~\bibnamefont {Shannon}},\ }\href@noop {} {}\bibinfo {note} {In
  preparation.}\BibitemShut {Stop}%
\bibitem [{\citenamefont {Jaubert}\ \emph {et~al.}(2013)\citenamefont
  {Jaubert}, \citenamefont {Harris}, \citenamefont {Fennell}, \citenamefont
  {Melko}, \citenamefont {Bramwell},\ and\ \citenamefont
  {Holdsworth}}]{jaubert13}%
  \BibitemOpen
  \bibfield  {author} {\bibinfo {author} {\bibfnamefont {L.~D.~C.}\
  \bibnamefont {Jaubert}}, \bibinfo {author} {\bibfnamefont {M.~J.}\
  \bibnamefont {Harris}}, \bibinfo {author} {\bibfnamefont {T.}~\bibnamefont
  {Fennell}}, \bibinfo {author} {\bibfnamefont {R.~G.}\ \bibnamefont {Melko}},
  \bibinfo {author} {\bibfnamefont {S.~T.}\ \bibnamefont {Bramwell}}, \ and\
  \bibinfo {author} {\bibfnamefont {P.~C.~W.}\ \bibnamefont {Holdsworth}},\
  }\bibfield  {title} {\enquote {\bibinfo {title} {{Topological-Sector
  Fluctuations and Curie-Law Crossover in Spin Ice}},}\ }\href {\doibase
  10.1103/PhysRevX.3.011014} {\bibfield  {journal} {\bibinfo  {journal} {Phys.
  Rev. X}\ }\textbf {\bibinfo {volume} {3}},\ \bibinfo {pages} {011014}
  (\bibinfo {year} {2013})}\BibitemShut {NoStop}%
\bibitem [{\citenamefont {Conlon}\ and\ \citenamefont
  {Chalker}(2009)}]{conlon09}%
  \BibitemOpen
  \bibfield  {author} {\bibinfo {author} {\bibfnamefont {P.~H.}\ \bibnamefont
  {Conlon}}\ and\ \bibinfo {author} {\bibfnamefont {J.~T.}\ \bibnamefont
  {Chalker}},\ }\bibfield  {title} {\enquote {\bibinfo {title} {{Spin Dynamics
  in Pyrochlore Heisenberg Antiferromagnets}},}\ }\href {\doibase
  10.1103/PhysRevLett.102.237206} {\bibfield  {journal} {\bibinfo  {journal}
  {Phys. Rev. Lett.}\ }\textbf {\bibinfo {volume} {102}},\ \bibinfo {pages}
  {237206} (\bibinfo {year} {2009})}\BibitemShut {NoStop}%
\bibitem [{\citenamefont {Henley}(2010)}]{henley10}%
  \BibitemOpen
  \bibfield  {author} {\bibinfo {author} {\bibfnamefont {C.~L.}\ \bibnamefont
  {Henley}},\ }\bibfield  {title} {\enquote {\bibinfo {title} {{The ``Coulomb
  Phase'' in Frustrated Systems}},}\ }\href {\doibase
  10.1146/annurev-conmatphys-070909-104138} {\bibfield  {journal} {\bibinfo
  {journal} {Annu. Rev. Condens. Matter Phys.}\ }\textbf {\bibinfo {volume}
  {1}},\ \bibinfo {pages} {179--2010} (\bibinfo {year} {2010})}\BibitemShut
  {NoStop}%
\bibitem [{\citenamefont {Andreev}\ and\ \citenamefont
  {Grishchuk}(1984)}]{andreev84}%
  \BibitemOpen
  \bibfield  {author} {\bibinfo {author} {\bibfnamefont {AF}~\bibnamefont
  {Andreev}}\ and\ \bibinfo {author} {\bibfnamefont {IA}~\bibnamefont
  {Grishchuk}},\ }\bibfield  {title} {\enquote {\bibinfo {title} {{Spin
  Nematics}},}\ }\href
  {http://www.jetp.ac.ru/cgi-bin/e/index/e/60/2/p267?a=list} {\bibfield
  {journal} {\bibinfo  {journal} {JETP}\ }\textbf {\bibinfo {volume} {87}},\
  \bibinfo {pages} {467--475} (\bibinfo {year} {1984})}\BibitemShut {NoStop}%
\bibitem [{\citenamefont {Chubukov}(1991)}]{chubukov91}%
  \BibitemOpen
  \bibfield  {author} {\bibinfo {author} {\bibfnamefont {A.~V.}\ \bibnamefont
  {Chubukov}},\ }\bibfield  {title} {\enquote {\bibinfo {title} {{Chiral,
  nematic, and dimer states in quantum spin chains}},}\ }\href {\doibase
  10.1103/PhysRevB.44.4693} {\bibfield  {journal} {\bibinfo  {journal} {Phys.
  Rev. B}\ }\textbf {\bibinfo {volume} {44}},\ \bibinfo {pages} {4693--4696}
  (\bibinfo {year} {1991})}\BibitemShut {NoStop}%
\bibitem [{\citenamefont {Shannon}\ \emph {et~al.}(2006)\citenamefont
  {Shannon}, \citenamefont {Momoi},\ and\ \citenamefont
  {Sindzingre}}]{shannon06}%
  \BibitemOpen
  \bibfield  {author} {\bibinfo {author} {\bibfnamefont {N.}~\bibnamefont
  {Shannon}}, \bibinfo {author} {\bibfnamefont {T.}~\bibnamefont {Momoi}}, \
  and\ \bibinfo {author} {\bibfnamefont {P.}~\bibnamefont {Sindzingre}},\
  }\bibfield  {title} {\enquote {\bibinfo {title} {{Nematic Order in Square
  Lattice Frustrated Ferromagnets}},}\ }\href {\doibase
  10.1103/PhysRevLett.96.027213} {\bibfield  {journal} {\bibinfo  {journal}
  {Phys. Rev. Lett.}\ }\textbf {\bibinfo {volume} {96}},\ \bibinfo {pages}
  {027213} (\bibinfo {year} {2006})}\BibitemShut {NoStop}%
\bibitem [{\citenamefont {Shannon}\ \emph {et~al.}(2010)\citenamefont
  {Shannon}, \citenamefont {Penc},\ and\ \citenamefont {Motome}}]{shannon10}%
  \BibitemOpen
  \bibfield  {author} {\bibinfo {author} {\bibfnamefont {N.}~\bibnamefont
  {Shannon}}, \bibinfo {author} {\bibfnamefont {K.}~\bibnamefont {Penc}}, \
  and\ \bibinfo {author} {\bibfnamefont {Y.}~\bibnamefont {Motome}},\
  }\bibfield  {title} {\enquote {\bibinfo {title} {{Nematic, vector-multipole,
  and plateau-liquid states in the classical $O(3)$ pyrochlore antiferromagnet
  with biquadratic interactions in applied magnetic field}},}\ }\href {\doibase
  10.1103/PhysRevB.81.184409} {\bibfield  {journal} {\bibinfo  {journal} {Phys.
  Rev. B}\ }\textbf {\bibinfo {volume} {81}},\ \bibinfo {pages} {184409}
  (\bibinfo {year} {2010})}\BibitemShut {NoStop}%
\bibitem [{\citenamefont {Smerald}\ \emph {et~al.}(2015)\citenamefont
  {Smerald}, \citenamefont {Ueda},\ and\ \citenamefont {Shannon}}]{smerald15}%
  \BibitemOpen
  \bibfield  {author} {\bibinfo {author} {\bibfnamefont {A.}~\bibnamefont
  {Smerald}}, \bibinfo {author} {\bibfnamefont {H.~T.}\ \bibnamefont {Ueda}}, \
  and\ \bibinfo {author} {\bibfnamefont {N.}~\bibnamefont {Shannon}},\
  }\bibfield  {title} {\enquote {\bibinfo {title} {{Theory of inelastic neutron
  scattering in a field-induced spin-nematic state}},}\ }\href {\doibase
  10.1103/PhysRevB.91.174402} {\bibfield  {journal} {\bibinfo  {journal} {Phys.
  Rev. B}\ }\textbf {\bibinfo {volume} {91}},\ \bibinfo {pages} {174402}
  (\bibinfo {year} {2015})}\BibitemShut {NoStop}%
\bibitem [{\citenamefont {Isakov}\ \emph {et~al.}(2005)\citenamefont {Isakov},
  \citenamefont {Moessner},\ and\ \citenamefont {Sondhi}}]{isakov05}%
  \BibitemOpen
  \bibfield  {author} {\bibinfo {author} {\bibfnamefont {S.~V.}\ \bibnamefont
  {Isakov}}, \bibinfo {author} {\bibfnamefont {R.}~\bibnamefont {Moessner}}, \
  and\ \bibinfo {author} {\bibfnamefont {S.~L.}\ \bibnamefont {Sondhi}},\
  }\bibfield  {title} {\enquote {\bibinfo {title} {{Why Spin Ice Obeys the Ice
  Rules}},}\ }\href {\doibase 10.1103/PhysRevLett.95.217201} {\bibfield
  {journal} {\bibinfo  {journal} {Phys. Rev. Lett.}\ }\textbf {\bibinfo
  {volume} {95}},\ \bibinfo {pages} {217201} (\bibinfo {year}
  {2005})}\BibitemShut {NoStop}%
\bibitem [{\citenamefont {Benton}\ \emph {et~al.}(2016)\citenamefont {Benton},
  \citenamefont {Jaubert}, \citenamefont {Yan},\ and\ \citenamefont
  {Shannon}}]{benton16-NatCommun7}%
  \BibitemOpen
  \bibfield  {author} {\bibinfo {author} {\bibfnamefont {O.}~\bibnamefont
  {Benton}}, \bibinfo {author} {\bibfnamefont {L.~D.~C.}\ \bibnamefont
  {Jaubert}}, \bibinfo {author} {\bibfnamefont {H.}~\bibnamefont {Yan}}, \ and\
  \bibinfo {author} {\bibfnamefont {N.}~\bibnamefont {Shannon}},\ }\bibfield
  {title} {\enquote {\bibinfo {title} {{A spin-liquid with pinch-line
  singularities on the pyrochlore lattice}},}\ }\href
  {https://www.nature.com/articles/ncomms11572} {\bibfield  {journal} {\bibinfo
   {journal} {Nat. Commun.}\ }\textbf {\bibinfo {volume} {7}},\ \bibinfo
  {pages} {11572} (\bibinfo {year} {2016})}\BibitemShut {NoStop}%
\bibitem [{\citenamefont {McClarty}\ \emph {et~al.}(2009)\citenamefont
  {McClarty}, \citenamefont {Curnoe},\ and\ \citenamefont
  {Gingras}}]{mcclarty09}%
  \BibitemOpen
  \bibfield  {author} {\bibinfo {author} {\bibfnamefont {P.~A.}\ \bibnamefont
  {McClarty}}, \bibinfo {author} {\bibfnamefont {S.~H.}\ \bibnamefont
  {Curnoe}}, \ and\ \bibinfo {author} {\bibfnamefont {M.~J.~P.}\ \bibnamefont
  {Gingras}},\ }\bibfield  {title} {\enquote {\bibinfo {title} {{Energetic
  selection of ordered states in a model of the
  $\mathrm{{Er}}_2\mathrm{{Ti}}_2\mathrm{{O}}_7$ frustrated pyrochlore XY
  antiferromagnet}},}\ }\href
  {http://stacks.iop.org/1742-6596/145/i=1/a=012032} {\bibfield  {journal}
  {\bibinfo  {journal} {J. Phys.: Conf. Ser.}\ }\textbf {\bibinfo {volume}
  {145}},\ \bibinfo {pages} {012032} (\bibinfo {year} {2009})}\BibitemShut
  {NoStop}%
\bibitem [{\citenamefont {Yan}\ \emph {et~al.}()\citenamefont {Yan},
  \citenamefont {Benton}, \citenamefont {Jaubert},\ and\ \citenamefont
  {Shannon}}]{yan-arXiv}%
  \BibitemOpen
  \bibfield  {author} {\bibinfo {author} {\bibfnamefont {H.}~\bibnamefont
  {Yan}}, \bibinfo {author} {\bibfnamefont {O.}~\bibnamefont {Benton}},
  \bibinfo {author} {\bibfnamefont {L.}~\bibnamefont {Jaubert}}, \ and\
  \bibinfo {author} {\bibfnamefont {N.}~\bibnamefont {Shannon}},\ }\bibfield
  {title} {\enquote {\bibinfo {title} {{Living on the edge: ground-state
  selection in quantum spin-ice pyrochlores}},}\ }\href
  {https://arxiv.org/abs/1311.3501} {\bibinfo  {journal} {arXiv:1311.3501}\
  }\BibitemShut {NoStop}%
\bibitem [{\citenamefont {Jaubert}\ \emph {et~al.}(2015)\citenamefont
  {Jaubert}, \citenamefont {Benton}, \citenamefont {Rau}, \citenamefont
  {Oitmaa}, \citenamefont {Singh}, \citenamefont {Shannon},\ and\ \citenamefont
  {Gingras}}]{jaubert15}%
  \BibitemOpen
\bibfield  {journal} {  }\bibfield  {author} {\bibinfo {author} {\bibfnamefont
  {L.~D.~C.}\ \bibnamefont {Jaubert}}, \bibinfo {author} {\bibfnamefont
  {O.}~\bibnamefont {Benton}}, \bibinfo {author} {\bibfnamefont {J.~G.}\
  \bibnamefont {Rau}}, \bibinfo {author} {\bibfnamefont {J.}~\bibnamefont
  {Oitmaa}}, \bibinfo {author} {\bibfnamefont {R.~R.~P.}\ \bibnamefont
  {Singh}}, \bibinfo {author} {\bibfnamefont {N.}~\bibnamefont {Shannon}}, \
  and\ \bibinfo {author} {\bibfnamefont {M.~J.~P.}\ \bibnamefont {Gingras}},\
  }\bibfield  {title} {\enquote {\bibinfo {title} {{Are Multiphase Competition
  and Order by Disorder the Keys to Understanding
  ${\mathrm{Yb}}_{2}{\mathrm{Ti}}_{2}{\mathrm{O}}_{7}$?}}}\ }\href {\doibase
  10.1103/PhysRevLett.115.267208} {\bibfield  {journal} {\bibinfo  {journal}
  {Phys. Rev. Lett.}\ }\textbf {\bibinfo {volume} {115}},\ \bibinfo {pages}
  {267208} (\bibinfo {year} {2015})}\BibitemShut {NoStop}%
\bibitem [{\citenamefont {Palmer}\ and\ \citenamefont
  {Chalker}(2000)}]{palmer00}%
  \BibitemOpen
  \bibfield  {author} {\bibinfo {author} {\bibfnamefont {S.~E.}\ \bibnamefont
  {Palmer}}\ and\ \bibinfo {author} {\bibfnamefont {J.~T.}\ \bibnamefont
  {Chalker}},\ }\bibfield  {title} {\enquote {\bibinfo {title} {{Order induced
  by dipolar interactions in a geometrically frustrated antiferromagnet}},}\
  }\href {\doibase 10.1103/PhysRevB.62.488} {\bibfield  {journal} {\bibinfo
  {journal} {Phys. Rev. B}\ }\textbf {\bibinfo {volume} {62}},\ \bibinfo
  {pages} {488--492} (\bibinfo {year} {2000})}\BibitemShut {NoStop}%
\bibitem [{\citenamefont {Taillefumier}\ \emph {et~al.}(2014)\citenamefont
  {Taillefumier}, \citenamefont {Robert}, \citenamefont {Henley}, \citenamefont
  {Moessner},\ and\ \citenamefont {Canals}}]{taillefumier14}%
  \BibitemOpen
  \bibfield  {author} {\bibinfo {author} {\bibfnamefont {M.}~\bibnamefont
  {Taillefumier}}, \bibinfo {author} {\bibfnamefont {J.}~\bibnamefont
  {Robert}}, \bibinfo {author} {\bibfnamefont {C.~L.}\ \bibnamefont {Henley}},
  \bibinfo {author} {\bibfnamefont {R.}~\bibnamefont {Moessner}}, \ and\
  \bibinfo {author} {\bibfnamefont {B.}~\bibnamefont {Canals}},\ }\bibfield
  {title} {\enquote {\bibinfo {title} {{Semiclassical spin dynamics of the
  antiferromagnetic Heisenberg model on the kagome lattice}},}\ }\href
  {\doibase 10.1103/PhysRevB.90.064419} {\bibfield  {journal} {\bibinfo
  {journal} {Phys. Rev. B}\ }\textbf {\bibinfo {volume} {90}},\ \bibinfo
  {pages} {064419} (\bibinfo {year} {2014})}\BibitemShut {NoStop}%
\bibitem [{\citenamefont {Knolle}\ \emph {et~al.}(2014)\citenamefont {Knolle},
  \citenamefont {Kovrizhin}, \citenamefont {Chalker},\ and\ \citenamefont
  {Moessner}}]{knolle14}%
  \BibitemOpen
  \bibfield  {author} {\bibinfo {author} {\bibfnamefont {J.}~\bibnamefont
  {Knolle}}, \bibinfo {author} {\bibfnamefont {D.~L.}\ \bibnamefont
  {Kovrizhin}}, \bibinfo {author} {\bibfnamefont {J.~T.}\ \bibnamefont
  {Chalker}}, \ and\ \bibinfo {author} {\bibfnamefont {R.}~\bibnamefont
  {Moessner}},\ }\bibfield  {title} {\enquote {\bibinfo {title} {{Dynamics of a
  Two-Dimensional Quantum Spin Liquid: Signatures of Emergent Majorana Fermions
  and Fluxes}},}\ }\href {\doibase 10.1103/PhysRevLett.112.207203} {\bibfield
  {journal} {\bibinfo  {journal} {Phys. Rev. Lett.}\ }\textbf {\bibinfo
  {volume} {112}},\ \bibinfo {pages} {207203} (\bibinfo {year}
  {2014})}\BibitemShut {NoStop}%
\bibitem [{\citenamefont {Punk}\ \emph {et~al.}(2014)\citenamefont {Punk},
  \citenamefont {Chowdhury},\ and\ \citenamefont {Sachdev}}]{punk14}%
  \BibitemOpen
  \bibfield  {author} {\bibinfo {author} {\bibfnamefont {M.}~\bibnamefont
  {Punk}}, \bibinfo {author} {\bibfnamefont {D.}~\bibnamefont {Chowdhury}}, \
  and\ \bibinfo {author} {\bibfnamefont {S.}~\bibnamefont {Sachdev}},\
  }\bibfield  {title} {\enquote {\bibinfo {title} {{Topological excitations and
  the dynamic structure factor of spin liquids on the kagome lattice}},}\
  }\href {http://dx.doi.org/10.1038/nphys2887} {\bibfield  {journal} {\bibinfo
  {journal} {Nat. Phys.}\ }\textbf {\bibinfo {volume} {10}},\ \bibinfo {pages}
  {289--293} (\bibinfo {year} {2014})}\BibitemShut {NoStop}%
\bibitem [{\citenamefont {Bieri}\ \emph {et~al.}(2015)\citenamefont {Bieri},
  \citenamefont {Messio}, \citenamefont {Bernu},\ and\ \citenamefont
  {Lhuillier}}]{bieri15}%
  \BibitemOpen
  \bibfield  {author} {\bibinfo {author} {\bibfnamefont {S.}~\bibnamefont
  {Bieri}}, \bibinfo {author} {\bibfnamefont {L.}~\bibnamefont {Messio}},
  \bibinfo {author} {\bibfnamefont {B.}~\bibnamefont {Bernu}}, \ and\ \bibinfo
  {author} {\bibfnamefont {C.}~\bibnamefont {Lhuillier}},\ }\bibfield  {title}
  {\enquote {\bibinfo {title} {{Gapless chiral spin liquid in a kagome
  Heisenberg model}},}\ }\href {\doibase 10.1103/PhysRevB.92.060407} {\bibfield
   {journal} {\bibinfo  {journal} {Phys. Rev. B}\ }\textbf {\bibinfo {volume}
  {92}},\ \bibinfo {pages} {060407} (\bibinfo {year} {2015})}\BibitemShut
  {NoStop}%
\bibitem [{\citenamefont {Han}\ \emph {et~al.}(2012)\citenamefont {Han},
  \citenamefont {Helton}, \citenamefont {Chu}, \citenamefont {Nocera},
  \citenamefont {Rodriguez-Rivera}, \citenamefont {Broholm},\ and\
  \citenamefont {Lee}}]{han12}%
  \BibitemOpen
  \bibfield  {author} {\bibinfo {author} {\bibfnamefont {T.~H.}\ \bibnamefont
  {Han}}, \bibinfo {author} {\bibfnamefont {J.~S.}\ \bibnamefont {Helton}},
  \bibinfo {author} {\bibfnamefont {S.}~\bibnamefont {Chu}}, \bibinfo {author}
  {\bibfnamefont {D.~G.}\ \bibnamefont {Nocera}}, \bibinfo {author}
  {\bibfnamefont {J.~A.}\ \bibnamefont {Rodriguez-Rivera}}, \bibinfo {author}
  {\bibfnamefont {C.}~\bibnamefont {Broholm}}, \ and\ \bibinfo {author}
  {\bibfnamefont {Y.~S.}\ \bibnamefont {Lee}},\ }\bibfield  {title} {\enquote
  {\bibinfo {title} {{Fractionalized excitations in the spin-liquid state of a
  kagome-lattice antiferromagnet}},}\ }\href
  {http://dx.doi.org/10.1038/nature11659} {\bibfield  {journal} {\bibinfo
  {journal} {Nature (London)}\ }\textbf {\bibinfo {volume} {492}},\ \bibinfo
  {pages} {406--410} (\bibinfo {year} {2012})}\BibitemShut {NoStop}%
\bibitem [{\citenamefont {Paddison}\ \emph {et~al.}(2017)\citenamefont
  {Paddison}, \citenamefont {Daum}, \citenamefont {Dun}, \citenamefont
  {Ehlers}, \citenamefont {Liu}, \citenamefont {Stone}, \citenamefont {Zhou},\
  and\ \citenamefont {Mourigal}}]{paddison17}%
  \BibitemOpen
  \bibfield  {author} {\bibinfo {author} {\bibfnamefont {J.~A.~M.}\
  \bibnamefont {Paddison}}, \bibinfo {author} {\bibfnamefont {M.}~\bibnamefont
  {Daum}}, \bibinfo {author} {\bibfnamefont {Z.~L.}\ \bibnamefont {Dun}},
  \bibinfo {author} {\bibfnamefont {G.}~\bibnamefont {Ehlers}}, \bibinfo
  {author} {\bibfnamefont {Y.}~\bibnamefont {Liu}}, \bibinfo {author}
  {\bibfnamefont {M.~B.}\ \bibnamefont {Stone}}, \bibinfo {author}
  {\bibfnamefont {H.}~\bibnamefont {Zhou}}, \ and\ \bibinfo {author}
  {\bibfnamefont {M.}~\bibnamefont {Mourigal}},\ }\bibfield  {title} {\enquote
  {\bibinfo {title} {{Continuous excitations of the triangular-lattice quantum
  spin liquid $\mathrm{{YbMgGaO}}_4$}},}\ }\href
  {http://dx.doi.org/10.1038/nphys3971} {\bibfield  {journal} {\bibinfo
  {journal} {Nat. Phys.}\ }\textbf {\bibinfo {volume} {13}},\ \bibinfo {pages}
  {117--122} (\bibinfo {year} {2017})}\BibitemShut {NoStop}%
\bibitem [{\citenamefont {Savary}\ and\ \citenamefont
  {Balents}(2017{\natexlab{b}})}]{savary17-RPP80}%
  \BibitemOpen
  \bibfield  {author} {\bibinfo {author} {\bibfnamefont {L.}~\bibnamefont
  {Savary}}\ and\ \bibinfo {author} {\bibfnamefont {L.}~\bibnamefont
  {Balents}},\ }\bibfield  {title} {\enquote {\bibinfo {title} {{Quantum spin
  liquids: a review}},}\ }\href
  {http://stacks.iop.org/0034-4885/80/i=1/a=016502} {\bibfield  {journal}
  {\bibinfo  {journal} {Rep. Prog. Phys.}\ }\textbf {\bibinfo {volume} {80}},\
  \bibinfo {pages} {016502} (\bibinfo {year} {2017}{\natexlab{b}})}\BibitemShut
  {NoStop}%
\bibitem [{\citenamefont {Savary}\ and\ \citenamefont {Senthil}()}]{Savary15a}%
  \BibitemOpen
  \bibfield  {author} {\bibinfo {author} {\bibfnamefont {L.}~\bibnamefont
  {Savary}}\ and\ \bibinfo {author} {\bibfnamefont {T.}~\bibnamefont
  {Senthil}},\ }\bibfield  {title} {\enquote {\bibinfo {title} {{Probing Hidden
  Orders with Resonant Inelastic X-Ray Scattering}},}\ }\href
  {https://arxiv.org/abs/1506.04752} {\bibinfo  {journal} {arXiv:1506.04752}\
  }\BibitemShut {NoStop}%
\bibitem [{\citenamefont {Smerald}\ and\ \citenamefont
  {Shannon}(2013)}]{smerald13}%
  \BibitemOpen
\bibfield  {journal} {  }\bibfield  {author} {\bibinfo {author} {\bibfnamefont
  {A.}~\bibnamefont {Smerald}}\ and\ \bibinfo {author} {\bibfnamefont
  {N.}~\bibnamefont {Shannon}},\ }\bibfield  {title} {\enquote {\bibinfo
  {title} {{Theory of spin excitations in a quantum spin-nematic state}},}\
  }\href {\doibase 10.1103/PhysRevB.88.184430} {\bibfield  {journal} {\bibinfo
  {journal} {Phys. Rev. B}\ }\textbf {\bibinfo {volume} {88}},\ \bibinfo
  {pages} {184430} (\bibinfo {year} {2013})}\BibitemShut {NoStop}%
\bibitem [{\citenamefont {Starykh}\ and\ \citenamefont
  {Balents}(2014)}]{starykh14}%
  \BibitemOpen
  \bibfield  {author} {\bibinfo {author} {\bibfnamefont {O.~A.}\ \bibnamefont
  {Starykh}}\ and\ \bibinfo {author} {\bibfnamefont {L.}~\bibnamefont
  {Balents}},\ }\bibfield  {title} {\enquote {\bibinfo {title} {{Excitations
  and quasi-one-dimensionality in field-induced nematic and spin density wave
  states}},}\ }\href {\doibase 10.1103/PhysRevB.89.104407} {\bibfield
  {journal} {\bibinfo  {journal} {Phys. Rev. B}\ }\textbf {\bibinfo {volume}
  {89}},\ \bibinfo {pages} {104407} (\bibinfo {year} {2014})}\BibitemShut
  {NoStop}%
\bibitem [{\citenamefont {Smerald}\ and\ \citenamefont
  {Shannon}(2016)}]{smerald16}%
  \BibitemOpen
  \bibfield  {author} {\bibinfo {author} {\bibfnamefont {A.}~\bibnamefont
  {Smerald}}\ and\ \bibinfo {author} {\bibfnamefont {N.}~\bibnamefont
  {Shannon}},\ }\bibfield  {title} {\enquote {\bibinfo {title} {{Theory of NMR
  $1/{T}_{1}$ relaxation in a quantum spin nematic in an applied magnetic
  field}},}\ }\href {\doibase 10.1103/PhysRevB.93.184419} {\bibfield  {journal}
  {\bibinfo  {journal} {Phys. Rev. B}\ }\textbf {\bibinfo {volume} {93}},\
  \bibinfo {pages} {184419} (\bibinfo {year} {2016})}\BibitemShut {NoStop}%
\bibitem [{\citenamefont {Tsunetsugu}\ and\ \citenamefont
  {Arikawa}(2006)}]{tsunetsugu06}%
  \BibitemOpen
  \bibfield  {author} {\bibinfo {author} {\bibfnamefont {H.}~\bibnamefont
  {Tsunetsugu}}\ and\ \bibinfo {author} {\bibfnamefont {M.}~\bibnamefont
  {Arikawa}},\ }\bibfield  {title} {\enquote {\bibinfo {title} {{Spin Nematic
  Phase in S=1 Triangular Antiferromagnets}},}\ }\href {\doibase
  10.1143/JPSJ.75.083701} {\bibfield  {journal} {\bibinfo  {journal} {J. Phys.
  Soc. Jpn}\ }\textbf {\bibinfo {volume} {75}},\ \bibinfo {pages} {083701}
  (\bibinfo {year} {2006})}\BibitemShut {NoStop}%
\bibitem [{\citenamefont {L\"auchli}\ \emph {et~al.}(2006)\citenamefont
  {L\"auchli}, \citenamefont {Mila},\ and\ \citenamefont {Penc}}]{laeuchli06}%
  \BibitemOpen
  \bibfield  {author} {\bibinfo {author} {\bibfnamefont {A.}~\bibnamefont
  {L\"auchli}}, \bibinfo {author} {\bibfnamefont {F.}~\bibnamefont {Mila}}, \
  and\ \bibinfo {author} {\bibfnamefont {K.}~\bibnamefont {Penc}},\ }\bibfield
  {title} {\enquote {\bibinfo {title} {{Quadrupolar Phases of the $S=1$
  Bilinear-Biquadratic Heisenberg Model on the Triangular Lattice}},}\ }\href
  {\doibase 10.1103/PhysRevLett.97.087205} {\bibfield  {journal} {\bibinfo
  {journal} {Phys. Rev. Lett.}\ }\textbf {\bibinfo {volume} {97}},\ \bibinfo
  {pages} {087205} (\bibinfo {year} {2006})}\BibitemShut {NoStop}%
\bibitem [{\citenamefont {V\"oll}\ and\ \citenamefont
  {Wessel}(2015)}]{voell15}%
  \BibitemOpen
  \bibfield  {author} {\bibinfo {author} {\bibfnamefont {A.}~\bibnamefont
  {V\"oll}}\ and\ \bibinfo {author} {\bibfnamefont {S.}~\bibnamefont
  {Wessel}},\ }\bibfield  {title} {\enquote {\bibinfo {title} {{Spin dynamics
  of the bilinear-biquadratic $S=1$ Heisenberg model on the triangular lattice:
  A quantum Monte Carlo study}},}\ }\href {\doibase 10.1103/PhysRevB.91.165128}
  {\bibfield  {journal} {\bibinfo  {journal} {Phys. Rev. B}\ }\textbf {\bibinfo
  {volume} {91}},\ \bibinfo {pages} {165128} (\bibinfo {year}
  {2015})}\BibitemShut {NoStop}%
\bibitem [{\citenamefont {Onishi}(2015)}]{onishi15}%
  \BibitemOpen
  \bibfield  {author} {\bibinfo {author} {\bibfnamefont {H.}~\bibnamefont
  {Onishi}},\ }\bibfield  {title} {\enquote {\bibinfo {title} {{Magnetic
  Excitations of Spin Nematic State in Frustrated Ferromagnetic Chain}},}\
  }\href {\doibase 10.7566/JPSJ.84.083702} {\bibfield  {journal} {\bibinfo
  {journal} {J. Phys. Soc. Jpn}\ }\textbf {\bibinfo {volume} {84}},\ \bibinfo
  {pages} {083702} (\bibinfo {year} {2015})}\BibitemShut {NoStop}%
\bibitem [{\citenamefont {Shindou}\ \emph {et~al.}(2013)\citenamefont
  {Shindou}, \citenamefont {Yunoki},\ and\ \citenamefont {Momoi}}]{shindou13}%
  \BibitemOpen
  \bibfield  {author} {\bibinfo {author} {\bibfnamefont {R.}~\bibnamefont
  {Shindou}}, \bibinfo {author} {\bibfnamefont {S.}~\bibnamefont {Yunoki}}, \
  and\ \bibinfo {author} {\bibfnamefont {T.}~\bibnamefont {Momoi}},\ }\bibfield
   {title} {\enquote {\bibinfo {title} {{Dynamical spin structure factors of
  quantum spin nematic states}},}\ }\href {\doibase 10.1103/PhysRevB.87.054429}
  {\bibfield  {journal} {\bibinfo  {journal} {Phys. Rev. B}\ }\textbf {\bibinfo
  {volume} {87}},\ \bibinfo {pages} {054429} (\bibinfo {year}
  {2013})}\BibitemShut {NoStop}%
\bibitem [{\citenamefont {Guitteny}\ \emph {et~al.}(2013)\citenamefont
  {Guitteny}, \citenamefont {Robert}, \citenamefont {Bonville}, \citenamefont
  {Ollivier}, \citenamefont {Decorse}, \citenamefont {Steffens}, \citenamefont
  {Boehm}, \citenamefont {Mutka}, \citenamefont {Mirebeau},\ and\ \citenamefont
  {Petit}}]{Guitteny2013}%
  \BibitemOpen
  \bibfield  {author} {\bibinfo {author} {\bibfnamefont {S.}~\bibnamefont
  {Guitteny}}, \bibinfo {author} {\bibfnamefont {J.}~\bibnamefont {Robert}},
  \bibinfo {author} {\bibfnamefont {P.}~\bibnamefont {Bonville}}, \bibinfo
  {author} {\bibfnamefont {J.}~\bibnamefont {Ollivier}}, \bibinfo {author}
  {\bibfnamefont {C.}~\bibnamefont {Decorse}}, \bibinfo {author} {\bibfnamefont
  {P.}~\bibnamefont {Steffens}}, \bibinfo {author} {\bibfnamefont
  {M.}~\bibnamefont {Boehm}}, \bibinfo {author} {\bibfnamefont
  {H.}~\bibnamefont {Mutka}}, \bibinfo {author} {\bibfnamefont
  {I.}~\bibnamefont {Mirebeau}}, \ and\ \bibinfo {author} {\bibfnamefont
  {S.}~\bibnamefont {Petit}},\ }\bibfield  {title} {\enquote {\bibinfo {title}
  {{Anisotropic Propagating Excitations and Quadrupolar Effects in
  ${\mathrm{Tb}}_{2}{\mathrm{Ti}}_{2}{\mathrm{O}}_{7}$}},}\ }\href {\doibase
  10.1103/PhysRevLett.111.087201} {\bibfield  {journal} {\bibinfo  {journal}
  {Phys. Rev. Lett.}\ }\textbf {\bibinfo {volume} {111}},\ \bibinfo {pages}
  {087201} (\bibinfo {year} {2013})}\BibitemShut {NoStop}%
\bibitem [{\citenamefont {Li}\ and\ \citenamefont {Chen}(2017)}]{Li17}%
  \BibitemOpen
  \bibfield  {author} {\bibinfo {author} {\bibfnamefont {Yao-Dong}\
  \bibnamefont {Li}}\ and\ \bibinfo {author} {\bibfnamefont {Gang}\
  \bibnamefont {Chen}},\ }\bibfield  {title} {\enquote {\bibinfo {title}
  {{Symmetry enriched U(1) topological orders for dipole-octupole doublets on a
  pyrochlore lattice}},}\ }\href {\doibase 10.1103/PhysRevB.95.041106}
  {\bibfield  {journal} {\bibinfo  {journal} {Phys. Rev. B}\ }\textbf {\bibinfo
  {volume} {95}},\ \bibinfo {pages} {041106} (\bibinfo {year}
  {2017})}\BibitemShut {NoStop}%
\bibitem [{\citenamefont {Benton}(2017)}]{benton-arXiv17}%
  \BibitemOpen
  \bibfield  {author} {\bibinfo {author} {\bibfnamefont {O.}~\bibnamefont
  {Benton}},\ }\bibfield  {title} {\enquote {\bibinfo {title} {{From quantum
  spin liquid to paramagnetic ground states in disordered non-Kramers
  pyrochlores}},}\ }\href {https://arxiv.org/abs/1706.09238} {\bibfield
  {journal} {\bibinfo  {journal} {arXiv:1706.09238}\ } (\bibinfo {year}
  {2017})}\BibitemShut {NoStop}%
\bibitem [{\citenamefont {Martin}\ \emph {et~al.}()\citenamefont {Martin},
  \citenamefont {Bonville}, \citenamefont {Lhotel}, \citenamefont {Guitteny},
  \citenamefont {Wildes}, \citenamefont {Decorse}, \citenamefont {Hatnean},
  \citenamefont {Balakrishnan}, \citenamefont {Mirebeau},\ and\ \citenamefont
  {Petit}}]{martin-arXiv17}%
  \BibitemOpen
  \bibfield  {author} {\bibinfo {author} {\bibfnamefont {N.}~\bibnamefont
  {Martin}}, \bibinfo {author} {\bibfnamefont {P.}~\bibnamefont {Bonville}},
  \bibinfo {author} {\bibfnamefont {E.}~\bibnamefont {Lhotel}}, \bibinfo
  {author} {\bibfnamefont {S.}~\bibnamefont {Guitteny}}, \bibinfo {author}
  {\bibfnamefont {A.}~\bibnamefont {Wildes}}, \bibinfo {author} {\bibfnamefont
  {C.}~\bibnamefont {Decorse}}, \bibinfo {author} {\bibfnamefont {M.~C.}\
  \bibnamefont {Hatnean}}, \bibinfo {author} {\bibfnamefont {G.}~\bibnamefont
  {Balakrishnan}}, \bibinfo {author} {\bibfnamefont {I.}~\bibnamefont
  {Mirebeau}}, \ and\ \bibinfo {author} {\bibfnamefont {S.}~\bibnamefont
  {Petit}},\ }\bibfield  {title} {\enquote {\bibinfo {title} {{Disorder and
  Quantum Spin Ice}},}\ }\href {https://arxiv.org/abs/1708.01845} {\bibinfo
  {journal} {arXiv:1708.01845}\ }\BibitemShut {NoStop}%
\bibitem [{\citenamefont {Dun}\ \emph {et~al.}(2014)\citenamefont {Dun},
  \citenamefont {Lee}, \citenamefont {Choi}, \citenamefont {Hallas},
  \citenamefont {Wiebe}, \citenamefont {Gardner}, \citenamefont {Arrighi},
  \citenamefont {Freitas}, \citenamefont {Arevalo-Lopez}, \citenamefont
  {Attfield}, \citenamefont {Zhou},\ and\ \citenamefont {Cheng}}]{dun14a}%
  \BibitemOpen
\bibfield  {journal} {  }\bibfield  {author} {\bibinfo {author} {\bibfnamefont
  {Z.~L.}\ \bibnamefont {Dun}}, \bibinfo {author} {\bibfnamefont
  {M.}~\bibnamefont {Lee}}, \bibinfo {author} {\bibfnamefont {E.~S.}\
  \bibnamefont {Choi}}, \bibinfo {author} {\bibfnamefont {A.~M.}\ \bibnamefont
  {Hallas}}, \bibinfo {author} {\bibfnamefont {C.~R.}\ \bibnamefont {Wiebe}},
  \bibinfo {author} {\bibfnamefont {J.~S.}\ \bibnamefont {Gardner}}, \bibinfo
  {author} {\bibfnamefont {E.}~\bibnamefont {Arrighi}}, \bibinfo {author}
  {\bibfnamefont {R.~S.}\ \bibnamefont {Freitas}}, \bibinfo {author}
  {\bibfnamefont {A.~M.}\ \bibnamefont {Arevalo-Lopez}}, \bibinfo {author}
  {\bibfnamefont {J.~P.}\ \bibnamefont {Attfield}}, \bibinfo {author}
  {\bibfnamefont {H.~D.}\ \bibnamefont {Zhou}}, \ and\ \bibinfo {author}
  {\bibfnamefont {J.~G.}\ \bibnamefont {Cheng}},\ }\bibfield  {title} {\enquote
  {\bibinfo {title} {{Chemical pressure effects on magnetism in the quantum
  spin liquid candidates $\mathrm{{Yb}}_2\mathrm{{X}}_2\mathrm{{O}}_7$
  ($\mathrm{X}=\mathrm{Sn}, \mathrm{Ti}, \mathrm{Ge}$)}},}\ }\href {\doibase
  10.1103/PhysRevB.89.064401} {\bibfield  {journal} {\bibinfo  {journal} {Phys.
  Rev. B}\ }\textbf {\bibinfo {volume} {89}},\ \bibinfo {pages} {064401}
  (\bibinfo {year} {2014})}\BibitemShut {NoStop}%
\bibitem [{\citenamefont {Wiebe}\ and\ \citenamefont
  {Hallas}(2015)}]{wiebe15a}%
  \BibitemOpen
  \bibfield  {author} {\bibinfo {author} {\bibfnamefont {C.~R.}\ \bibnamefont
  {Wiebe}}\ and\ \bibinfo {author} {\bibfnamefont {A.~M.}\ \bibnamefont
  {Hallas}},\ }\bibfield  {title} {\enquote {\bibinfo {title} {{Frustration
  under pressure: Exotic magnetism in new pyrochlore oxides}},}\ }\href
  {\doibase http://dx.doi.org/10.1063/1.4916020} {\bibfield  {journal}
  {\bibinfo  {journal} {APL Materials}\ }\textbf {\bibinfo {volume} {3}},\
  \bibinfo {pages} {041519} (\bibinfo {year} {2015})}\BibitemShut {NoStop}%
\bibitem [{\citenamefont {Matsuhira}\ \emph {et~al.}(2004)\citenamefont
  {Matsuhira}, \citenamefont {Sekine}, \citenamefont {Paulsen},\ and\
  \citenamefont {Hinatsu}}]{Matsuhira2004}%
  \BibitemOpen
  \bibfield  {author} {\bibinfo {author} {\bibfnamefont {K.}~\bibnamefont
  {Matsuhira}}, \bibinfo {author} {\bibfnamefont {C.}~\bibnamefont {Sekine}},
  \bibinfo {author} {\bibfnamefont {C.}~\bibnamefont {Paulsen}}, \ and\
  \bibinfo {author} {\bibfnamefont {Y.}~\bibnamefont {Hinatsu}},\ }\bibfield
  {title} {\enquote {\bibinfo {title} {{Low-temperature magnetic properties of
  the geometrically frustrated pyrochlore
  $\mathrm{{Pr}}_2\mathrm{{Sn}}_2\mathrm{{O}}_7$}},}\ }\href {\doibase
  http://doi.org/10.1016/j.jmmm.2003.12.500} {\bibfield  {journal} {\bibinfo
  {journal} {J. Magn. Magn. Mater.}\ }\textbf {\bibinfo {volume} {272--276}},\
  \bibinfo {pages} {E981 -- E982} (\bibinfo {year} {2004})}\BibitemShut
  {NoStop}%
\bibitem [{\citenamefont {Robert}\ \emph {et~al.}(2015)\citenamefont {Robert},
  \citenamefont {Lhotel}, \citenamefont {Remenyi}, \citenamefont {Sahling},
  \citenamefont {Mirebeau}, \citenamefont {Decorse}, \citenamefont {Canals},\
  and\ \citenamefont {Petit}}]{Robert2015}%
  \BibitemOpen
  \bibfield  {author} {\bibinfo {author} {\bibfnamefont {J.}~\bibnamefont
  {Robert}}, \bibinfo {author} {\bibfnamefont {E.}~\bibnamefont {Lhotel}},
  \bibinfo {author} {\bibfnamefont {G.}~\bibnamefont {Remenyi}}, \bibinfo
  {author} {\bibfnamefont {S.}~\bibnamefont {Sahling}}, \bibinfo {author}
  {\bibfnamefont {I.}~\bibnamefont {Mirebeau}}, \bibinfo {author}
  {\bibfnamefont {C.}~\bibnamefont {Decorse}}, \bibinfo {author} {\bibfnamefont
  {B.}~\bibnamefont {Canals}}, \ and\ \bibinfo {author} {\bibfnamefont
  {S.}~\bibnamefont {Petit}},\ }\bibfield  {title} {\enquote {\bibinfo {title}
  {{Spin dynamics in the presence of competing ferromagnetic and
  antiferromagnetic correlations in
  ${\mathrm{Yb}}_{2}{\mathrm{Ti}}_{2}{\mathrm{O}}_{7}$}},}\ }\href {\doibase
  10.1103/PhysRevB.92.064425} {\bibfield  {journal} {\bibinfo  {journal} {Phys.
  Rev. B}\ }\textbf {\bibinfo {volume} {92}},\ \bibinfo {pages} {064425}
  (\bibinfo {year} {2015})}\BibitemShut {NoStop}%
\bibitem [{\citenamefont {Thompson}\ \emph {et~al.}(2017)\citenamefont
  {Thompson}, \citenamefont {McClarty}, \citenamefont {Prabhakaran},
  \citenamefont {Cabrera}, \citenamefont {Guidi},\ and\ \citenamefont
  {Coldea}}]{Thompson17}%
  \BibitemOpen
  \bibfield  {author} {\bibinfo {author} {\bibfnamefont {J.~D.}\ \bibnamefont
  {Thompson}}, \bibinfo {author} {\bibfnamefont {P.~A.}\ \bibnamefont
  {McClarty}}, \bibinfo {author} {\bibfnamefont {D.}~\bibnamefont
  {Prabhakaran}}, \bibinfo {author} {\bibfnamefont {I.}~\bibnamefont
  {Cabrera}}, \bibinfo {author} {\bibfnamefont {T.}~\bibnamefont {Guidi}}, \
  and\ \bibinfo {author} {\bibfnamefont {R.}~\bibnamefont {Coldea}},\
  }\bibfield  {title} {\enquote {\bibinfo {title} {{Quasiparticle Breakdown and
  Spin Hamiltonian of the Frustrated Quantum Pyrochlore Yb$_{2}$Ti$_{2}$O$_{7}$
  in a Magnetic Field}},}\ }\href {\doibase 10.1103/PhysRevLett.119.057203}
  {\bibfield  {journal} {\bibinfo  {journal} {Phys. Rev. Lett.}\ }\textbf
  {\bibinfo {volume} {119}},\ \bibinfo {pages} {057203} (\bibinfo {year}
  {2017})}\BibitemShut {NoStop}%
\bibitem [{\citenamefont {Ross}\ \emph {et~al.}(2009)\citenamefont {Ross},
  \citenamefont {Ruff}, \citenamefont {Adams}, \citenamefont {Gardner},
  \citenamefont {Dabkowska}, \citenamefont {Qiu}, \citenamefont {Copley},\ and\
  \citenamefont {Gaulin}}]{Ross2009}%
  \BibitemOpen
  \bibfield  {author} {\bibinfo {author} {\bibfnamefont {K.~A.}\ \bibnamefont
  {Ross}}, \bibinfo {author} {\bibfnamefont {J.~P.~C.}\ \bibnamefont {Ruff}},
  \bibinfo {author} {\bibfnamefont {C.~P.}\ \bibnamefont {Adams}}, \bibinfo
  {author} {\bibfnamefont {J.~S.}\ \bibnamefont {Gardner}}, \bibinfo {author}
  {\bibfnamefont {H.~A.}\ \bibnamefont {Dabkowska}}, \bibinfo {author}
  {\bibfnamefont {Y.}~\bibnamefont {Qiu}}, \bibinfo {author} {\bibfnamefont
  {J.~R.~D.}\ \bibnamefont {Copley}}, \ and\ \bibinfo {author} {\bibfnamefont
  {B.~D.}\ \bibnamefont {Gaulin}},\ }\bibfield  {title} {\enquote {\bibinfo
  {title} {{Two-Dimensional Kagome Correlations and Field Induced Order in the
  Ferromagnetic $XY$ Pyrochlore
  ${\mathrm{Yb}}_{2}{\mathrm{Ti}}_{2}{\mathrm{O}}_{7}$}},}\ }\href {\doibase
  10.1103/PhysRevLett.103.227202} {\bibfield  {journal} {\bibinfo  {journal}
  {Phys. Rev. Lett.}\ }\textbf {\bibinfo {volume} {103}},\ \bibinfo {pages}
  {227202} (\bibinfo {year} {2009})}\BibitemShut {NoStop}%
\bibitem [{\citenamefont {Maisuradze}\ \emph {et~al.}(2015)\citenamefont
  {Maisuradze}, \citenamefont {Dalmas~de R\'eotier}, \citenamefont {Yaouanc},
  \citenamefont {Forget}, \citenamefont {Baines},\ and\ \citenamefont
  {King}}]{Maisuradze2015}%
  \BibitemOpen
  \bibfield  {author} {\bibinfo {author} {\bibfnamefont {A.}~\bibnamefont
  {Maisuradze}}, \bibinfo {author} {\bibfnamefont {P.}~\bibnamefont {Dalmas~de
  R\'eotier}}, \bibinfo {author} {\bibfnamefont {A.}~\bibnamefont {Yaouanc}},
  \bibinfo {author} {\bibfnamefont {A.}~\bibnamefont {Forget}}, \bibinfo
  {author} {\bibfnamefont {C.}~\bibnamefont {Baines}}, \ and\ \bibinfo {author}
  {\bibfnamefont {P.~J.~C.}\ \bibnamefont {King}},\ }\bibfield  {title}
  {\enquote {\bibinfo {title} {{Anomalously slow spin dynamics and short-range
  correlations in the quantum spin ice systems
  ${\mathrm{Yb}}_{2}{\mathrm{Ti}}_{2}{\mathrm{O}}_{7}$ and
  ${\mathrm{Yb}}_{2}{\mathrm{Sn}}_{2}{\mathrm{O}}_{7}$}},}\ }\href {\doibase
  10.1103/PhysRevB.92.094424} {\bibfield  {journal} {\bibinfo  {journal} {Phys.
  Rev. B}\ }\textbf {\bibinfo {volume} {92}},\ \bibinfo {pages} {094424}
  (\bibinfo {year} {2015})}\BibitemShut {NoStop}%
\bibitem [{\citenamefont {Gaudet}\ \emph {et~al.}(2016)\citenamefont {Gaudet},
  \citenamefont {Ross}, \citenamefont {Kermarrec}, \citenamefont {Butch},
  \citenamefont {Ehlers}, \citenamefont {Dabkowska},\ and\ \citenamefont
  {Gaulin}}]{gaudet16}%
  \BibitemOpen
  \bibfield  {author} {\bibinfo {author} {\bibfnamefont {J.}~\bibnamefont
  {Gaudet}}, \bibinfo {author} {\bibfnamefont {K.~A.}\ \bibnamefont {Ross}},
  \bibinfo {author} {\bibfnamefont {E.}~\bibnamefont {Kermarrec}}, \bibinfo
  {author} {\bibfnamefont {N.~P.}\ \bibnamefont {Butch}}, \bibinfo {author}
  {\bibfnamefont {G.}~\bibnamefont {Ehlers}}, \bibinfo {author} {\bibfnamefont
  {H.~A.}\ \bibnamefont {Dabkowska}}, \ and\ \bibinfo {author} {\bibfnamefont
  {B.~D.}\ \bibnamefont {Gaulin}},\ }\bibfield  {title} {\enquote {\bibinfo
  {title} {{Gapless quantum excitations from an icelike splayed ferromagnetic
  ground state in stoichiometric
  $\mathrm{{Yb}}_2\mathrm{{Ti}}_2\mathrm{{O}}_7$}},}\ }\href {\doibase
  10.1103/PhysRevB.93.064406} {\bibfield  {journal} {\bibinfo  {journal} {Phys.
  Rev. B}\ }\textbf {\bibinfo {volume} {93}},\ \bibinfo {pages} {064406}
  (\bibinfo {year} {2016})}\BibitemShut {NoStop}%
\bibitem [{\citenamefont {Yasui}\ \emph {et~al.}(2003)\citenamefont {Yasui},
  \citenamefont {Soda}, \citenamefont {Iikubo}, \citenamefont {Ito},
  \citenamefont {Sato}, \citenamefont {Hamaguchi}, \citenamefont {Matsushita},
  \citenamefont {Wada}, \citenamefont {Takeuchi}, \citenamefont {Aso},\ and\
  \citenamefont {Kakurai}}]{Yasui2003a}%
  \BibitemOpen
  \bibfield  {author} {\bibinfo {author} {\bibfnamefont {Y.}~\bibnamefont
  {Yasui}}, \bibinfo {author} {\bibfnamefont {M.}~\bibnamefont {Soda}},
  \bibinfo {author} {\bibfnamefont {S.}~\bibnamefont {Iikubo}}, \bibinfo
  {author} {\bibfnamefont {M.}~\bibnamefont {Ito}}, \bibinfo {author}
  {\bibfnamefont {M.}~\bibnamefont {Sato}}, \bibinfo {author} {\bibfnamefont
  {N.}~\bibnamefont {Hamaguchi}}, \bibinfo {author} {\bibfnamefont
  {T.}~\bibnamefont {Matsushita}}, \bibinfo {author} {\bibfnamefont
  {N.}~\bibnamefont {Wada}}, \bibinfo {author} {\bibfnamefont {T.}~\bibnamefont
  {Takeuchi}}, \bibinfo {author} {\bibfnamefont {N.}~\bibnamefont {Aso}}, \
  and\ \bibinfo {author} {\bibfnamefont {K.}~\bibnamefont {Kakurai}},\
  }\bibfield  {title} {\enquote {\bibinfo {title} {{Ferromagnetic Transition of
  Pyrochlore Compound $\mathrm{{Yb}}_2\mathrm{{Ti}}_2\mathrm{{O}}_7$}},}\
  }\href {\doibase 10.1143/JPSJ.72.3014} {\bibfield  {journal} {\bibinfo
  {journal} {J. Phys. Soc. Jpn}\ }\textbf {\bibinfo {volume} {72}},\ \bibinfo
  {pages} {3014--3015} (\bibinfo {year} {2003})}\BibitemShut {NoStop}%
\bibitem [{\citenamefont {Lhotel}\ \emph {et~al.}(2014)\citenamefont {Lhotel},
  \citenamefont {Giblin}, \citenamefont {Lees}, \citenamefont {Balakrishnan},
  \citenamefont {Chang},\ and\ \citenamefont {Yasui}}]{lhotel14}%
  \BibitemOpen
  \bibfield  {author} {\bibinfo {author} {\bibfnamefont {E.}~\bibnamefont
  {Lhotel}}, \bibinfo {author} {\bibfnamefont {S.~R.}\ \bibnamefont {Giblin}},
  \bibinfo {author} {\bibfnamefont {M.~R.}\ \bibnamefont {Lees}}, \bibinfo
  {author} {\bibfnamefont {G.}~\bibnamefont {Balakrishnan}}, \bibinfo {author}
  {\bibfnamefont {L.~J.}\ \bibnamefont {Chang}}, \ and\ \bibinfo {author}
  {\bibfnamefont {Y.}~\bibnamefont {Yasui}},\ }\bibfield  {title} {\enquote
  {\bibinfo {title} {{First-order magnetic transition in
  ${\mathrm{Yb}}_{2}{\mathrm{Ti}}_{2}{\mathrm{O}}_{7}$}},}\ }\href {\doibase
  10.1103/PhysRevB.89.224419} {\bibfield  {journal} {\bibinfo  {journal} {Phys.
  Rev. B}\ }\textbf {\bibinfo {volume} {89}},\ \bibinfo {pages} {224419}
  (\bibinfo {year} {2014})}\BibitemShut {NoStop}%
\bibitem [{\citenamefont {Hallas}\ \emph
  {et~al.}(2016{\natexlab{b}})\citenamefont {Hallas}, \citenamefont {Gaudet},
  \citenamefont {Wilson}, \citenamefont {Munsie}, \citenamefont {Aczel},
  \citenamefont {Stone}, \citenamefont {Freitas}, \citenamefont
  {Arevalo-Lopez}, \citenamefont {Attfield}, \citenamefont {Tachibana},
  \citenamefont {Wiebe}, \citenamefont {Luke},\ and\ \citenamefont
  {Gaulin}}]{hallas16-PRB93.104405}%
  \BibitemOpen
  \bibfield  {author} {\bibinfo {author} {\bibfnamefont {A.~M.}\ \bibnamefont
  {Hallas}}, \bibinfo {author} {\bibfnamefont {J.}~\bibnamefont {Gaudet}},
  \bibinfo {author} {\bibfnamefont {M.~N.}\ \bibnamefont {Wilson}}, \bibinfo
  {author} {\bibfnamefont {T.~J.}\ \bibnamefont {Munsie}}, \bibinfo {author}
  {\bibfnamefont {A.~A.}\ \bibnamefont {Aczel}}, \bibinfo {author}
  {\bibfnamefont {M.~B.}\ \bibnamefont {Stone}}, \bibinfo {author}
  {\bibfnamefont {R.~S.}\ \bibnamefont {Freitas}}, \bibinfo {author}
  {\bibfnamefont {A.~M.}\ \bibnamefont {Arevalo-Lopez}}, \bibinfo {author}
  {\bibfnamefont {J.~P.}\ \bibnamefont {Attfield}}, \bibinfo {author}
  {\bibfnamefont {M.}~\bibnamefont {Tachibana}}, \bibinfo {author}
  {\bibfnamefont {C.~R.}\ \bibnamefont {Wiebe}}, \bibinfo {author}
  {\bibfnamefont {G.~M.}\ \bibnamefont {Luke}}, \ and\ \bibinfo {author}
  {\bibfnamefont {B.~D.}\ \bibnamefont {Gaulin}},\ }\bibfield  {title}
  {\enquote {\bibinfo {title} {{XY antiferromagnetic ground state in the
  effective $S=\frac{1}{2}$ pyrochlore
  ${\mathrm{Yb}}_{2}{\mathrm{Ge}}_{2}{\mathrm{O}}_{7}$}},}\ }\href
  {http://link.aps.org/doi/10.1103/PhysRevB.93.104405} {\bibfield  {journal}
  {\bibinfo  {journal} {Phys. Rev. B}\ }\textbf {\bibinfo {volume} {93}},\
  \bibinfo {pages} {104405} (\bibinfo {year} {2016}{\natexlab{b}})}\BibitemShut
  {NoStop}%
\bibitem [{\citenamefont {Powell}(2015)}]{Powell2015}%
  \BibitemOpen
  \bibfield  {author} {\bibinfo {author} {\bibfnamefont {S.}~\bibnamefont
  {Powell}},\ }\bibfield  {title} {\enquote {\bibinfo {title} {{Ferromagnetic
  Coulomb phase in classical spin ice}},}\ }\href {\doibase
  10.1103/PhysRevB.91.094431} {\bibfield  {journal} {\bibinfo  {journal} {Phys.
  Rev. B}\ }\textbf {\bibinfo {volume} {91}},\ \bibinfo {pages} {094431}
  (\bibinfo {year} {2015})}\BibitemShut {NoStop}%
\bibitem [{\citenamefont {Brooks-Bartlett}\ \emph {et~al.}(2014)\citenamefont
  {Brooks-Bartlett}, \citenamefont {Banks}, \citenamefont {Jaubert},
  \citenamefont {Harman-Clarke},\ and\ \citenamefont
  {Holdsworth}}]{Brooks-Bartlett2014}%
  \BibitemOpen
  \bibfield  {author} {\bibinfo {author} {\bibfnamefont {M.~E.}\ \bibnamefont
  {Brooks-Bartlett}}, \bibinfo {author} {\bibfnamefont {S.~T.}\ \bibnamefont
  {Banks}}, \bibinfo {author} {\bibfnamefont {L.~D~C}\ \bibnamefont {Jaubert}},
  \bibinfo {author} {\bibfnamefont {A.}~\bibnamefont {Harman-Clarke}}, \ and\
  \bibinfo {author} {\bibfnamefont {P.~C~W}\ \bibnamefont {Holdsworth}},\
  }\bibfield  {title} {\enquote {\bibinfo {title} {{Magnetic-Moment
  Fragmentation and Monopole Crystallization}},}\ }\href
  {https://journals.aps.org/prx/abstract/10.1103/PhysRevX.4.011007} {\bibfield
  {journal} {\bibinfo  {journal} {Phys. Rev. X}\ }\textbf {\bibinfo {volume}
  {4}},\ \bibinfo {pages} {011007} (\bibinfo {year} {2014})}\BibitemShut
  {NoStop}%
\bibitem [{\citenamefont {Benton}(2016)}]{Benton16}%
  \BibitemOpen
  \bibfield  {author} {\bibinfo {author} {\bibfnamefont {O.}~\bibnamefont
  {Benton}},\ }\bibfield  {title} {\enquote {\bibinfo {title} {{Quantum origins
  of moment fragmentation in
  ${\mathrm{Nd}}_{2}{\mathrm{Zr}}_{2}{\mathrm{O}}_{7}$}},}\ }\href {\doibase
  10.1103/PhysRevB.94.104430} {\bibfield  {journal} {\bibinfo  {journal} {Phys.
  Rev. B}\ }\textbf {\bibinfo {volume} {94}},\ \bibinfo {pages} {104430}
  (\bibinfo {year} {2016})}\BibitemShut {NoStop}%
\bibitem [{\citenamefont {Lefran{\c c}ois}\ \emph {et~al.}(2017)\citenamefont
  {Lefran{\c c}ois}, \citenamefont {Cathelin}, \citenamefont {Lhotel},
  \citenamefont {Robert}, \citenamefont {Lejay}, \citenamefont {Colin},
  \citenamefont {Canals}, \citenamefont {Damay}, \citenamefont {Ollivier},
  \citenamefont {F{\aa}k}, \citenamefont {Chapon}, \citenamefont {Ballou},\
  and\ \citenamefont {Simonet}}]{Lefrancois17}%
  \BibitemOpen
  \bibfield  {author} {\bibinfo {author} {\bibfnamefont {E.}~\bibnamefont
  {Lefran{\c c}ois}}, \bibinfo {author} {\bibfnamefont {V.}~\bibnamefont
  {Cathelin}}, \bibinfo {author} {\bibfnamefont {E.}~\bibnamefont {Lhotel}},
  \bibinfo {author} {\bibfnamefont {J.}~\bibnamefont {Robert}}, \bibinfo
  {author} {\bibfnamefont {P.}~\bibnamefont {Lejay}}, \bibinfo {author}
  {\bibfnamefont {C.~V.}\ \bibnamefont {Colin}}, \bibinfo {author}
  {\bibfnamefont {B.}~\bibnamefont {Canals}}, \bibinfo {author} {\bibfnamefont
  {F.}~\bibnamefont {Damay}}, \bibinfo {author} {\bibfnamefont
  {J.}~\bibnamefont {Ollivier}}, \bibinfo {author} {\bibfnamefont
  {B.}~\bibnamefont {F{\aa}k}}, \bibinfo {author} {\bibfnamefont {L.~C.}\
  \bibnamefont {Chapon}}, \bibinfo {author} {\bibfnamefont {R.}~\bibnamefont
  {Ballou}}, \ and\ \bibinfo {author} {\bibfnamefont {V.}~\bibnamefont
  {Simonet}},\ }\bibfield  {title} {\enquote {\bibinfo {title} {{Fragmentation
  in spin ice from magnetic charge injection}},}\ }\href {\doibase
  10.1038/s41467-017-00277-1} {\bibfield  {journal} {\bibinfo  {journal} {Nat.
  Commun.}\ }\textbf {\bibinfo {volume} {8}},\ \bibinfo {pages} {209} (\bibinfo
  {year} {2017})}\BibitemShut {NoStop}%
\bibitem [{\citenamefont {Ross}\ \emph {et~al.}(2016)\citenamefont {Ross},
  \citenamefont {Krizan}, \citenamefont {Rodriguez-Rivera}, \citenamefont
  {Cava},\ and\ \citenamefont {Broholm}}]{ross16}%
  \BibitemOpen
  \bibfield  {author} {\bibinfo {author} {\bibfnamefont {K.~A.}\ \bibnamefont
  {Ross}}, \bibinfo {author} {\bibfnamefont {J.~W.}\ \bibnamefont {Krizan}},
  \bibinfo {author} {\bibfnamefont {J.~A.}\ \bibnamefont {Rodriguez-Rivera}},
  \bibinfo {author} {\bibfnamefont {R.~J.}\ \bibnamefont {Cava}}, \ and\
  \bibinfo {author} {\bibfnamefont {C.~L.}\ \bibnamefont {Broholm}},\
  }\bibfield  {title} {\enquote {\bibinfo {title} {{Static and dynamic
  $XY$-like short-range order in a frustrated magnet with exchange
  disorder}},}\ }\href {\doibase 10.1103/PhysRevB.93.014433} {\bibfield
  {journal} {\bibinfo  {journal} {Phys. Rev. B}\ }\textbf {\bibinfo {volume}
  {93}},\ \bibinfo {pages} {014433} (\bibinfo {year} {2016})}\BibitemShut
  {NoStop}%
\bibitem [{\citenamefont {Ross}\ \emph {et~al.}(2017)\citenamefont {Ross},
  \citenamefont {Brown}, \citenamefont {Cava}, \citenamefont {Krizan},
  \citenamefont {Nagler}, \citenamefont {Rodriguez-Rivera},\ and\ \citenamefont
  {Stone}}]{ross17}%
  \BibitemOpen
  \bibfield  {author} {\bibinfo {author} {\bibfnamefont {K.~A.}\ \bibnamefont
  {Ross}}, \bibinfo {author} {\bibfnamefont {J.~M.}\ \bibnamefont {Brown}},
  \bibinfo {author} {\bibfnamefont {R.~J.}\ \bibnamefont {Cava}}, \bibinfo
  {author} {\bibfnamefont {J.~W.}\ \bibnamefont {Krizan}}, \bibinfo {author}
  {\bibfnamefont {S.~E.}\ \bibnamefont {Nagler}}, \bibinfo {author}
  {\bibfnamefont {J.~A.}\ \bibnamefont {Rodriguez-Rivera}}, \ and\ \bibinfo
  {author} {\bibfnamefont {M.~B.}\ \bibnamefont {Stone}},\ }\bibfield  {title}
  {\enquote {\bibinfo {title} {{Single-ion properties of the
  ${S}_{\mathrm{eff}}$ = $\frac{1}{2}$ XY antiferromagnetic pyrochlores
  $\mathrm{Na}{A}^{\ensuremath{'}}{\mathrm{Co}}_{2}{\mathrm{F}}_{7}$
  (${A}^{\ensuremath{'}}={\mathrm{Ca}}^{2+}, {\mathrm{Sr}}^{2+}$)}},}\ }\href
  {\doibase 10.1103/PhysRevB.95.144414} {\bibfield  {journal} {\bibinfo
  {journal} {Phys. Rev. B}\ }\textbf {\bibinfo {volume} {95}},\ \bibinfo
  {pages} {144414} (\bibinfo {year} {2017})}\BibitemShut {NoStop}%
\bibitem [{\citenamefont {Ross}\ \emph {et~al.}(2012)\citenamefont {Ross},
  \citenamefont {Proffen}, \citenamefont {Dabkowska}, \citenamefont {Quilliam},
  \citenamefont {Yaraskavitch}, \citenamefont {Kycia},\ and\ \citenamefont
  {Gaulin}}]{ross12}%
  \BibitemOpen
  \bibfield  {author} {\bibinfo {author} {\bibfnamefont {K.~A.}\ \bibnamefont
  {Ross}}, \bibinfo {author} {\bibfnamefont {Th.}\ \bibnamefont {Proffen}},
  \bibinfo {author} {\bibfnamefont {H.~A.}\ \bibnamefont {Dabkowska}}, \bibinfo
  {author} {\bibfnamefont {J.~A.}\ \bibnamefont {Quilliam}}, \bibinfo {author}
  {\bibfnamefont {L.~R.}\ \bibnamefont {Yaraskavitch}}, \bibinfo {author}
  {\bibfnamefont {J.~B.}\ \bibnamefont {Kycia}}, \ and\ \bibinfo {author}
  {\bibfnamefont {B.~D.}\ \bibnamefont {Gaulin}},\ }\bibfield  {title}
  {\enquote {\bibinfo {title} {{Lightly stuffed pyrochlore structure of
  single-crystalline $\mathrm{{Yb}}_2\mathrm{{Ti}}_2\mathrm{{O}}_7$ grown by
  the optical floating zone technique}},}\ }\href {\doibase
  10.1103/PhysRevB.86.174424} {\bibfield  {journal} {\bibinfo  {journal} {Phys.
  Rev. B}\ }\textbf {\bibinfo {volume} {86}},\ \bibinfo {pages} {174424}
  (\bibinfo {year} {2012})}\BibitemShut {NoStop}%
\bibitem [{\citenamefont {Mostaed}\ \emph {et~al.}(2017)\citenamefont
  {Mostaed}, \citenamefont {Balakrishnan}, \citenamefont {Lees}, \citenamefont
  {Yasui}, \citenamefont {Chang},\ and\ \citenamefont {Beanland}}]{mostaed17}%
  \BibitemOpen
  \bibfield  {author} {\bibinfo {author} {\bibfnamefont {A.}~\bibnamefont
  {Mostaed}}, \bibinfo {author} {\bibfnamefont {G.}~\bibnamefont
  {Balakrishnan}}, \bibinfo {author} {\bibfnamefont {M.~R.}\ \bibnamefont
  {Lees}}, \bibinfo {author} {\bibfnamefont {Y.}~\bibnamefont {Yasui}},
  \bibinfo {author} {\bibfnamefont {L.~J.}\ \bibnamefont {Chang}}, \ and\
  \bibinfo {author} {\bibfnamefont {R.}~\bibnamefont {Beanland}},\ }\bibfield
  {title} {\enquote {\bibinfo {title} {{Atomic structure study of the
  pyrochlore
  $\mathrm{Y}{\mathrm{b}}_{2}\mathrm{T}{\mathrm{i}}_{2}{\mathrm{O}}_{7}$ and
  its relationship with low-temperature magnetic order}},}\ }\href {\doibase
  10.1103/PhysRevB.95.094431} {\bibfield  {journal} {\bibinfo  {journal} {Phys.
  Rev. B}\ }\textbf {\bibinfo {volume} {95}},\ \bibinfo {pages} {094431}
  (\bibinfo {year} {2017})}\BibitemShut {NoStop}%
\bibitem [{\citenamefont {Kermarrec}\ \emph {et~al.}(2014)\citenamefont
  {Kermarrec}, \citenamefont {Zorko}, \citenamefont {Bert}, \citenamefont
  {Colman}, \citenamefont {Koteswararao}, \citenamefont {Bouquet},
  \citenamefont {Bonville}, \citenamefont {Hillier}, \citenamefont {Amato},
  \citenamefont {van Tol}, \citenamefont {Ozarowski}, \citenamefont {Wills},\
  and\ \citenamefont {Mendels}}]{kermarrec14}%
  \BibitemOpen
  \bibfield  {author} {\bibinfo {author} {\bibfnamefont {E.}~\bibnamefont
  {Kermarrec}}, \bibinfo {author} {\bibfnamefont {A.}~\bibnamefont {Zorko}},
  \bibinfo {author} {\bibfnamefont {F.}~\bibnamefont {Bert}}, \bibinfo {author}
  {\bibfnamefont {R.~H.}\ \bibnamefont {Colman}}, \bibinfo {author}
  {\bibfnamefont {B.}~\bibnamefont {Koteswararao}}, \bibinfo {author}
  {\bibfnamefont {F.}~\bibnamefont {Bouquet}}, \bibinfo {author} {\bibfnamefont
  {P.}~\bibnamefont {Bonville}}, \bibinfo {author} {\bibfnamefont
  {A.}~\bibnamefont {Hillier}}, \bibinfo {author} {\bibfnamefont
  {A.}~\bibnamefont {Amato}}, \bibinfo {author} {\bibfnamefont
  {J.}~\bibnamefont {van Tol}}, \bibinfo {author} {\bibfnamefont
  {A.}~\bibnamefont {Ozarowski}}, \bibinfo {author} {\bibfnamefont {A.~S.}\
  \bibnamefont {Wills}}, \ and\ \bibinfo {author} {\bibfnamefont
  {P.}~\bibnamefont {Mendels}},\ }\bibfield  {title} {\enquote {\bibinfo
  {title} {{Spin dynamics and disorder effects in the $S=\frac{1}{2}$ kagome
  Heisenberg spin-liquid phase of kapellasite}},}\ }\href {\doibase
  10.1103/PhysRevB.90.205103} {\bibfield  {journal} {\bibinfo  {journal} {Phys.
  Rev. B}\ }\textbf {\bibinfo {volume} {90}},\ \bibinfo {pages} {205103}
  (\bibinfo {year} {2014})}\BibitemShut {NoStop}%
\bibitem [{\citenamefont {Han}\ \emph {et~al.}(2016)\citenamefont {Han},
  \citenamefont {Norman}, \citenamefont {Wen}, \citenamefont
  {Rodriguez-Rivera}, \citenamefont {Helton}, \citenamefont {Broholm},\ and\
  \citenamefont {Lee}}]{han16}%
  \BibitemOpen
  \bibfield  {author} {\bibinfo {author} {\bibfnamefont {T.-H.}\ \bibnamefont
  {Han}}, \bibinfo {author} {\bibfnamefont {M.~R.}\ \bibnamefont {Norman}},
  \bibinfo {author} {\bibfnamefont {J.-J.}\ \bibnamefont {Wen}}, \bibinfo
  {author} {\bibfnamefont {J.~A.}\ \bibnamefont {Rodriguez-Rivera}}, \bibinfo
  {author} {\bibfnamefont {J.~S.}\ \bibnamefont {Helton}}, \bibinfo {author}
  {\bibfnamefont {C.}~\bibnamefont {Broholm}}, \ and\ \bibinfo {author}
  {\bibfnamefont {Y.~S.}\ \bibnamefont {Lee}},\ }\bibfield  {title} {\enquote
  {\bibinfo {title} {{Correlated impurities and intrinsic spin-liquid physics
  in the kagome material herbertsmithite}},}\ }\href {\doibase
  10.1103/PhysRevB.94.060409} {\bibfield  {journal} {\bibinfo  {journal} {Phys.
  Rev. B}\ }\textbf {\bibinfo {volume} {94}},\ \bibinfo {pages} {060409}
  (\bibinfo {year} {2016})}\BibitemShut {NoStop}%
\bibitem [{\citenamefont {Norman}(2016)}]{norman16}%
  \BibitemOpen
  \bibfield  {author} {\bibinfo {author} {\bibfnamefont {M.~R.}\ \bibnamefont
  {Norman}},\ }\bibfield  {title} {\enquote {\bibinfo {title} {{{\it
  Colloquium}: Herbertsmithite and the search for the quantum spin liquid}},}\
  }\href {\doibase 10.1103/RevModPhys.88.041002} {\bibfield  {journal}
  {\bibinfo  {journal} {Rev. Mod. Phys.}\ }\textbf {\bibinfo {volume} {88}},\
  \bibinfo {pages} {041002} (\bibinfo {year} {2016})}\BibitemShut {NoStop}%
\bibitem [{\citenamefont {Li}\ \emph {et~al.}(2017)\citenamefont {Li},
  \citenamefont {Adroja}, \citenamefont {Bewley}, \citenamefont {Voneshen},
  \citenamefont {Tsirlin}, \citenamefont {Gegenwart},\ and\ \citenamefont
  {Zhang}}]{li17-PRL118}%
  \BibitemOpen
  \bibfield  {author} {\bibinfo {author} {\bibfnamefont {Y.}~\bibnamefont
  {Li}}, \bibinfo {author} {\bibfnamefont {D.}~\bibnamefont {Adroja}}, \bibinfo
  {author} {\bibfnamefont {R.~I.}\ \bibnamefont {Bewley}}, \bibinfo {author}
  {\bibfnamefont {D.}~\bibnamefont {Voneshen}}, \bibinfo {author}
  {\bibfnamefont {A.~A.}\ \bibnamefont {Tsirlin}}, \bibinfo {author}
  {\bibfnamefont {P.}~\bibnamefont {Gegenwart}}, \ and\ \bibinfo {author}
  {\bibfnamefont {Q.}~\bibnamefont {Zhang}},\ }\bibfield  {title} {\enquote
  {\bibinfo {title} {{Crystalline Electric-Field Randomness in the Triangular
  Lattice Spin-Liquid ${\mathrm{YbMgGaO}}_{4}$}},}\ }\href {\doibase
  10.1103/PhysRevLett.118.107202} {\bibfield  {journal} {\bibinfo  {journal}
  {Phys. Rev. Lett.}\ }\textbf {\bibinfo {volume} {118}},\ \bibinfo {pages}
  {107202} (\bibinfo {year} {2017})}\BibitemShut {NoStop}%
\bibitem [{\citenamefont {Bellier-Castella}\ \emph {et~al.}(2001)\citenamefont
  {Bellier-Castella}, \citenamefont {Gingras}, \citenamefont {Holdsworth},\
  and\ \citenamefont {Moessner}}]{bellier01}%
  \BibitemOpen
  \bibfield  {author} {\bibinfo {author} {\bibfnamefont {L.}~\bibnamefont
  {Bellier-Castella}}, \bibinfo {author} {\bibfnamefont {M.~J.~P.}\
  \bibnamefont {Gingras}}, \bibinfo {author} {\bibfnamefont {P.~C.~W.}\
  \bibnamefont {Holdsworth}}, \ and\ \bibinfo {author} {\bibfnamefont
  {R.}~\bibnamefont {Moessner}},\ }\bibfield  {title} {\enquote {\bibinfo
  {title} {{Frustrated order by disorder: The pyrochlore anti-ferromagnet with
  bond disorder}},}\ }\href {\doibase 10.1139/p01-098} {\bibfield  {journal}
  {\bibinfo  {journal} {Can. J. Phys.}\ }\textbf {\bibinfo {volume} {79}},\
  \bibinfo {pages} {1365--1371} (\bibinfo {year} {2001})}\BibitemShut {NoStop}%
\bibitem [{\citenamefont {Saunders}\ and\ \citenamefont
  {Chalker}(2007)}]{saunders07}%
  \BibitemOpen
  \bibfield  {author} {\bibinfo {author} {\bibfnamefont {T.~E.}\ \bibnamefont
  {Saunders}}\ and\ \bibinfo {author} {\bibfnamefont {J.~T.}\ \bibnamefont
  {Chalker}},\ }\bibfield  {title} {\enquote {\bibinfo {title} {{Spin Freezing
  in Geometrically Frustrated Antiferromagnets with Weak Disorder}},}\ }\href
  {\doibase 10.1103/PhysRevLett.98.157201} {\bibfield  {journal} {\bibinfo
  {journal} {Phys. Rev. Lett.}\ }\textbf {\bibinfo {volume} {98}},\ \bibinfo
  {pages} {157201} (\bibinfo {year} {2007})}\BibitemShut {NoStop}%
\bibitem [{\citenamefont {Tsunetsugu}(2001)}]{tsunetsugu01}%
  \BibitemOpen
  \bibfield  {author} {\bibinfo {author} {\bibfnamefont {H.}~\bibnamefont
  {Tsunetsugu}},\ }\bibfield  {title} {\enquote {\bibinfo {title}
  {{Antiferromagnetic Quantum Spins on the Pyrochlore Lattice}},}\ }\href
  {\doibase 10.1143/JPSJ.70.640} {\bibfield  {journal} {\bibinfo  {journal} {J.
  Phys. Soc. Jpn}\ }\textbf {\bibinfo {volume} {70}},\ \bibinfo {pages}
  {640--643} (\bibinfo {year} {2001})}\BibitemShut {NoStop}%
\bibitem [{\citenamefont {Ueda}\ \emph {et~al.}(2016)\citenamefont {Ueda},
  \citenamefont {Akagi},\ and\ \citenamefont {Shannon}}]{ueda16-PRA93}%
  \BibitemOpen
  \bibfield  {author} {\bibinfo {author} {\bibfnamefont {H.~T.}\ \bibnamefont
  {Ueda}}, \bibinfo {author} {\bibfnamefont {Y.}~\bibnamefont {Akagi}}, \ and\
  \bibinfo {author} {\bibfnamefont {N.}~\bibnamefont {Shannon}},\ }\bibfield
  {title} {\enquote {\bibinfo {title} {Quantum solitons with emergent
  interactions in a model of cold atoms on the triangular lattice},}\ }\href
  {\doibase 10.1103/PhysRevA.93.021606} {\bibfield  {journal} {\bibinfo
  {journal} {Phys. Rev. A}\ }\textbf {\bibinfo {volume} {93}},\ \bibinfo
  {pages} {021606} (\bibinfo {year} {2016})}\BibitemShut {NoStop}%
\bibitem [{\citenamefont {Singh}\ and\ \citenamefont
  {Oitmaa}()}]{rajiv-unpublished}%
  \BibitemOpen
  \bibfield  {author} {\bibinfo {author} {\bibfnamefont {R.~R.~P.}\
  \bibnamefont {Singh}}\ and\ \bibinfo {author} {\bibfnamefont
  {J.}~\bibnamefont {Oitmaa}},\ }\href@noop {} {\bibinfo  {journal}
  {unpublished}\ }\BibitemShut {NoStop}%
\end{thebibliography}%


\end{document}